\documentclass[12pt,a4paper]{article}

\usepackage{amsmath, amssymb}
\usepackage{geometry}
\usepackage{booktabs}
\usepackage{graphicx}
\usepackage{physics}
\usepackage{array}
\usepackage{xcolor}
\usepackage{tikz}
\usetikzlibrary{calc}
\usepackage{float}
\usepackage[numbers,sort&compress]{natbib}
\usepackage{pdflscape}
\usepackage[utf8]{inputenc}
\usepackage{authblk}
\usepackage{longtable}
\usepackage{enumitem}
\usepackage{multirow}
\usepackage{hyperref}
\newcommand{\doi}[1]{\url{https://doi.org/#1}}

\hypersetup{
    colorlinks=true,
    linkcolor=blue!60!black,
    urlcolor=blue!60!black,
    citecolor=blue!60!black
}

\title{\textbf{Producing and Measuring Circularly Polarized X-rays:\\
A Jones and Mueller--Stokes Treatment of Crystal Phase Plates}\\
Everything you always wanted to know about X-ray polarization, but were afraid to ask}
\author[1*]{Manuel Sanchez del Rio}
\author[1]{Carsten Detlefs}

\affil[1]{ESRF, Grenoble, France}

\affil[*]{\textit{\small srio@esrf.eu}}

\date{\today}

\begin{document}

\maketitle

\begin{abstract}
Converting a linearly polarized X-ray beam into a circularly polarized one, and
then verifying the polarization state actually delivered, are standard operations
at synchrotron radiation facilities. The key optical element is a
\textbf{quarter-wave plate (QWP)}, a birefringent crystal that introduces a
$90^\circ$ ($\lambda/4$) phase retardation between the two orthogonal components
of the electric field.

This paper presents a self-contained theoretical treatment of X-ray
polarization optics. We first establish the Jones and Mueller--Stokes
formalisms and their connection via the Pauli matrices, fixing the sign and
handedness conventions explicitly and reconciling them with those used
elsewhere in the literature. From there we derive the Jones and Mueller
matrices of a general retarder at an arbitrary orientation angle $\chi$ ---
including the QWP that converts horizontal linear polarization into
circular polarization of either handedness --- and of linear polarizers,
and describe the experimental Stokes polarimetry protocol used at
synchrotron beamlines. These results are then combined to treat the general
crystal phase plate, characterized by two complex reflection (or
transmission) amplitudes $r_s$ and $r_p$, obtaining the degree of circular
polarization delivered for an arbitrary incident beam as a function of
$\chi$ and the phase retardation $\delta$. Finally, we turn to
the practical implementation of these elements with perfect-crystal
diffraction: the residual polarization impurity of a Brewster-angle
polarizer and the role of ray tracing in evaluating it, and the choice of
diffraction geometry --- Bragg or Laue, diffracted or transmitted beam ---
for obtaining a significant, controllable phase shift. The formalism
presented here is the mathematical background of the treatment of
polarization in the ray-tracing code \texttt{SHADOW4}.
\end{abstract}

\newpage

{\footnotesize\tableofcontents}

\newpage

\section{Introduction}
\label{sec:introduction}

Synchrotron X-ray beams produced by planar undulators, or by bending magnets
in the orbit plane, are naturally linearly polarized in the horizontal plane
of the storage ring. Many X-ray
spectroscopy and scattering experiments --- X-ray magnetic circular dichroism (XMCD),
resonant magnetic X-ray scattering~\cite{blume1988,blume1994}, chiral diffraction
--- require circularly polarized beams. The standard approach is to insert a
\textbf{crystal phase plate} in the beam, which acts as the X-ray analogue of a
quarter-wave plate~\cite{detlefs2012,hirano1993,giles1994}.

To convert horizontal linear polarization to circular polarization, the crystal
must be oriented with its birefringence axis at $\chi = 45^\circ$ to the incoming
polarization direction, so that the $\sigma$- and $\pi$-polarization components
are equally excited; tuning the crystal angle then gives them a relative phase
shift of $\pi/2$. The optical layout is:

\begin{center}
\begin{tikzpicture}[>=stealth, thick, node distance=3.5cm]
  \node (source) [draw, rounded corners, fill=gray!15,
        minimum width=2.2cm, minimum height=0.8cm] {X-ray source};
  \node (lp) [draw, rounded corners, fill=blue!15,
        minimum width=2.2cm, minimum height=0.8cm, right of=source]
        {Linear (H) beam};
  \node (qwp) [draw, rounded corners, fill=green!15,
        minimum width=2.2cm, minimum height=0.8cm, right of=lp]
        {Crystal @ $\chi=45^\circ$};
  \node (out) [draw, rounded corners, fill=orange!15,
        minimum width=2.2cm, minimum height=0.8cm, right of=qwp]
        {Circular (RCP)$^{\dagger}$};
  \draw[->] (source) -- (lp);
  \draw[->] (lp)     -- (qwp);
  \draw[->] (qwp)    -- (out);
\end{tikzpicture}

{\small $^{\dagger}$For a retardation $\delta=+\pi/2$; left circular (LCP) for $\delta=-\pi/2$.}
\end{center}

The optical properties of perfect crystals offer a powerful mechanism for
manipulating polarization through birefringence near a Bragg reflection. The
$\sigma$- and $\pi$-polarization components, along $\hat{x}$ and $\hat{y}$ of
the crystal frame, experience different complex amplitudes $r_s$ and $r_p$
(reflection or transmission), whose relative magnitude and phase can be
tuned by rocking the crystal angle. The phase retardation $\delta = \arg(r_p) -
\arg(r_s)$ reaches $\pm\pi/2$ in the transmission (forward-diffracted)
geometry at an angular offset from the Bragg peak set by the crystal
thickness (e.g., about three Darwin half-widths for a 50\,$\mu$m Si 220
plate at 10\,keV, Section~\ref{sec:phase_shifter_geometry}), providing the ideal quarter-wave condition.

In practice, thin (tens to a few hundred micrometres) perfect crystals (silicon or diamond)
are used as phase plates at the hard X-ray beamlines of many synchrotron
sources, for example ESRF~\cite{bouchenoire2003}, APS~\cite{lang1995},
SPring-8~\cite{suzuki2013}, the Swiss Light Source~\cite{apseros2024}, Diamond Light Source~\cite{diamondI16} and
PETRA~III~\cite{francoual2013}, because they can be tuned to a phase shift
close to $\pi/2$ with high and nearly equal $\sigma$ and $\pi$ transmission,
and offer excellent mechanical stability~\cite{giles1994,hirano1993}.

\section{Jones and Mueller--Stokes Calculus Treatment}

\subsection{Jones Vector and Stokes Parameters}

We start from the physical, real-valued electric field of a fully coherent
monochromatic plane wave propagating along $\hat{z}$, written exactly as in
Eq.~(1) of Ref.~\cite{detlefs2012}:
\begin{equation}
    \boldsymbol{\mathcal{E}}(t, \mathbf{r})
    = \Re\!\left[\left(E_x\,\hat{x} + E_y\,\hat{y}\right)
      e^{-i(\omega t - k z)}\right]
    \label{eq:physical_wave}
\end{equation}
where the coordinate system is right-handed,
$\hat{x}\times\hat{y}=\hat{z}=\hat{k}$, and the complex amplitudes carry the
absolute phases, $E_{x,y} = |E_{x,y}|\,e^{i\phi_{x,y}}$. Writing
$\tau = \omega t - kz$ for the phase, the real field components are
\begin{equation}
    \mathcal{E}_x(t) = |E_x|\cos(\tau - \phi_x), \qquad
    \mathcal{E}_y(t) = |E_y|\cos(\tau - \phi_y)
    \label{eq:real_components}
\end{equation}
which reproduces Eq.~(2) of Ref.~\cite{detlefs2012} with $V_1 = E_x$,
$V_2 = E_y$, $\hat{\epsilon}_1 = \hat{x}$, $\hat{\epsilon}_2 = \hat{y}$.

The choice of the time factor $e^{-i\omega t}$ in
Eq.~(\ref{eq:physical_wave}) is a genuine convention (the complex conjugate
representation, $e^{+i\omega t}$, is equally common in the literature and
describes the same physical field). It must be stated explicitly because
every statement about the \emph{sign} of the circular polarization --- and
therefore about handedness --- ultimately traces back to it: conjugating the
representation flips the sign of $S_3$ defined below while leaving $S_0$,
$S_1$, $S_2$ unchanged.

The \textbf{Jones vector}~\cite{jones1941} is the complex envelope of
Eq.~(\ref{eq:physical_wave}), i.e.\ the pair of complex amplitudes:
\begin{equation}
    \mathbf{E} = \begin{pmatrix} E_x \\ E_y \end{pmatrix}
    = \begin{pmatrix} |E_x|\,e^{i\phi_x} \\ |E_y|\,e^{i\phi_y} \end{pmatrix}
\end{equation}
where $E_x$ and $E_y$ are the complex amplitudes of the horizontal ($\hat{x}$)
and vertical ($\hat{y}$) field components, and $\phi_x$, $\phi_y$ are their
absolute phases.

The four real, observable \textbf{Stokes parameters}~\cite{born2002,collett1993} are defined as:
\begin{align}
    S_0 &= |E_x|^2 + |E_y|^2 \label{eq:S0def}\\
    S_1 &= |E_x|^2 - |E_y|^2 \label{eq:S1def}\\
    S_2 &= 2\,\mathrm{Re}(E_x E_y^*)
     & &= 2\,|E_x| |E_y|  \cos(\phi_y - \phi_x)
    \label{eq:S2def}\\
    S_3 &= 2\,\mathrm{Im}(E_x^* E_y) 
    & &= 2\,|E_x| |E_y| \sin(\phi_y - \phi_x).
    \label{eq:S3def}
\end{align}

$S_0$ is the total intensity, $S_1$ measures linear horizontal/vertical
polarization preference, $S_2$ measures $\pm45^\circ$ linear polarization, and
$S_3$ measures circular polarization content.
Note that the Stokes parameters depend only on the relative phase, $\Delta \phi = \phi_y - \phi_x$, of the Jones vector.

A beam described by a single Jones vector is always fully polarized, $S_0^2 = S_1^2+S_2^2+S_3^2$.

\begin{table}[H]
\centering
\begin{tabular}{clll}
\toprule
\textbf{Parameter} & \textbf{Measures} & \multicolumn{2}{l}{\textbf{Extreme values}} \\
\midrule
$S_0$ & Total intensity       & $S_0 \geq 0$ always & conserved through lossless optics \\
$S_1$ & H vs.\ V linear       & $+S_0$: horiz.\ (H) & $-S_0$: vertical (V) \\
$S_2$ & $\pm45^\circ$ linear  & $+S_0$: $+45^\circ$ & $-S_0$: $-45^\circ$ \\
$S_3$ & Circular content      & $+S_0$: LCP         & $-S_0$: RCP \\
\bottomrule
\end{tabular}
\caption{Physical interpretation of the Stokes parameters. $S_1$, $S_2$, $S_3$ are not normalized: they range from $-S_0$ to $+S_0$ (Eq.~\ref{eq:DOCP_def}).}
\label{tab:stokes}
\end{table}

In addition to the Stokes parameters $S_{0,1,2,3}$, we define the normalized
\textbf{Poincar\'{e}--Stokes parameters}, as in Refs.~\cite{detlefs2012,mazzoli2007},
\begin{equation}
    P_1 = \frac{S_1}{S_0}, \qquad
    P_2 = \frac{S_2}{S_0}, \qquad
    P_3 = \frac{S_3}{S_0}, \qquad -1 \le P_k \le 1 .
    \label{eq:Pdef}
\end{equation}
They describe the polarization state independently of the intensity, whereas
the Stokes parameters are proportional to the beam intensity $S_0$. For the
lossless retarders of Section~\ref{sec:general_retarders} the intensity is
conserved and we simply set $S_0 = 1$, so that $S_k = P_k$. Polarizers
(Section~\ref{sec:general_polarizers}), polarimetry
(Section~\ref{sec:experimental}) and crystals
(Section~\ref{sec:crystal_plate}) change the intensity, and there the two
sets must be kept apart.

\subsection{Pauli Matrix Representation and Density Matrix}

Experimental beams are typically an incoherent superposition of many Jones vectors. The Stokes parameters are then the statistical averages of Eqs.~(\ref{eq:S0def})--(\ref{eq:S3def}):

\begin{align}
    S_0 &= \left< |E_x|^2 + |E_y|^2  \right> \label{eq:S0def2}\\
    S_1 &= \left<|E_x|^2 - |E_y|^2 \right> \label{eq:S1def2}\\
    S_2 &= \left< 2\,\mathrm{Re}(E_x E_y^*) \right>
     & &= \left< 2\,|E_x| |E_y|  \cos(\phi_y - \phi_x) \right>
    \label{eq:S2def2}\\
    S_3 &= \left< 2\,\mathrm{Im}(E_x^* E_y) \right>
    & &= \left< 2\,|E_x| |E_y| \sin(\phi_y - \phi_x) \right>,
    \label{eq:S3def2}
\end{align}
where the average is taken over all Jones vectors in the beam.

Mathematically, this can be expressed in the form of a  \textbf{density matrix} $\boldsymbol{\rho} = \langle\mathbf{E}\mathbf{E}^\dagger\rangle$
(called the \emph{coherency matrix} in classical optics~\cite{born2002}),
where $\langle\cdots\rangle$ is a time or ensemble average. This
provides a $2\times2$ Hermitian representation of the polarization state:
\begin{equation}
    \boldsymbol{\rho} =
    \begin{pmatrix}
        \langle E_x E_x^*\rangle & \langle E_x E_y^*\rangle \\
        \langle E_y E_x^*\rangle & \langle E_y E_y^*\rangle
    \end{pmatrix}.
\end{equation}

The four \textbf{Pauli matrices} (including the $2\times2$ identity) form a
complete orthogonal basis, with real coefficients, for all $2\times2$
Hermitian matrices:
\begin{equation}
    \boldsymbol{\sigma}_0 = \begin{pmatrix}1&0\\0&1\end{pmatrix},\quad
    \boldsymbol{\sigma}_1 = \begin{pmatrix}1&0\\0&-1\end{pmatrix},\quad
    \boldsymbol{\sigma}_2 = \begin{pmatrix}0&1\\1&0\end{pmatrix},\quad
    \boldsymbol{\sigma}_3 = \begin{pmatrix}0&-i\\i&0\end{pmatrix}
    \label{eq:Pauli_def}
\end{equation}

 The density matrix can then be written as

\begin{equation}
    \boldsymbol{\rho}
    = \frac{1}{2}\sum_{k=0}^{3} S_k\,\boldsymbol{\sigma}_k
\end{equation}

The Pauli matrices satisfy the orthogonality relation
$\mathrm{Tr}(\boldsymbol{\sigma}_m\boldsymbol{\sigma}_n) = 2\delta_{mn}$
and the product rule
$\boldsymbol{\sigma}_j\boldsymbol{\sigma}_k = \delta_{jk}\boldsymbol{\sigma}_0
+ i\varepsilon_{jkl}\boldsymbol{\sigma}_l$ ($j,k,l \in \{1,2,3\}$),
where $\varepsilon_{jkl}$ is the Levi-Civita symbol.

The Stokes parameters are then the projections of $\boldsymbol{\rho}$ onto the
Pauli basis via the trace inner product:
\begin{equation}
    {S_k = \mathrm{Tr}(\boldsymbol{\sigma}_k\,\boldsymbol{\rho}), \quad k=0,1,2,3}
    \label{eq:Sk_trace}
\end{equation}

This compact formula reproduces Eqs.~(\ref{eq:S0def})--(\ref{eq:S3def}) exactly
and is the foundation of the Jones-to-Mueller conversion used throughout this
document (see Section~\ref{sec:JM_principle}). Conversely, expanding the sum
$\boldsymbol{\rho} = \tfrac{1}{2}\sum_k S_k\boldsymbol{\sigma}_k$ gives the
density matrix explicitly in terms of the Stokes parameters:
\begin{equation}
    \boldsymbol{\rho} = \frac{1}{2}
    \begin{pmatrix}
        S_0 + S_1 & S_2 - i S_3 \\
        S_2 + i S_3 & S_0 - S_1
    \end{pmatrix}
    \label{eq:rho_from_Stokes}
\end{equation}
a form that provides a quick consistency check of all sign conventions: the
imaginary part of the off-diagonal element carries $S_3$, with
$\rho_{21} = \langle E_y E_x^*\rangle = \tfrac{1}{2}(S_2 + iS_3)$.

An example of a polarization state that cannot be described by a single Jones vector is an unpolarized beam, $\mathbf{S}_{\mathrm{unpolarized}}=(1,0,0,0)^T$ or $\boldsymbol{\rho}_{\mathrm{unpolarized}}=\dfrac{1}{2}\boldsymbol{\sigma}_0$.
At least two Jones vectors are needed to produce this, for example two mutually incoherent, equal-intensity orthogonal polarizations such as horizontal and vertical, $\pm 45^\circ$ linear, or left- and right-circular,

\begin{align}
    \mathbf{S}_{\mathrm{unpolarized}}
    &=
    \frac{1}{2} \mathbf{S}_{\mathrm{horizontal}}
    +
    \frac{1}{2} \mathbf{S}_{\mathrm{vertical}};
    &
    \begin{pmatrix} 1 \\ 0 \\ 0 \\ 0 \end{pmatrix}
    &=
    \frac{1}{2}
    \begin{pmatrix} 1 \\ 1 \\ 0 \\ 0\end{pmatrix}
    +
    \frac{1}{2}
    \begin{pmatrix} 1 \\ -1 \\ 0 \\ 0\end{pmatrix}
    \\
    &=
    \frac{1}{2} \mathbf{S}_{\mathrm{45}^\circ}
    +
    \frac{1}{2} \mathbf{S}_{\mathrm{-45}^\circ}
    &    
    &=
    \frac{1}{2}
    \begin{pmatrix} 1 \\ 0 \\ 1 \\ 0\end{pmatrix}
    +
    \frac{1}{2}
    \begin{pmatrix} 1 \\ 0 \\ -1 \\ 0\end{pmatrix}
    \\
    &=
    \frac{1}{2} \mathbf{S}_{\mathrm{LCP}}
    +
    \frac{1}{2} \mathbf{S}_{\mathrm{RCP}}
    &
    &=
    \frac{1}{2}
    \begin{pmatrix} 1 \\ 0 \\ 0 \\ 1\end{pmatrix}
    +
    \frac{1}{2}
    \begin{pmatrix} 1 \\ 0 \\ 0 \\ -1\end{pmatrix}
\end{align}

\subsection{Sign and Handedness Conventions}
\label{sec:sign_convention}

The link between the algebraic sign of $S_3$ and the physical handedness of
circular polarization is fixed by a chain of independent conventions ---
the time factor, the definition of $S_3$ itself, the Pauli-matrix assignment,
and the naming of ``left''/``right'' handedness --- none of which is universal
in the literature, and the resulting ambiguities are a well-documented source
of errors. This section fixes that chain once, explicitly, and verifies it
in two independent ways: algebraically (via the Pauli trace) and geometrically
(via the time-domain field vector).

\paragraph{The sign of $S_3$.}
Both $S_3 = 2\,\mathrm{Im}(E_x^*E_y)$ and its negative,
$2\,\mathrm{Im}(E_xE_y^*)$, appear in standard references, differing by an
overall sign; conjugating the analytic-signal representation
($e^{-i\omega t} \leftrightarrow e^{+i\omega t}$) describes the same physical
field while flipping this sign. Here we adopt $S_3 = 2\,\mathrm{Im}(E_x^*E_y)$
(Eq.~\ref{eq:S3def}) because, with the Pauli matrices of
Eq.~(\ref{eq:Pauli_def}), it is the choice for which the Pauli trace
projection $S_k = \mathrm{Tr}(\boldsymbol{\sigma}_k \boldsymbol{\rho})$ of
Eq.~(\ref{eq:Sk_trace}) holds for all $k$; all Mueller matrices in this
paper are built on it.

\paragraph{A cross-check on $\boldsymbol{\sigma}_3$.}
Ref.~\cite{detlefs2012}, whose handedness convention this paper follows,
prints the third Pauli matrix as
$\bigl[\begin{smallmatrix}0&i\\-i&0\end{smallmatrix}\bigr]$, i.e.\ with the
sign opposite to Eq.~(\ref{eq:Pauli_def}). Its own definition of $P_3$ (its
Eq.~(10)) and its own worked example are reproduced by the matrix adopted
here, not by the one printed. For the left circularly polarized state
$\mathbf{V} = (1,i)^T/\sqrt{2}$, to which that reference explicitly assigns
$P_3 = +1$, its Eq.~(10) returns $+1$; the trace formula
$P_3 = \mathrm{Tr}(\boldsymbol{\sigma}_3\boldsymbol{\rho})/S_0$ evaluated with
$\boldsymbol{\sigma}_3$ of Eq.~(\ref{eq:Pauli_def}) also returns $+1$, whereas
the same formula evaluated with the printed matrix returns $-1$. The same
situation recurs for the rotation matrix of that reference
(Section~\ref{sec:rotation}): in both cases the conventions used here agree
with its defining equations, its results and its worked examples, and it is
the individual printed matrix that stands apart. This is precisely why the
chain is fixed here by explicit equations rather than by citation.

\paragraph{Polar form.}
Writing the complex amplitudes in polar form, $E_{x,y} = |E_{x,y}|\,e^{i\phi_{x,y}}$,
and defining the \textbf{relative phase} $\Delta\phi \equiv \phi_y - \phi_x$, the
last two Stokes parameters take the equivalent, and physically more transparent,
form:
\begin{equation}
    S_2 = 2|E_x||E_y|\cos\Delta\phi, \qquad
    S_3 = 2|E_x||E_y|\sin\Delta\phi
    \label{eq:S23_polar}
\end{equation}
($S_0$ and $S_1$ depend on the amplitudes $|E_x|,|E_y|$ alone and are
independent of $\Delta\phi$.) Hence $S_3>0$ if and only if
$0 < \Delta\phi < \pi$, with the purely circular case at $\Delta\phi = +\pi/2$
and $|E_x|=|E_y|$. Throughout the paper $\Delta\phi$ denotes only this
relative phase of the two field components of the \emph{beam}, measured in
the laboratory axes $\hat{x}$, $\hat{y}$. The phase delay that an optical
\emph{element} imposes between the field components along its own axes is
its \textbf{retardation}, always written $\delta$: for the ideal retarder of
Section~\ref{sec:retarder} and, as $\delta = \arg(r_p) - \arg(r_s)$
(Eq.~\ref{eq:Dphi_def}), for the crystal of Section~\ref{sec:crystal_plate}.
The two coincide only when the element's axes are the laboratory axes. A
simple example shows the difference. A horizontally polarized beam has
$E_y = 0$, so its $\Delta\phi$ is undefined (and irrelevant, $S_2 = S_3 =
0$). A quarter-wave plate ($\delta = +\pi/2$) with its axes at $\chi =
45^\circ$ splits this field into two equal components along its own axes,
delays one of them by $\pi/2$, and recombines them; in the laboratory axes
the output (Section~\ref{sec:QWP45}) has $|E_x| = |E_y|$ and $\Delta\phi =
-\pi/2$, i.e.\ RCP. The same plate with its axes along $\hat{x}$, $\hat{y}$
($\chi = 0$) would leave the horizontal beam unchanged, whatever $\delta$.

\paragraph{Anchor state and handedness naming.}
With this choice, and with the time convention of Eq.~(\ref{eq:physical_wave}),
the state $\mathbf{E} = (1, i)^T/\sqrt{2}$ ($\Delta\phi=+\pi/2$) has
$S_3 = +1$: its field vector rotates counter-clockwise when looking toward
the source, i.e.\ it is \emph{left} circular polarization (LCP) in the
classical optics naming, exactly as in Ref.~\cite{detlefs2012}. Conversely
$\mathbf{E} = (1, -i)^T/\sqrt{2}$ has $S_3 = -1$ and is right circular
polarization (RCP).

\paragraph{Geometric verification.}
Equation~(\ref{eq:S23_polar}) lets the handedness claim above be checked
independently of any specific state. Set $\phi_x = 0$ (the global phase is
irrelevant) and $z=0$ in Eq.~(\ref{eq:real_components}); this gives the
sense of rotation of the field, which Eq.~(\ref{eq:S23_polar}) links to the
sign of $S_3$:
\begin{equation}
    \mathcal{E}_x(t) = |E_x|\cos\omega t, \qquad
    \mathcal{E}_y(t) = |E_y|\cos(\omega t - \Delta\phi)
\end{equation}
For $\Delta\phi = +\pi/2$, $\mathcal{E}_y(t) = |E_y|\sin\omega t$: the tip of the
field vector starts on $+\hat{x}$ at $t=0$ and sweeps toward $+\hat{y}$ as $t$
increases, i.e.\ a counter-clockwise rotation in the standard mathematical sense,
for the $(x,y)$ plane drawn with $x$ to the right, $y$ upward, and $z$ (the
propagation direction $\hat{k}$) pointing \emph{out of the page toward the
viewer}. This is exactly the geometry of Fig.~1 of Ref.~\cite{detlefs2012}, where
$\hat{k} = \hat{e}_3$ points toward the reader's eye. Because $\hat{k}$ points out
of the page \emph{at} the viewer, the viewer necessarily stands downstream of the
source and looks back \emph{toward} it (along $-\hat{z}$): the figure, read as
drawn, already depicts the ``looking toward the source'' vantage point, with no
additional flip required. Hence $\Delta\phi = +\pi/2$ (equivalently $S_3=+1$) is
counter-clockwise looking toward the source, i.e.\ LCP --- confirming the
handedness assignment above from the time-domain field of
Eq.~(\ref{eq:real_components}), independently of the Pauli/density-matrix
argument used to fix the sign of $S_3$ itself.

\paragraph{Conventions in the literature.}
This dependence on viewing direction is exactly why the convention must be
stated explicitly, and it is by no means universal. Handedness \emph{naming}
is an independent choice on top of the sign of $S_3$: classical optics names
the rotation sense as seen by an observer positioned downstream, with $\hat k$
pointing at them, looking back toward the source, as above~\cite{born2002},
while the IEEE/antenna and particle-physics helicity conventions instead place
the observer at the source looking \emph{along} $\hat{k}$ --- the same ``up''
direction but the opposite facing, which mirrors left and right and reverses
the sense of rotation, relabelling the identical physical state with the
opposite name. Jackson~\cite{jackson1999} explicitly warns that his
``positive helicity'' corresponds to \emph{left} circular polarization in the
traditional optics nomenclature. The representation freedom in the
Pauli-matrix assignment was analyzed early on by McMaster~\cite{mcmaster1961},
and in radio astronomy the confusion was severe enough that the IAU fixed the
sign of Stokes $V$ by resolution, whose reconciliation with the IEEE
definitions required a dedicated paper~\cite{hamaker1996}. In the X-ray
magnetic scattering literature, the Blume--Gibbs density-matrix
formulation~\cite{blume1988} orders the polarization components differently
again. For these reasons no single formula can be quoted safely without its
companions; in this paper the complete chain is fixed once ---
Eq.~(\ref{eq:physical_wave}), Eqs.~(\ref{eq:S0def})--(\ref{eq:S3def}),
Eq.~(\ref{eq:Pauli_def}), the geometry of Fig.~1 of Ref.~\cite{detlefs2012},
and the anchor state $(1, i)^T/\sqrt{2} \mapsto S_3 = +1$ (LCP) --- and
everything else in this paper is derived from it.

\subsection{Degree of Polarization and Poincar\'{e} Sphere}

The \textbf{degree of polarization} (DOP) quantifies how far a beam departs from
a completely unpolarized state:
\begin{equation}
    \mathrm{DOP} = \frac{\sqrt{S_1^2 + S_2^2 + S_3^2}}{S_0}  = \sqrt{P_1^2+P_2^2+P_3^2}\in [0,\,1]
\end{equation}
$\mathrm{DOP}=1$ for fully polarized light; $\mathrm{DOP}=0$ for
completely unpolarized light. The degrees of linear and circular polarization are:
\begin{equation}
    \mathrm{DOLP} = \frac{\sqrt{S_1^2+S_2^2}}{S_0}= \sqrt{P_1^2+P_2^2}, \qquad
    \mathrm{DOCP} \equiv P_C = \frac{S_3}{S_0} = P_3 \in [-1,\,+1]
    \label{eq:DOCP_def}
\end{equation}

\begin{figure}[H]
\centering
\begin{tikzpicture}[scale=1.05, >=stealth, thick]
    \shade[ball color=gray!30] (0,0) circle (2.2);
    \draw (0,0) circle (2.2);

    \draw[thick] (0,0) ellipse (2.2 and 0.75);

    \draw[->] (0,-2.6) -- (0,2.65) node[above] {$P_3$};
    \draw[->] (-2.6,0) -- (2.7,0) node[right] {$P_1$};
    \draw[->] (0,0) -- (1.9,0.65) node[above right] {$P_2$};

    \fill (2.2,0) circle (1.3pt);
    \node[right] at (2.2,0.3) {H};
    \fill (-2.2,0) circle (1.3pt);
    \node[left] at (-2.2,0.3) {V};
    \fill (1.56,0.53) circle (1.3pt);
    \node[below=2pt] at (1.56,0.53) {$+45^\circ$};
    \fill (-1.56,-0.53) circle (1.3pt);
    \node[below] at (-1.56,-0.53) {$-45^\circ$};

    \draw[<->, red] (1.95,-0.35) -- (2.45,-0.35);
    \draw[<->, red] (1.42,0.39) -- (1.70,0.67);

    \fill (0,2.2) circle (1.5pt);
    \draw[->, red] ([shift=(40:0.22)]0.75,2.3) arc (40:340:0.22);
    \node[right] at (1.0,2.3) {\textbf{LCP} ($P_3=+1$)};

    \fill (0,-2.2) circle (1.5pt);
    \draw[->, red] ([shift=(340:0.22)]0.75,-2.3) arc (340:40:0.22);
    \node[right] at (1.0,-2.3) {\textbf{RCP} ($P_3=-1$)};
\end{tikzpicture}
\caption{The Poincar\'{e} sphere (unit radius, axes $P_1$, $P_2$, $P_3$ of
Eq.~\ref{eq:Pdef}) in the conventions of
Section~\ref{sec:sign_convention}. The equator ($P_3=0$) contains all linear
polarization states, with $+P_1$ (H, $0^\circ$) opposite $-P_1$ (V,
$90^\circ$), and $+P_2$ ($+45^\circ$) opposite $-P_2$ ($-45^\circ$); the red
double-headed arrows mark the field oscillation direction for the H and
$+45^\circ$ states. The north pole ($P_3=+1$) is LCP, with the field vector
rotating counter-clockwise looking toward the source; the south pole
($P_3=-1$) is RCP, rotating clockwise --- the small red arcs are a mnemonic
for the sense of rotation at each pole. Other references place these axes
differently (see the discussion of conventions in
Section~\ref{sec:sign_convention}).}
\label{fig:poincare}
\end{figure}
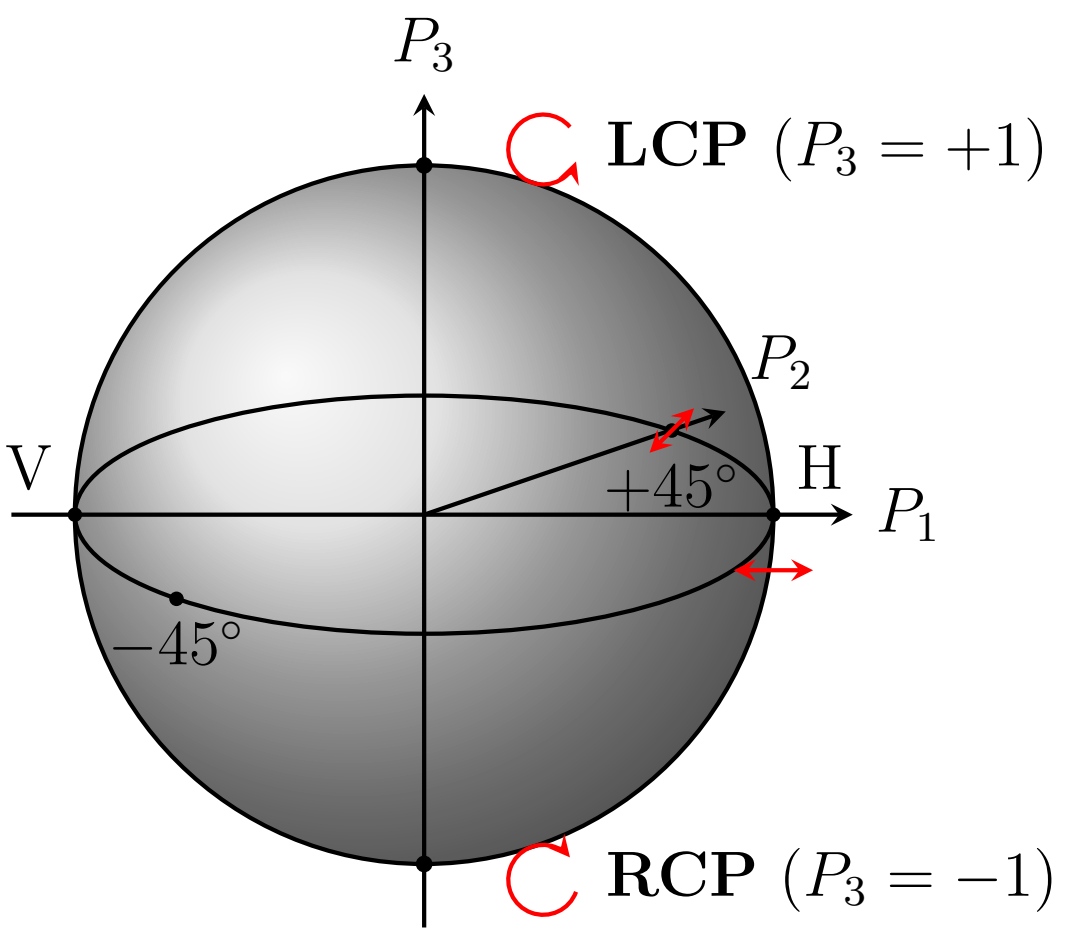

The \textbf{Poincar\'{e} sphere} provides a geometric representation of the
normalized parameters: a polarization state is the point $(P_1, P_2, P_3)$
of Eq.~(\ref{eq:Pdef}), and the sphere has unit radius. Fully polarized
states ($\mathrm{DOP}=1$, $P_1^2+P_2^2+P_3^2=1$) lie on its surface,
partially polarized states inside it, at the distance $\mathrm{DOP}$ from the
center.
\begin{itemize}
    \item \textbf{North pole} $(0,0,+1)$: left circular polarization (LCP)
    \item \textbf{South pole} $(0,0,-1)$: right circular polarization (RCP)
    \item \textbf{Equator} ($P_3=0$, $P_1^2+P_2^2=1$): all linear polarization states, $\mathrm{DOLP}=1$
    \item $+P_1$ axis: horizontal linear; $-P_1$ axis: vertical linear
    \item $+P_2$ axis: $+45^\circ$ linear; $-P_2$ axis: $-45^\circ$ linear
    \item \textbf{Origin} ($P_1=P_2=P_3=0$): unpolarized, $\mathrm{DOP}=0$
\end{itemize}
Lossless retarders act as rotations on the Poincar\'{e} sphere;
dichroic elements (e.g., polarizers) move the states toward the point of their
transmission axis; depolarizing elements contract the states toward the
origin.
Figure~\ref{fig:poincare} shows the sphere with the conventions used here.

\subsection{Summary of Key Polarization States}

Table~\ref{tab:states} collects the three equivalent representations ---
Jones vector, Stokes vector, and density matrix --- of the basic fully
polarized states used throughout this paper, in the conventions of
Section~\ref{sec:sign_convention}.

\begin{table}[H]
\centering
\renewcommand{\arraystretch}{1.45}
\setlength{\tabcolsep}{4pt}
\begin{tabular}{lccc}
\toprule
State & Jones $\mathbf{E}$ & Stokes $\mathbf{S}^T$ & Density matrix $\boldsymbol{\rho}$ \\
\midrule
Linear horizontal ($0^\circ$) &
$\begin{pmatrix}1\\0\end{pmatrix}$ &
$(1,\,1,\,0,\,0)$ &
$\begin{pmatrix}1&0\\0&0\end{pmatrix}$ \\
Linear horizontal ($180^\circ$) &
$\begin{pmatrix}-1\\0\end{pmatrix}$ &
$(1,\,1,\,0,\,0)$ &
$\begin{pmatrix}1&0\\0&0\end{pmatrix}$ \\
\midrule
Linear $+45^\circ$ &
$\dfrac{1}{\sqrt{2}}\begin{pmatrix}1\\1\end{pmatrix}$ &
$(1,\,0,\,1,\,0)$ &
$\dfrac{1}{2}\begin{pmatrix}1&1\\1&1\end{pmatrix}$ \\
Linear $225^\circ$ &
$\dfrac{1}{\sqrt{2}}\begin{pmatrix}-1\\-1\end{pmatrix}$ &
$(1,\,0,\,1,\,0)$ &
$\dfrac{1}{2}\begin{pmatrix}1&1\\1&1\end{pmatrix}$ \\
\midrule
Linear vertical ($90^\circ$) &
$\begin{pmatrix}0\\1\end{pmatrix}$ &
$(1,\,-1,\,0,\,0)$ &
$\begin{pmatrix}0&0\\0&1\end{pmatrix}$ \\
\midrule
Linear $-45^\circ$ &
$\dfrac{1}{\sqrt{2}}\begin{pmatrix}1\\-1\end{pmatrix}$ &
$(1,\,0,\,-1,\,0)$ &
$\dfrac{1}{2}\begin{pmatrix}1&-1\\-1&1\end{pmatrix}$ \\
Linear $135^\circ$ &
$\dfrac{1}{\sqrt{2}}\begin{pmatrix}-1\\1\end{pmatrix}$ &
$(1,\,0,\,-1,\,0)$ &
$\dfrac{1}{2}\begin{pmatrix}1&-1\\-1&1\end{pmatrix}$ \\
\midrule
Linear $\theta$ &
$\begin{pmatrix} \cos(\theta) \\ \sin(\theta) \end{pmatrix}$ &
$(1, \cos(2\theta), \sin(2\theta), 0)$ &
$\dfrac{1}{2}\begin{pmatrix} 2 \cos^2(\theta) & \sin(2\theta) \\ \sin(2\theta) & 2\sin^2(\theta) \end{pmatrix}$\\
\midrule
Right circular (RCP) &
$\dfrac{1}{\sqrt{2}}\begin{pmatrix}1\\-i\end{pmatrix}$ &
$(1,\,0,\,0,\,-1)$ &
$\dfrac{1}{2}\begin{pmatrix}1&i\\-i&1\end{pmatrix}$ \\
\midrule
Left circular (LCP) &
$\dfrac{1}{\sqrt{2}}\begin{pmatrix}1\\i\end{pmatrix}$ &
$(1,\,0,\,0,\,+1)$ &
$\dfrac{1}{2}\begin{pmatrix}1&-i\\i&1\end{pmatrix}$ \\
\midrule
Unpolarized (DOP=0) & 
n.a. & 
$(1,\,0,\,0,\,0)$ & 
$\dfrac{1}{2}\begin{pmatrix}1&0\\0&1\end{pmatrix}$ \\
\bottomrule
\end{tabular}
\caption{Basic fully polarized states and the unpolarized state (normalized, $S_0=1$) in the three
equivalent representations, using the conventions of
Section~\ref{sec:sign_convention}
($e^{-i(\omega t - kz)}$ time factor; $S_3 = 2\,\mathrm{Im}(E_x^*E_y)$;
handedness named looking toward the source). Jones vectors are defined up to
an arbitrary global phase: all vectors $e^{i\alpha}\mathbf{E}$ have the same
Stokes vector and density matrix. In particular, linear polarization at
$\theta$ and $\theta+180^\circ$ is the same state,
$(\cos(\theta+\pi), \sin(\theta+\pi)) = -(\cos(\theta), \sin(\theta))$,
so that, e.g., $0^\circ$ and $180^\circ$, $+45^\circ$ and $225^\circ$, or
$-45^\circ$ and $135^\circ$ coincide. The density matrices follow from
$\boldsymbol{\rho} = \mathbf{E}\mathbf{E}^\dagger$ or equivalently from
Eq.~(\ref{eq:rho_from_Stokes}).}
\label{tab:states}
\end{table}
\subsection{Jones Matrices and Mueller Matrices}
\label{sec:JM_principle}

An optical element transforms the polarization state of a beam. In the three
formalisms:

\begin{itemize}
    \item In the \textbf{Jones formalism}, a $2\times2$ complex matrix
          $\mathbf{J}$ transforms the Jones vector:
          $\mathbf{E}' = \mathbf{J}\,\mathbf{E}$.
          This is valid for fully polarized beams only.

    \item In the \textbf{density matrix formalism}, the same Jones matrix $\mathbf{J}$ is applied, $\boldsymbol{\rho}'= \mathbf{J} \boldsymbol{\rho} \mathbf{J}^\dagger$. This is valid for fully and partially polarized beams; an element that is not described by a single Jones matrix is handled by averaging $\boldsymbol{\rho}'$ over the processes (see below).

    \item In the \textbf{Mueller--Stokes formalism}~\cite{mueller1948,chipman2018},
    a real $4\times4$ matrix
    $\mathbf{M}$ transforms the Stokes vector:
    $\mathbf{S}' = \mathbf{M}\,\mathbf{S}$.
    This is valid for any degree of coherence and polarization, and can describe the statistical effect of several optical processes, e.g.\ in the case of inhomogeneous optical elements.
\end{itemize}

For a non-depolarizing element (a single Jones matrix) the three formalisms are equivalent, the Jones form being restricted to fully polarized beams. The Mueller matrix
corresponding to a Jones matrix $\mathbf{J}$ is obtained via the Pauli trace
formula. Since the density matrix transforms as
$\boldsymbol{\rho}' = \mathbf{J}\boldsymbol{\rho}\mathbf{J}^\dagger$,
projecting onto the Pauli basis (Eq.~\ref{eq:Sk_trace}) gives each Mueller
matrix element as:
\begin{equation}
    {
    M_{mn} = \frac{1}{2}\,\mathrm{Tr}\!\left(
    \boldsymbol{\sigma}_m\,\mathbf{J}\,\boldsymbol{\sigma}_n\,\mathbf{J}^\dagger
    \right), \qquad m,n \in \{0,1,2,3\}}
    \label{eq:PauliTrace}
\end{equation}

When several optical processes $\mathbf{J}_i$ occur simultaneously, e.g.~in an inhomogeneous optical medium or due to the divergence or polychromaticity of the beam, the output density matrix and Mueller matrices are averaged over the processes
\begin{align}
    \left< \boldsymbol{\rho}' \right>
    &=
    \left< \mathbf{J}_i \boldsymbol{\rho} \mathbf{J}_i^\dagger \right>
    \\
    \left<M_{mn}\right>
    & =
    \frac{1}{2} \mathrm{Tr}\left(
    \boldsymbol{\sigma}_m \left<
        \mathbf{J}_i \boldsymbol{\sigma}_n \mathbf{J}_i^\dagger
    \right>
    \right)
\end{align}

Not every Mueller matrix can be generated by a single Jones matrix. An example is the Mueller matrix of a depolarizer,
\begin{align}
    \mathbf{M}
    &=
    \begin{pmatrix} 
    1 & 0 & 0 & 0 \\
    0 & 0 & 0 & 0 \\
    0 & 0 & 0 & 0 \\
    0 & 0 & 0 & 0
    \end{pmatrix}
\end{align}

When several optical elements are placed in series, their Jones matrices (or
Mueller matrices) are multiplied in order from right to left:
\begin{equation}
    \mathbf{J}_{\mathrm{total}} = \mathbf{J}_N \cdots \mathbf{J}_2\,\mathbf{J}_1,
    \qquad
    \mathbf{M}_{\mathrm{total}} = \mathbf{M}_N \cdots \mathbf{M}_2\,\mathbf{M}_1
\end{equation}

\subsection{Rotation of an Optical Element by Angle $\chi$}
\label{sec:rotation}

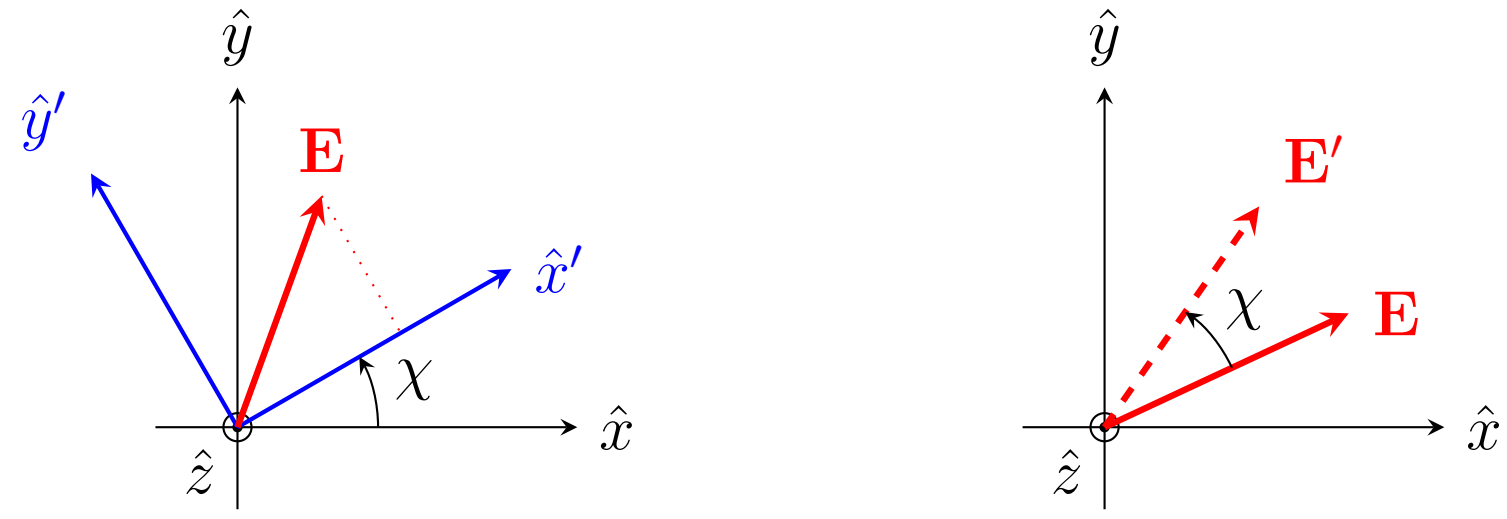
\begin{figure}[H]
\centering
\begin{tikzpicture}[scale=1.55, >=stealth]
  \begin{scope}
    \draw[->] (-0.35,0) -- (1.45,0) node[right] {$\hat{x}$};
    \draw[->] (0,-0.35) -- (0,1.45) node[above] {$\hat{y}$};
    \draw (0,0) circle (0.06); \fill (0,0) circle (0.022);
    \node[below left] at (0,0) {$\hat{z}$};
    \draw[->, blue, thick] (0,0) -- (30:1.35) node[right] {$\hat{x}'$};
    \draw[->, blue, thick] (0,0) -- (120:1.25) node[above left] {$\hat{y}'$};
    \draw[->] (0.6,0) arc (0:30:0.6);
    \node at (15:0.78) {$\chi$};
    \draw[->, red, very thick] (0,0) -- (70:1.05) node[above] {$\mathbf{E}$};
    \draw[red, dotted] (70:1.05) -- ($(0,0)!(70:1.05)!(30:1.35)$);
    \node[align=center, font=\small] at (0.55,-0.75)
      {(a) element axes rotated by $+\chi$:\\ $\mathbf{E}_{\mathrm{comp}} = \mathbf{R}_J(\chi)\,\mathbf{E}$};
  \end{scope}
  \begin{scope}[xshift=3.7cm]
    \draw[->] (-0.35,0) -- (1.45,0) node[right] {$\hat{x}$};
    \draw[->] (0,-0.35) -- (0,1.45) node[above] {$\hat{y}$};
    \draw (0,0) circle (0.06); \fill (0,0) circle (0.022);
    \node[below left] at (0,0) {$\hat{z}$};
    \draw[->, red, very thick] (0,0) -- (25:1.15) node[right] {$\mathbf{E}$};
    \draw[->, red, very thick, dashed] (0,0) -- (55:1.15) node[above right] {$\mathbf{E}'$};
    \draw[->] (25:0.6) arc (25:55:0.6);
    \node at (40:0.78) {$\chi$};
    \node[align=center, font=\small] at (0.55,-0.75)
      {(b) field rotated by $+\chi$:\\ $\mathbf{E}' = \mathbf{R}_J(-\chi)\,\mathbf{E}$};
  \end{scope}
\end{tikzpicture}
\caption{Rotations about the beam axis $\hat{z}$ (pointing out of the page,
toward the viewer). (a) Passive rotation, used throughout this paper: the
field $\mathbf{E}$ is fixed and the element's axes $\hat{x}'$, $\hat{y}'$
(e.g.\ the fast and slow axes of a retarder, or $\sigma$ and $\pi$ of a
crystal) are turned by $+\chi$ from $\hat{x}$ toward $\hat{y}$; the
components of $\mathbf{E}$ along them are $\mathbf{R}_J(\chi)\,\mathbf{E}$,
Eq.~(\ref{eq:Rchi}). (b) Active rotation: the axes are fixed and the field
itself is turned by $+\chi$, $\mathbf{E}' = \mathbf{R}_{\mathrm{act}}(\chi)\,
\mathbf{E} = \mathbf{R}_J(-\chi)\,\mathbf{E}$.}
\label{fig:rotation}
\end{figure}

Optical elements are characterized in their own frame, where their Jones
matrix is simplest (diagonal for a retarder or a crystal, a projector onto
$\hat{x}$ for a polarizer), and are then rotated about the beam axis. This
section works out how vectors and matrices transform under such a rotation,
starting, as for the rest of the paper, from the Jones vector.
Figure~\ref{fig:rotation} shows the geometry.

\paragraph{Jones vector and density matrix.}
Let the element's axes $\hat{x}'$, $\hat{y}'$ be turned by $+\chi$ from
$\hat{x}$ toward $\hat{y}$ (Fig.~\ref{fig:rotation}a). A fixed field with
laboratory components $\mathbf{E}$ has, along the element's axes, the
components
\begin{equation}
    \mathbf{E}_{\mathrm{comp}} = \mathbf{R}_J(\chi)\,\mathbf{E}, \qquad
    \mathbf{R}_J(\chi) =
    \begin{pmatrix} \cos\chi & \sin\chi \\ -\sin\chi & \cos\chi \end{pmatrix}
    \label{eq:Rchi}
\end{equation}
($\hat{x}' = (\cos\chi, \sin\chi)$, $\hat{y}' = (-\sin\chi, \cos\chi)$ in
the laboratory frame), and $\mathbf{R}_J(\chi)^{-1} = \mathbf{R}_J(\chi)^T =
\mathbf{R}_J(-\chi)$ takes components back from the element to the laboratory.
The density matrix $\boldsymbol{\rho} = \langle\mathbf{E}\mathbf{E}^\dagger\rangle$
transforms accordingly; because $\mathbf{R}_J$ is the same for every member
of the ensemble, it can be taken out of the average (and, being real,
$\mathbf{R}_J^\dagger = \mathbf{R}_J^T$), so the transformation holds
unchanged for partially polarized beams:
\begin{equation}
    \boldsymbol{\rho}_{\mathrm{comp}}
    = \langle \mathbf{R}_J(\chi)\mathbf{E}\,\mathbf{E}^\dagger\mathbf{R}_J(\chi)^T\rangle
    = \mathbf{R}_J(\chi)\,\boldsymbol{\rho}\,\mathbf{R}_J(-\chi)
    \label{eq:rho_rotation}
\end{equation}

\paragraph{Stokes vector.}
Inserting Eq.~(\ref{eq:rho_rotation}) into Eq.~(\ref{eq:Sk_trace}) gives the
Stokes parameters in the rotated frame, $\mathbf{S}_{\mathrm{comp}} =
\mathbf{R}_M(\chi)\,\mathbf{S}$, with the Mueller rotation matrix
\begin{equation}
    {
    \mathbf{R}_M(\chi) =
    \begin{pmatrix}
        1 & 0            &  0           & 0 \\
        0 & \cos 2\chi   &  \sin 2\chi  & 0 \\
        0 & -\sin 2\chi  &  \cos 2\chi  & 0 \\
        0 & 0            &  0           & 1
    \end{pmatrix}
    }
    \label{eq:RM}
\end{equation}
Equivalently, $\mathbf{R}_M(\chi)$ is simply the Mueller matrix of the Jones
matrix $\mathbf{R}_J(\chi)$, obtained with the Pauli trace formula,
Eq.~(\ref{eq:PauliTrace}). $S_0$ (total intensity) and $S_3$ (circular
content) are invariant, while $(S_1, S_2)$ rotate as a two-dimensional
vector. A rotation by $\chi$ of the axes appears as a rotation by $2\chi$ in
Stokes space, i.e.\ on the Poincar\'{e} sphere: the Stokes vector turns
through twice the angle of the axes. The reason is that a rotation by
$180^\circ$ only changes the sign of both field components,
$\mathbf{R}_J(\chi+\pi) = -\mathbf{R}_J(\chi)$, and a common sign (an absolute
phase) does not change the Stokes parameters, which are quadratic in the
field. The Stokes parameters, and hence $\mathbf{R}_M$, must therefore have
period $180^\circ$ in $\chi$; being quadratic in $\cos\chi$ and $\sin\chi$,
they depend on $2\chi$ only.

\paragraph{Rotated element.}
When an optical element with Jones matrix $\mathbf{J}^{(0)}$ in its own
frame is physically rotated by $\chi$ about the beam axis, its Jones matrix
in the laboratory frame follows in three steps: project the incident field
onto the element's axes, $\mathbf{R}_J(\chi)\,\mathbf{E}_{\mathrm{in}}$;
apply $\mathbf{J}^{(0)}$; and project the result back onto the laboratory
axes with $\mathbf{R}_J(-\chi)$. Hence
$\mathbf{E}_{\mathrm{out}} = \mathbf{R}_J(-\chi)\,\mathbf{J}^{(0)}\,
\mathbf{R}_J(\chi)\,\mathbf{E}_{\mathrm{in}}$, i.e.
\begin{equation}
    \mathbf{J}^{(\chi)} = \mathbf{R}_J(-\chi)\cdot\mathbf{J}^{(0)}\cdot\mathbf{R}_J(\chi)
    \label{eq:Jones_rotation}
\end{equation}
The same three steps applied to the density matrix with
Eq.~(\ref{eq:rho_rotation}) give $\boldsymbol{\rho}_{\mathrm{out}} =
\mathbf{J}^{(\chi)}\,\boldsymbol{\rho}_{\mathrm{in}}\,\mathbf{J}^{(\chi)\dagger}$,
the general rule for a non-depolarizing element. In Stokes space, the Pauli
trace formula is multiplicative: the Mueller matrix of a product of Jones
matrices is the product of their Mueller matrices (it describes the same
sequence of operations on $\boldsymbol{\rho}$). Since the Mueller matrix of
$\mathbf{R}_J(\chi)$ is $\mathbf{R}_M(\chi)$, Eq.~(\ref{eq:Jones_rotation})
translates term by term into
\begin{equation}
    \mathbf{M}^{(\chi)} =
    \mathbf{R}_M(-\chi)\cdot\mathbf{M}^{(0)}\cdot\mathbf{R}_M(\chi)
    \label{eq:Mueller_rotation}
\end{equation}

\paragraph{The opposite ordering.}
The sandwich is sometimes written with the opposite ordering,
$\mathbf{R}_J(\chi)\,\mathbf{J}^{(0)}\,\mathbf{R}_J(-\chi)$. This is not a
different result: it is Eq.~(\ref{eq:Jones_rotation}) for the rotation
$-\chi$. It arises when $\mathbf{R}_J(\chi)$ is read as acting on the field
itself, $\mathbf{E} \to \mathbf{R}_J(\chi)\,\mathbf{E}$: with the matrix of
Eq.~(\ref{eq:Rchi}) this turns the field by $-\chi$ (Fig.~\ref{fig:rotation}b
with $\chi \to -\chi$). Turning field and element together by the same
angle, $\mathbf{E} \to \mathbf{R}_J(\chi)\mathbf{E}$ and
$\mathbf{J} \to \mathbf{R}_J(\chi)\,\mathbf{J}\,\mathbf{R}_J(-\chi)$, leaves the
physics unchanged, since $\mathbf{R}_J(\chi)\mathbf{J}\mathbf{R}_J(-\chi)\,
\mathbf{R}_J(\chi)\mathbf{E} = \mathbf{R}_J(\chi)\,(\mathbf{J}\mathbf{E})$: the
output is simply turned with the rest. Conversely, applying the unrotated
$\mathbf{J}^{(0)}$ to a field turned by $\mathbf{R}_J(\chi)$ and turning the
result back gives Eq.~(\ref{eq:Jones_rotation}) again. What matters is to
use one convention consistently: combined with the passive matrix of
Eq.~(\ref{eq:Rchi}), the opposite ordering describes the element turned by
$-\chi$, and mixing the two readings flips the sign of every $\sin 2\chi$
term.

\paragraph{Checking the sense of rotation.}
A useful test must distinguish $+\chi$ from $-\chi$. Rotating a horizontal
polarizer by $\pm 90^\circ$ does not: both give a vertical polarizer,
because $\mathbf{R}_J(+90^\circ) = -\mathbf{R}_J(-90^\circ)$ and the sign
cancels in the sandwich of Eq.~(\ref{eq:Jones_rotation}): any element is
invariant under $\chi \to \chi + 180^\circ$. A rotation by $\chi = +45^\circ$ does. With
$\mathbf{J}^{(0)} = \bigl[\begin{smallmatrix}1&0\\0&0\end{smallmatrix}\bigr]$,
Eq.~(\ref{eq:Jones_rotation}) gives
\begin{equation}
    \mathbf{J}^{(+45^\circ)} = \frac{1}{2}\begin{pmatrix}1&1\\1&1\end{pmatrix},
    \qquad
    \mathbf{J}^{(-45^\circ)} = \frac{1}{2}\begin{pmatrix}1&-1\\-1&1\end{pmatrix}
\end{equation}
The first transmits $+45^\circ$ light, $(1,1)^T/\sqrt{2}$, unchanged and
blocks $-45^\circ$ light, $(1,-1)^T/\sqrt{2}$, as a polarizer whose
transmission axis is turned from $\hat{x}$ toward $\hat{y}$ must; the second
does the opposite.

\paragraph{Convention note.}
$\mathbf{R}_J(\chi)$ as defined in Eq.~(\ref{eq:Rchi}) is the \emph{passive}
rotation matrix: it expresses the components of a fixed field in a coordinate
frame rotated by $+\chi$ (from $+\hat{x}$ toward $+\hat{y}$). It is the
transpose (inverse) of the standard counter-clockwise \emph{active} rotation
matrix
$\bigl[\begin{smallmatrix}\cos\chi&-\sin\chi\\\sin\chi&\cos\chi\end{smallmatrix}\bigr]$
used in some optics references. With the active matrix the sandwich for a
rotated element reads $\mathbf{R}_{\mathrm{act}}(\chi)\,\mathbf{J}^{(0)}\,
\mathbf{R}_{\mathrm{act}}(-\chi)$ --- identical physics, opposite letter
ordering. Table~\ref{tab:passive_active} summarizes the two conventions.

\begin{table}[H]
\centering
\renewcommand{\arraystretch}{1.45}
\setlength{\tabcolsep}{5pt}
\begin{tabular}{@{}>{\raggedright\arraybackslash}p{2.7cm}
                  >{\raggedright\arraybackslash}p{5.2cm}
                  >{\raggedright\arraybackslash}p{5.2cm}@{}}
\toprule
& \textbf{Passive} $\mathbf{R}_J(\chi)$ \newline
\textbf{(used throughout this paper)} &
\textbf{Active} $\mathbf{R}_{\mathrm{act}}(\chi)$ \newline
(shown for comparison only) \\
\midrule
\multicolumn{3}{@{}l}{\emph{Differences}} \\
\midrule
Matrix &
$\begin{pmatrix}\cos\chi&\sin\chi\\-\sin\chi&\cos\chi\end{pmatrix}$ \newline
Eq.~(\ref{eq:Rchi}) &
$\begin{pmatrix}\cos\chi&-\sin\chi\\\sin\chi&\cos\chi\end{pmatrix}$ \newline
(not used here) \\
What moves &
The \emph{axes} turn by $+\chi$; the field is held fixed (change of basis) &
The \emph{field} turns by $+\chi$; the axes are held fixed \\
Acts on components as &
$\mathbf{E}_{\mathrm{comp}} = \mathbf{R}_J(\chi)\,\mathbf{E}_{\mathrm{lab}}$
(same field, new axes) &
$\mathbf{E}' = \mathbf{R}_{\mathrm{act}}(\chi)\,\mathbf{E}$
(new field, same axes) \\
Rotated-element sandwich &
$\mathbf{J}^{(\chi)} = \mathbf{R}_J(-\chi)\,\mathbf{J}^{(0)}\,\mathbf{R}_J(\chi)$
\newline Eq.~(\ref{eq:Jones_rotation}) &
$\mathbf{J}^{(\chi)} = \mathbf{R}_{\mathrm{act}}(\chi)\,\mathbf{J}^{(0)}\,
\mathbf{R}_{\mathrm{act}}(-\chi)$ \\
Mueller counterpart &
$\mathbf{R}_M(\chi)$, Eq.~(\ref{eq:RM}); rotated element, Eq.~(\ref{eq:Mueller_rotation}) &
--- \\
Upper-right entry & $+\sin\chi$ & $-\sin\chi$ \\
\midrule
\multicolumn{3}{@{}l}{\emph{Similarities}} \\
\midrule
Algebraic &
\multicolumn{2}{@{}>{\raggedright\arraybackslash}p{10.6cm}@{}}{Both real, orthogonal, $\det = +1$ (proper rotations of SO(2)); each is the other's inverse and transpose:
$\mathbf{R}_{\mathrm{act}}(\chi) = \mathbf{R}_J(\chi)^{-1} = \mathbf{R}_J(\chi)^{T} = \mathbf{R}_J(-\chi)$} \\
Physical content &
\multicolumn{2}{@{}>{\raggedright\arraybackslash}p{10.6cm}@{}}{Identical when each is used self-consistently: same $\mathbf{J}^{(\chi)}$, same Mueller rotation $\mathbf{R}_M(\chi)$ of Eq.~(\ref{eq:RM}), same observable Stokes transformation} \\
Invariants &
\multicolumn{2}{@{}>{\raggedright\arraybackslash}p{10.6cm}@{}}{Both leave $S_0$ and $S_3$ unchanged, rotate $(S_1,S_2)$ by $\mp2\chi$ (passive/active), and preserve the degree of polarization} \\
\bottomrule
\end{tabular}
\caption{The two rotation-matrix conventions for the sandwich of
Eq.~(\ref{eq:Jones_rotation}): inverse bookkeeping choices,
$\mathbf{R}_{\mathrm{act}}(\chi) = \mathbf{R}_J(-\chi)$, that are safe
individually but must not be mixed. This paper uses the passive
convention exclusively.}
\label{tab:passive_active}
\end{table}

In either case, positive $\chi$ in Eq.~(\ref{eq:Jones_rotation})
means that the element's $\hat{x}'$ ($\sigma$) axis --- the fast axis for the
retarders of Section~\ref{sec:general_retarders} --- is rotated from
$+\hat{x}$ toward $+\hat{y}$; this is the same sense as the polarizer angle $\theta$ in
Section~\ref{sec:experimental}, whose transmission axis is $(\cos\theta, \sin\theta)$. Reversing
the sign convention of the sandwich would flip the sign of $\sin 2\chi$
(equivalently of $S_2$ and $S_3$ generated from horizontal input) everywhere
below. When importing Jones matrices from other sources, always check which
sign convention $\mathbf{R}_J(\chi)$ uses.

Ref.~\cite{detlefs2012} illustrates the point. The $2\times2$ matrix printed
just above its Eq.~(24), $R(\eta) =
\bigl[\begin{smallmatrix}\cos\eta&-\sin\eta\\\sin\eta&\cos\eta\end{smallmatrix}\bigr]$,
has the \emph{active} form of Table~\ref{tab:passive_active}, whereas the
transformation it is introduced to derive --- its Eqs.~(24)--(25) for the
Poincar\'e--Stokes parameters --- is the one generated by the \emph{passive}
matrix, and agrees term by term with the Mueller rotation
$\mathbf{R}_M(\chi)$ of Eq.~(\ref{eq:RM}) above, i.e.\ with the component
transformation of Eq.~(\ref{eq:Rchi}). Its Fig.~3, which draws the $B$ axes
turned counter-clockwise from the $A$ axes, implies the passive form as well.
One evaluation settles which is which: for $\eta = 45^\circ$ and
$\mathbf{V}_A = (1,0)^T$ (i.e.\ $P_1 = 1$, $P_2 = 0$), Eq.~(25) of
Ref.~\cite{detlefs2012} gives $P_{2,B} = -1$, which is what
Eq.~(\ref{eq:Rchi}) returns, while the matrix as printed returns $P_{2,B} =
+1$. The conventions used here therefore agree with the \emph{results} of
Ref.~\cite{detlefs2012}; the discrepancy is confined to that printed matrix.

\section{General Retarders}
\label{sec:general_retarders}

Building on the rotation formalism above, this section works out the Jones
and Mueller matrices of a general phase retarder, specializes them to the
quarter-wave plate, which at $\chi=\pm45^\circ$ converts linear to circular
polarization, and to the half-wave plate, and summarizes the key matrices
obtained.

\subsection{Jones and Mueller Matrices of a General Retarder}
\label{sec:retarder}

\paragraph{What is a retarder, and why the name.}
A \textbf{retarder} --- also called a \emph{wave plate} or \emph{phase
plate} --- is a birefringent optical element cut so that light polarized
along one of two orthogonal directions in its cross-section experiences a
different refractive index than light polarized along the other, i.e.~along the two axes of the elliptical cross-section of the index ellipsoid normal to the propagation direction.\footnote{See
e.g.~the overview of wave plates at
\url{https://en.wikipedia.org/wiki/Waveplate} and the description of the index ellipsoid at \url{https://en.wikipedia.org/wiki/Index_ellipsoid}.} Since the phase velocity of
light in a medium is $v = c/n$, the component travelling along the
higher-index axis falls behind, i.e.\ is \emph{retarded}, relative to the
component on the lower-index axis; the relative phase lag accumulated after
propagating through the plate is the \textbf{retardation} $\delta$, which
gives the element its name. The axis with the \emph{lower} refractive index
(higher phase velocity) is the \textbf{fast axis}; the orthogonal axis, with
the \emph{higher} refractive index (lower phase velocity), is the
\textbf{slow axis}. For a plate of thickness $d$ with indices
$n_{\mathrm{slow}} > n_{\mathrm{fast}}$ along the two axes, the retardation is
$\delta = (2\pi/\lambda)\,(n_{\mathrm{slow}} - n_{\mathrm{fast}})\,d$, with
$\lambda$ the vacuum wavelength; a \textbf{quarter-wave plate} (QWP) is cut to
give $\delta = \pi/2$ and a \textbf{half-wave plate} (HWP) $\delta = \pi$.

\paragraph{Effect on the field.}
With the fast axis aligned along $\hat{x}$ ($\chi = 0$) and using the complex
amplitudes of Eq.~(\ref{eq:physical_wave}), a retarder leaves the fast-axis
component $E_x$ unchanged while the slow-axis component picks up the
retardation as an extra phase, $E_y \to E_y\,e^{i\delta}$. The sign is fixed
by the time factor $e^{-i\omega t}$: a wave $e^{i(knz-\omega t)}$ that
crosses a plate of thickness $d$ gains the phase $knd$, larger on the slow
(higher-index) axis, so with the fast axis along $\hat{x}$ the retardation
$\delta = k(n_{\mathrm{slow}}-n_{\mathrm{fast}})d$ is positive. The formulas
below hold for either sign of $\delta$: $\delta < 0$ simply means that
$\hat{x}$ is the slow axis, which is the same plate rotated by $90^\circ$
(exchanging the two axes changes the sign of $\delta$).
\begin{equation}
    \begin{pmatrix} E_x \\ E_y \end{pmatrix}
    \;\xrightarrow{\ \text{retarder}\ }\;
    \begin{pmatrix} E_x \\ E_y\,e^{i\delta} \end{pmatrix} .
\end{equation}
A real plate also multiplies \emph{both} components by a common factor,
the phase $e^{ikn_{\mathrm{fast}}d}$ accumulated along the fast axis (and,
for an absorbing plate, a common attenuation). Being common to $E_x$ and
$E_y$, this global phase does not change the polarization state (it drops out
of the Stokes parameters and of the density matrix; see Table~\ref{tab:states}),
and it is omitted here and below.
In the time domain (Eq.~\ref{eq:real_components}) this shifts the
slow-axis oscillation to
\begin{equation*}
\mathcal{E}_y(t) = |E_y|\cos(\tau - \phi_y - \delta) ,
\end{equation*}
it reaches any given point of its cycle a time $\delta/\omega$
\emph{later} than it would without the retarder (for in-phase input, as in
Figure~\ref{fig:retarder}, a delay $\delta/\omega$ behind the fast-axis
component), which is the literal, time-domain meaning of ``retardation.'' Horizontal or vertical input ($E_xE_y=0$) is
unaffected, since only one component is present; input at $\pm45^\circ$
($E_x = \pm E_y$) is the sensitive case, where a QWP's quarter-cycle lag
turns the linear oscillation of $(\mathcal{E}_x,\mathcal{E}_y)$ into a
circular one (the configuration worked out in detail, for a horizontal input
and a rotated QWP, in Sections~\ref{sec:QWP45} and \ref{sec:QWPm45}; the
left-handed case drawn in Figure~\ref{fig:retarder} corresponds to the
latter).

\begin{figure}[H]
\centering
\begin{tikzpicture}[>=stealth, scale=0.95]

\begin{scope}[shift={(1.5,0)}]
    \draw[->, gray] (-1.35,0) -- (1.35,0) node[right, font=\small] {$x$};
    \draw[->, gray] (0,-1.35) -- (0,1.4) node[above, font=\small] {$y$};
    \draw[<->, red, very thick] (-0.95,-0.95) -- (0.95,0.95);
    \node[font=\small, align=center] at (0,-2.1)
        {(a) in: linear\\at $45^\circ$};
\end{scope}

\draw[->, very thick] (3.2,0) -- (4.3,0);

\begin{scope}[shift={(5.7,0)}]
    \draw[fill=gray!18, draw=black] (-0.72,-1.35) -- (0.72,-0.95)
        -- (0.72,1.4) -- (-0.72,1.0) -- cycle;
    \draw[->, red, very thick] (0,0.05) -- (0.95,0.32)
        node[right, font=\small] {fast};
    \draw[->, blue, very thick] (0,0.05) -- (0,1.3)
        node[above, font=\small] {slow};
    \node[font=\small, align=center] at (0,-2.1)
        {(b) retarder,\\retardation $\delta$};
\end{scope}

\draw[->, very thick] (7.9,0) -- (9.0,0);

\begin{scope}[shift={(10.4,0)}]
    \draw[->, gray] (-1.35,0) -- (1.35,0) node[right, font=\small] {$x$};
    \draw[->, gray] (0,-1.35) -- (0,1.4) node[above, font=\small] {$y$};
    \draw[red, very thick] (0,0) circle (0.95);
    \draw[->, red, very thick] (0.95,0) arc (0:80:0.95);
    \node[font=\small, align=center] at (0,-2.1)
        {(c) out: circular\\($\delta=\pi/2$)};
\end{scope}

\begin{scope}[shift={(1.2,-5.6)}]
    \draw[->, thick] (-0.3,0) -- (9.9,0) node[right, font=\small] {$\tau$};
    \draw[->, thick] (0,-1.35) -- (0,1.5) node[above, font=\small] {$\mathcal{E}$};

    \draw[red, very thick, domain=0:9.4, samples=160, smooth, variable=\t]
        plot ({\t},{1.05*cos((\t) r)});
    \node[red, font=\small] at (9.75,1.15) {$\mathcal{E}_x$};

    \draw[blue, very thick, dashed, domain=0:9.4, samples=160, smooth, variable=\t]
        plot ({\t},{1.05*cos((\t-1.5708) r)});
    \node[blue, font=\small] at (9.75,0.35) {$\mathcal{E}_y$};

    \draw[gray, dotted] (0,1.05) -- (0,-1.75);
    \draw[gray, dotted] (1.5708,1.05) -- (1.5708,-1.75);
    \draw[<->, thick] (0,-1.6) -- (1.5708,-1.6);
    \node[font=\small, anchor=west] at (1.75,-1.6) {$\delta/\omega$};

    \node[font=\small] at (4.7,-2.5)
        {(d) the slow-axis component lags by the retardation $\delta$};
\end{scope}

\end{tikzpicture}
\caption{How a retarder converts linear into circular polarization, for input
at $45^\circ$ to the fast axis. Top row, the transformation: (a) before the
plate the two components are in phase, so the field vector oscillates along a
straight line at $45^\circ$; (b) the plate delays the slow-axis component
relative to the fast-axis one by the retardation $\delta$; (c) after a
quarter-wave plate ($\delta=\pi/2$) the two components are in quadrature and
the tip of the field vector traces a circle. Bottom, (d): the same statement
in the time domain --- the slow-axis component $\mathcal{E}_y$ (blue, dashed)
reaches each point of its cycle a time $\delta/\omega$ after the fast-axis
component $\mathcal{E}_x$ (red, solid), which is Eq.~(\ref{eq:real_components})
with $\phi_y \to \phi_y + \delta$. Only the \emph{relative} phase matters:
horizontal or vertical input, having a single component, is unaffected. The
transverse planes are drawn looking \emph{toward the source}, the convention
of Section~\ref{sec:sign_convention}, so the counter-clockwise sense in (c) is
left circular polarization; this is the same configuration as the
$\chi=-45^\circ$ case of Section~\ref{sec:QWPm45}, seen in the frame of the
plate rather than of the laboratory.}
\label{fig:retarder}
\end{figure}

\paragraph{Jones and Mueller matrices.}
The aligned Jones matrix is therefore
\begin{equation}
    \mathbf{J}_{\mathrm{ret}}^{(0)} =
    \begin{pmatrix} 1 & 0 \\ 0 & e^{i\delta} \end{pmatrix} .
    \label{eq:Jret0}
\end{equation}
Applying the Pauli
trace formula (Eq.~\ref{eq:PauliTrace}) to $\mathbf{J}_{\mathrm{ret}}^{(0)}$,
the corresponding aligned Mueller matrix of the general retarder is:
\begin{equation}
    \mathbf{M}_{\mathrm{ret}}^{(0)} =
    \begin{pmatrix}
        1 & 0 & 0 & 0 \\
        0 & 1 & 0 & 0 \\
        0 & 0 & \cos\delta & -\sin\delta \\
        0 & 0 & \sin\delta & \cos\delta
    \end{pmatrix}
    \label{eq:Mret0}
\end{equation}
i.e.\ a rotation by the angle $\delta$ in the $(S_2, S_3)$ plane of the
Poincar\'{e} sphere, leaving $S_0$ and $S_1$ untouched. Since $S_1$ and
$S_2^2+S_3^2$ are conserved, so is the degree of polarization: a retarder
changes the polarization state but never depolarizes. The degree of
\emph{linear} polarization, $\sqrt{S_1^2+S_2^2}/S_0$, is not conserved:
$\pm45^\circ$ linear polarization ($S_2$) is converted into circular
polarization ($S_3$), completely for $\delta = \pm\pi/2$ ($\sin\delta =
\pm1$, $\cos\delta = 0$).
Specific values of interest are $\delta = \pm \pi/2$ (quarter-wave plate) and $\delta = \pi$ (half-wave plate). We discuss these two cases in detail below.

For a general retarder with fast axis rotated about the beam axis to angle $\chi$, applying
Eq.~(\ref{eq:Jones_rotation}) to $\mathbf{J}_{\mathrm{ret}}^{(0)}$ gives the full Jones matrix:
\begin{equation}
    \mathbf{J}_{\mathrm{ret}}^{(\chi,\delta)}
    =
    \begin{pmatrix}
        \cos^2\chi + e^{i\delta}\sin^2\chi
        &
        \bigl(1 - e^{i\delta}\bigr)\sin\chi\cos\chi
        \\
        \bigl(1 - e^{i\delta}\bigr)\sin\chi\cos\chi
        &
        \sin^2\chi + e^{i\delta}\cos^2\chi
    \end{pmatrix}
    \label{eq:Jret_general}
\end{equation}
or, in the Pauli basis,
$\mathbf{J}_{\mathrm{ret}}^{(\chi,\delta)} = \frac{1+e^{i\delta}}{2}\,
\boldsymbol{\sigma}_0 + \frac{1-e^{i\delta}}{2}\,\bigl(\cos 2\chi\,
\boldsymbol{\sigma}_1 + \sin 2\chi\,\boldsymbol{\sigma}_2\bigr) =
e^{i\delta/2}\bigl[\cos\tfrac{\delta}{2}\,\boldsymbol{\sigma}_0 -
i\sin\tfrac{\delta}{2}\,\bigl(\cos 2\chi\,\boldsymbol{\sigma}_1 +
\sin 2\chi\,\boldsymbol{\sigma}_2\bigr)\bigr]$, which gives the quarter- and
half-wave plates below at a glance and has the same structure as the crystal
matrix of Section~\ref{sec:crystal_plate} (Eq.~\ref{eq:Jchi_Pauli} with
$r_s = 1$, $r_p = e^{i\delta}$).

Similarly, applying
Eq.~(\ref{eq:Mueller_rotation}) to $\mathbf{M}_{\mathrm{ret}}^{(0)}$ gives the
full Mueller matrix:
\begin{equation}
    \mathbf{M}_{\mathrm{ret}}^{(\chi,\delta)} =
    \begin{pmatrix}
        1 & 0 & 0 & 0 \\[4pt]
        0 & \cos^2(2\chi) + \sin^2(2\chi)\cos(\delta)
          & \sin(4\chi) \sin^2\left(\frac{\delta}{2}\right)
          & \sin(2\chi)\sin(\delta) \\[4pt]
        0 & \sin(4\chi) \sin^2\left(\frac{\delta}{2}\right)
          & \sin^2(2\chi) + \cos^2(2\chi) \cos(\delta)
          & -\cos(2\chi) \sin(\delta) \\[4pt]
        0 & -\sin(2\chi) \sin(\delta)
          & \cos(2\chi) \sin(\delta)
          & \cos(\delta)
    \end{pmatrix}
    \label{eq:Mret_general}
\end{equation}

\subsection{Quarter-wave plate}

For a \textbf{quarter-wave plate} (QWP) with
$\delta = \pi/2$ we have
\begin{equation}
    \mathbf{J}_{\mathrm{QWP}}^{(0)} =
    \begin{pmatrix} 1 & 0 \\ 0 & i \end{pmatrix} .
    \label{eq:JQWP0}
\end{equation}

Setting $\delta = \pi/2$ in Eq.~(\ref{eq:Mret0}) (equivalently, using
Eq.~\ref{eq:PauliTrace}), the corresponding aligned Mueller matrix is:
\begin{equation}
    {
    \mathbf{M}_{\mathrm{QWP}}^{(0)} =
    \begin{pmatrix}
        1 & 0 &  0 &  0 \\
        0 & 1 &  0 &  0 \\
        0 & 0 &  0 &  -1 \\
        0 & 0 & +1 &  0
    \end{pmatrix}
    }
    \label{eq:MQWP0}
\end{equation}

This matrix leaves $S_0$ and $S_1$ unchanged (H/V polarization unaffected since
the fast axis is along $\hat{x}$) and gives $S_2' = -S_3$, $S_3' = S_2$: it
converts $\pm45^\circ$ linear polarization into circular polarization.

For a quarter-wave plate with fast axis rotated to angle $\chi$, setting $\delta=\pi/2$ in Eq.~(\ref{eq:Jret_general}) gives the full Jones matrix:
\begin{equation}
    \mathbf{J}_{\mathrm{QWP}}^{(\chi)}
    =
    \frac{e^{i\pi/4}}{\sqrt{2}}
    \begin{pmatrix}
        1 - i \cos(2\chi) 
        &
        -i \sin(2\chi)
        \\
        -i \sin(2\chi)
        &
        1 + i \cos(2\chi)
    \end{pmatrix}
    \label{eq:JQWP_general}
\end{equation}

Setting $\delta=\pi/2$ in Eq.~(\ref{eq:Mret_general}) gives the full Mueller matrix:
\begin{equation}
    \mathbf{M}_{\mathrm{QWP}}^{(\chi)} =
    \begin{pmatrix}
        1 & 0 & 0 & 0 \\[4pt]
        0 & \cos^2(2\chi)
          & \frac{1}{2}\sin(4\chi)
          & \sin(2\chi)\\[4pt]
        0 & \frac{1}{2}\sin(4\chi)
          & \sin^2(2\chi)
          & -\cos(2\chi)\\[4pt]
        0 & -\sin 2\chi
          & \cos 2\chi
          & 0
    \end{pmatrix}
    \label{eq:MQWP_general}
\end{equation}

One verifies immediately: $\chi=0$ recovers Eq.~(\ref{eq:MQWP0})
\checkmark.

\subsubsection{QWP applied to general linear polarization}
\label{sec:QWPlinear}

Applying the quarter-wave plate to a linearly polarized beam at angle $\theta$,
$\boldsymbol{E}_{\mathrm{in}}(\theta) = (\cos(\theta), \sin(\theta))^T$ and $\boldsymbol{S}_{\mathrm{in}}(\theta) = (1, \cos(2\theta), \sin(2\theta), 0)^T$ yields

\begin{align}
    \mathbf{E}_{\mathrm{out}}
    &=
    \frac{e^{i\pi/4}}{\sqrt{2}}
    \begin{pmatrix}
        \cos(\theta) - i \cos(\theta-2\chi) \\
        \sin(\theta) + i \sin(\theta-2\chi)
    \end{pmatrix}
    \\
    \mathbf{S}_{\mathrm{out}}
    &=
    \begin{pmatrix}
    1 \\
    \cos(2\theta - 2 \chi) \cos(2\chi) \\
    \cos(2\theta - 2 \chi) \sin(2\chi) \\
    \sin(2\theta - 2 \chi)
    \end{pmatrix}
\end{align}

The degree of linear polarization is
\begin{equation}
    \mathrm{DOLP}
    =
    \left| \cos(2\theta - 2 \chi) \right|
\end{equation}

Specifically, for $\chi=\theta$ or $\chi = \theta+\pi/2$ the linear polarization remains unchanged, and for $\chi=\theta+\pi/4$ or $\chi=\theta + 3 \pi/4$ the resulting polarization is circular: right-handed ($S_3=-S_0$) for $\chi=\theta+\pi/4$, left-handed for $\chi=\theta+3\pi/4$. 

The polarization of a bending-magnet or planar-undulator beam on axis is linear horizontal, $\theta=0$, such that circular polarization is produced for $\chi=\pi/4$ ($45^\circ$) or $\chi=3\pi/4$ ($135^\circ$, equivalently $-45^\circ$). We consider these cases in further detail below.

\subsubsection{QWP at $\chi = 45^\circ$: Jones and Mueller Matrices}
\label{sec:QWP45}

Applying the rotation sandwich (Eq.~\ref{eq:Jones_rotation}) to
$\mathbf{J}_{\mathrm{QWP}}^{(0)}$ with $\chi = 45^\circ$:
\begin{equation}
    \mathbf{J}_{\mathrm{QWP}}^{(45^\circ)} =
    \mathbf{R}_J(-45^\circ)\cdot
    \begin{pmatrix}1&0\\0&i\end{pmatrix}
    \cdot\mathbf{R}_J(45^\circ)
    = \frac{e^{i\pi/4}}{\sqrt{2}}
    \begin{pmatrix} 1 & -i \\ -i & 1 \end{pmatrix}
    \label{eq:JQWP45}
\end{equation}

Setting $\delta=\pi/2$ and $\chi=45^\circ$ in Eq.~(\ref{eq:Mret_general}):
\begin{equation}
    {
    \mathbf{M}_{\mathrm{QWP}}^{(45^\circ)} =
    \begin{pmatrix}
        1 &  0 & 0 & 0 \\
        0 &  0 & 0 & 1 \\
        0 &  0 & 1 & 0 \\
        0 & -1 & 0 & 0
    \end{pmatrix}
    }
    \label{eq:MQWP45}
\end{equation}

This matrix gives $S_1' = S_3$ and $S_3' = -S_1$, while leaving $S_0$ and $S_2$
unchanged. It converts horizontal linear polarization ($S_1=1, S_3=0$) into
right circular polarization ($S_1=0, S_3=-1$).

Equivalently, applying $\mathbf{R}_M(-45^\circ)\,\mathbf{M}_{\mathrm{QWP}}^{(0)}\,\mathbf{R}_M(45^\circ)$
with $\mathbf{R}_M(\pm45^\circ)$ evaluated at $2\chi = \pm90^\circ$ yields the same result as Eq.~(\ref{eq:MQWP45}). \checkmark

\paragraph{Horizontal linear input, right circular output.}

The Jones vector of a horizontally polarized beam (normalized, $E_0 = 1$) is:
\begin{equation}
    \mathbf{E}_{\mathrm{in}} = \begin{pmatrix} 1 \\ 0 \end{pmatrix}
\end{equation}

Applying Eqs.~(\ref{eq:S0def})--(\ref{eq:S3def}):
\begin{equation}
    \mathbf{S}_{\mathrm{in}} =
    \begin{pmatrix} S_0 \\ S_1 \\ S_2 \\ S_3 \end{pmatrix} =
    \begin{pmatrix} 1 \\ 1 \\ 0 \\ 0 \end{pmatrix}
    \label{eq:Sin}
\end{equation}

This state lies on the equator of the Poincar\'{e} sphere at the point
$(P_1, P_2, P_3) = (1, 0, 0)$.

\paragraph{Output beam via the Jones route.}

Applying Eq.~(\ref{eq:JQWP45}) to the input Jones vector:
\begin{equation}
    \mathbf{E}_{\mathrm{out}} =
    \mathbf{J}_{\mathrm{QWP}}^{(45^\circ)}\,\mathbf{E}_{\mathrm{in}}
    = \frac{e^{i\pi/4}}{\sqrt{2}}
    \begin{pmatrix} 1 & -i \\ -i & 1 \end{pmatrix}
    \begin{pmatrix} 1 \\ 0 \end{pmatrix}
    = \frac{e^{i\pi/4}}{\sqrt{2}}
    \begin{pmatrix} 1 \\ -i \end{pmatrix}
\end{equation}

Dropping the global phase $e^{i\pi/4}$, which no observable sees:
\begin{equation}
    {
    \mathbf{E}_{\mathrm{out}} = \frac{1}{\sqrt{2}}\begin{pmatrix} 1 \\ -i \end{pmatrix}
    }
    \label{eq:Eout_RCP}
\end{equation}

This is the canonical Jones vector for right circular polarization. Computing the
Stokes parameters from Eqs.~(\ref{eq:S0def})--(\ref{eq:S3def}):
\begin{align}
    S_0 &= \Bigl|\tfrac{1}{\sqrt{2}}\Bigr|^2 + \Bigl|\tfrac{-i}{\sqrt{2}}\Bigr|^2
         = \tfrac{1}{2} + \tfrac{1}{2} = 1 \\[4pt]
    S_1 &= \Bigl|\tfrac{1}{\sqrt{2}}\Bigr|^2 - \Bigl|\tfrac{-i}{\sqrt{2}}\Bigr|^2
         = \tfrac{1}{2} - \tfrac{1}{2} = 0 \\[4pt]
    S_2 &= 2\,\mathrm{Re}\!\left(\Bigl(\tfrac{1}{\sqrt{2}}\Bigr)^{\!*}\cdot\tfrac{-i}{\sqrt{2}}\right)
         = 2\,\mathrm{Re}\!\left(\tfrac{-i}{2}\right) = 0 \\[4pt]
    S_3 &= 2\,\mathrm{Im}\!\left(\Bigl(\tfrac{1}{\sqrt{2}}\Bigr)^{\!*}\cdot\tfrac{-i}{\sqrt{2}}\right)
         = 2\,\mathrm{Im}\!\left(\tfrac{-i}{2}\right) = -1
\end{align}

The corresponding density matrix, from
$\boldsymbol{\rho} = \mathbf{E}_{\mathrm{out}}\mathbf{E}_{\mathrm{out}}^\dagger$
or equivalently from Eq.~(\ref{eq:rho_from_Stokes}) with
$\mathbf{S} = (1,0,0,-1)^T$, is:
\begin{equation}
    \boldsymbol{\rho}_{\mathrm{RCP}} = \frac{1}{2}
    \begin{pmatrix} 1 & i \\ -i & 1 \end{pmatrix}
    \label{eq:rho_RCP}
\end{equation}

\paragraph{Output beam via the Mueller route.}

Applying $\mathbf{M}_{\mathrm{QWP}}^{(45^\circ)}$ (Eq.~\ref{eq:MQWP45}) to
$\mathbf{S}_{\mathrm{in}}$ (Eq.~\ref{eq:Sin}):
\begin{equation}
    \mathbf{S}_{\mathrm{out}} =
    \begin{pmatrix}
        1 &  0 & 0 & 0 \\
        0 &  0 & 0 & 1 \\
        0 &  0 & 1 & 0 \\
        0 & -1 & 0 & 0
    \end{pmatrix}
    \begin{pmatrix} 1 \\ 1 \\ 0 \\ 0 \end{pmatrix}
    = \begin{pmatrix} 1 \\ 0 \\ 0 \\ -1 \end{pmatrix}
\end{equation}

The Mueller route reproduces exactly the Stokes vector obtained from the Jones
route: both give $S_3 = -1$, i.e.\ right circular polarization in the
convention of Section~\ref{sec:sign_convention}. This agreement is automatic because
Eq.~(\ref{eq:S3def}) and the Pauli trace formula (Eq.~\ref{eq:PauliTrace}) use
the same sign convention.

\paragraph{Time-domain picture.}

In real space and time, inserting $\mathbf{E}_{\mathrm{out}}$ of
Eq.~(\ref{eq:Eout_RCP}) into Eqs.~(\ref{eq:physical_wave})--(\ref{eq:real_components})
(at $z=0$, with $\phi_x = 0$, $\phi_y = -\pi/2$), the output field components are:
\begin{equation}
    \mathcal{E}_x(t) = \frac{1}{\sqrt{2}}\cos\omega t, \qquad
    \mathcal{E}_y(t) = -\frac{1}{\sqrt{2}}\sin\omega t
\end{equation}

The tip of the electric field vector traces a \textbf{circle} of radius
$1/\sqrt{2}$ in the $(x,y)$ plane, confirming circular polarization. The
rotation is clockwise when viewed looking toward the source --- right circular
polarization (RCP) in the classical optics naming adopted here
(Section~\ref{sec:sign_convention} and Ref.~\cite{detlefs2012}).

\paragraph{Summary: both routes give RCP.}

\begin{equation}
    {
    \mathbf{S}_{\mathrm{out}} =
    \begin{pmatrix} 1 \\ 0 \\ 0 \\ -1 \end{pmatrix}
    \quad\text{(RCP), \quad DOCP} = -1
    }
\end{equation}

The state sits at the \textbf{south pole} of the Poincar\'{e} sphere.

\subsubsection{QWP at $\chi = -45^\circ$ to get Left Circular Polarization}
\label{sec:QWPm45}

Rotating the QWP to $\chi = -45^\circ$ reverses the sign of the off-diagonal
terms in the Jones matrix:
\begin{equation}
    \mathbf{J}_{\mathrm{QWP}}^{(-45^\circ)} =
    \frac{e^{i\pi/4}}{\sqrt{2}}\begin{pmatrix} 1 & i \\ i & 1 \end{pmatrix}
    \quad\Longrightarrow\quad
    \mathbf{E}_{\mathrm{out}} = \frac{e^{i\pi/4}}{\sqrt{2}}\begin{pmatrix} 1 \\ i \end{pmatrix},
    \quad
    \mathbf{S}_{\mathrm{out}} = \begin{pmatrix}1\\0\\0\\+1\end{pmatrix}
    \quad\text{(LCP)}
\end{equation}

The corresponding Mueller matrix is:
\begin{equation}
    \mathbf{M}_{\mathrm{QWP}}^{(-45^\circ)} =
    \begin{pmatrix}
        1 & 0 &  0 &  0 \\
        0 & 0 &  0 & -1 \\
        0 & 0 &  1 &  0 \\
        0 & 1 &  0 &  0
    \end{pmatrix}
    \label{eq:MQWPm45}
\end{equation}

which maps horizontal linear ($S_1=1$) to $S_3=+1$ (LCP), as expected. The
density matrix of the LCP state, from
$\boldsymbol{\rho} = \mathbf{E}_{\mathrm{out}}\mathbf{E}_{\mathrm{out}}^\dagger$
or from Eq.~(\ref{eq:rho_from_Stokes}) with $\mathbf{S} = (1,0,0,+1)^T$, is:
\begin{equation}
    \boldsymbol{\rho}_{\mathrm{LCP}} = \frac{1}{2}
    \begin{pmatrix} 1 & -i \\ i & 1 \end{pmatrix}
    \label{eq:rho_LCP}
\end{equation}
Note that $\boldsymbol{\rho}_{\mathrm{RCP}}$ and $\boldsymbol{\rho}_{\mathrm{LCP}}$
differ only in the sign of the imaginary off-diagonal element, which carries
$S_3$ (Eq.~\ref{eq:rho_from_Stokes}).

\subsubsection{QWP with incident circular polarization}

For incident circular polarization with $\mathbf{E}_{\mathrm{in}} = \frac{1}{\sqrt{2}} \begin{pmatrix} 1 \\ \pm i \end{pmatrix}$, or equivalently $\mathbf{S}_{\mathrm{in}}=(1,0,0,\pm 1)$, Eqs.~(\ref{eq:JQWP_general}) and (\ref{eq:MQWP_general}) yield

\begin{align}
    \mathbf{E}_{\mathrm{out}}
    &=
    \frac{e^{i\pi/4}}{2}
    \begin{pmatrix}
    1 - i e^{\pm 2i \chi} \\
    \pm(i - e^{\pm 2i \chi})
    \end{pmatrix}
    =
    \frac{1+i}{2}\,\frac{1}{\sqrt{2}}
    \begin{pmatrix} 1 \\ \pm i \end{pmatrix}
    + \frac{1-i}{2}\,e^{\pm 2i\chi}\,\frac{1}{\sqrt{2}}
    \begin{pmatrix} 1 \\ \mp i \end{pmatrix}
\\
    \mathbf{S}_{\mathrm{out}}
    & =
    \begin{pmatrix}
    1 & 0 & 0 & 0 \\
    0 & \cos^2(2\chi) & \frac{1}{2} \sin(4\chi) & \sin(2\chi) \\
    0 & \frac{1}{2} \sin(4\chi) & \sin^2(2\chi) & -\cos(2\chi) \\
    0 & -\sin(2\chi) & \cos(2\chi) & 0
    \end{pmatrix}
    \begin{pmatrix} 1 \\ 0 \\ 0 \\ \pm 1 \end{pmatrix}
    =
    \begin{pmatrix}
    1 \\ \pm \sin(2\chi) \\ \mp \cos(2\chi) \\ 0
    \end{pmatrix}
\end{align}

We see that the quarter-wave plate turns circularly polarized light into linearly polarized light, with the plane of polarization rotated to $\theta = \chi \mp \pi/4$. The second form of $\mathbf{E}_{\mathrm{out}}$ shows how: the output is the superposition, with equal amplitudes, of the incident circular state and of the state of opposite handedness, which is a linear polarization whose direction is set by the relative phase $e^{\pm 2i\chi}$.

Consequently, two quarter-wave plates, one oriented at $\chi=\pm \pi/4$ and one at $\chi=\theta \mp \pi/4$, can be combined to rotate the plane of linear polarization from 0 (horizontal) to $\theta$.

\subsection{Half-wave plate}
\label{sec:HWP}

For a \textbf{half-wave plate} (HWP), setting $\delta=\pi$ in Eq.~(\ref{eq:Jret0}) yields the Jones matrix

\begin{equation}
    \mathbf{J}^{(0)}_{\mathrm{HWP}}
    =
    \begin{pmatrix}1 & 0 \\ 0 & -1 \end{pmatrix} = \boldsymbol{\sigma}_1 .
\end{equation}

Setting $\delta = \pi$ in Eq.~(\ref{eq:Mret0}) (equivalently, using
Eq.~\ref{eq:PauliTrace}), the corresponding aligned Mueller matrix is:
\begin{equation}
    {
    \mathbf{M}_{\mathrm{HWP}}^{(0)} =
    \begin{pmatrix}
        1 & 0 &  0 &  0 \\
        0 & 1 &  0 &  0 \\
        0 & 0 & -1 &  0 \\
        0 & 0 &  0 &  -1
    \end{pmatrix} .
    }
    \label{eq:MHWP0}
\end{equation}

Again, this matrix leaves $S_0$ and $S_1$ unchanged (H/V polarization unaffected since
the fast axis is along $\hat{x}$) and gives $S_2' = -S_2$, $S_3' = -S_3$: it
exchanges $+45^\circ$ and $-45^\circ$ linear polarization and reverses the
handedness of circular polarization.

For a half-wave plate with fast axis rotated to angle $\chi$, setting $\delta=\pi$ in Eq.~(\ref{eq:Jret_general}) gives the full Jones matrix:
\begin{equation}
    \mathbf{J}_{\mathrm{HWP}}^{(\chi)}
    =
    \begin{pmatrix}
        \cos(2\chi)
        &
        \sin(2\chi)
        \\
        \sin(2\chi)
        &
        -\cos(2\chi)
    \end{pmatrix}
    \label{eq:JHWP_general}
\end{equation}

Likewise, $\delta=\pi$ in Eq.~(\ref{eq:Mret_general}) gives the full Mueller matrix:
\begin{equation}
    \mathbf{M}_{\mathrm{HWP}}^{(\chi)} =
    \begin{pmatrix}
        1 & 0 & 0 & 0 \\[4pt]
        0 & \cos(4\chi)
          & \sin(4\chi)
          & 0\\[4pt]
        0 & \sin(4\chi)
          & -\cos(4\chi)
          & 0\\[4pt]
        0 & 0 & 0 & -1
    \end{pmatrix}
    \label{eq:MHWP_general}
\end{equation}

One verifies immediately: $\chi=0$ recovers Eq.~(\ref{eq:MHWP0})
\checkmark.

Furthermore, we see that $\mathrm{DOP}_{\mathrm{in}} = \mathrm{DOP}_{\mathrm{out}}$ and $\mathrm{DOLP}_{\mathrm{in}}=\mathrm{DOLP}_{\mathrm{out}}$, i.e.\ the degrees of total and linear polarization are conserved, but the direction of linear polarization is reflected about the fast axis.
Circular polarization is again inverted, $S'_3 = -S_3$.

Applying Eqs.~(\ref{eq:JHWP_general}) and (\ref{eq:MHWP_general}) to linear polarization at a general angle $\theta_{\mathrm{in}}$ yields:

\begin{align}
    \mathbf{E}_{\mathrm{out}}
    &=
    \begin{pmatrix}
        \cos(2\chi - \theta_{\mathrm{in}})  \\
        \sin(2\chi - \theta_{\mathrm{in}})
    \end{pmatrix}
    \\
    \mathbf{S}_{\mathrm{out}}
    &=
    \begin{pmatrix}
        1 \\
        \cos(4\chi - 2\theta_{\mathrm{in}})  \\
        \sin(4\chi - 2\theta_{\mathrm{in}}) \\
        0
    \end{pmatrix}
\end{align}

The half-wave plate reflects incident linear polarization at angle $\theta_{\mathrm{in}}$ about its fast axis, to $\theta_{\mathrm{out}} = 2\chi - \theta_{\mathrm{in}}$.

\subsection{Summary of Key Jones and Mueller Matrices}

\begin{table}[H]
\centering
\small
\renewcommand{\arraystretch}{1.8}
\begin{tabular}{p{3.5cm}p{4.5cm}p{5.0cm}}
\toprule
\textbf{Element} & \textbf{Jones matrix} $\mathbf{J}$ & \textbf{Mueller matrix} $\mathbf{M}$ \\
\midrule

QWP, fast axis at $\chi=0^\circ$ &
$\begin{pmatrix}1&0\\0&i\end{pmatrix}$ &
$\begin{pmatrix}1&0&0&0\\0&1&0&0\\0&0&0&-1\\0&0&+1&0\end{pmatrix}$ \\[14pt]

QWP, fast axis at $\chi=+45^\circ$ &
$\dfrac{e^{i\pi/4}}{\sqrt{2}}\!\begin{pmatrix}1&{-i}\\{-i}&1\end{pmatrix}$ &
$\begin{pmatrix}1&0&0&0\\0&0&0&1\\0&0&1&0\\0&{-1}&0&0\end{pmatrix}$ \\[14pt]

QWP, fast axis at $\chi=-45^\circ$ &
$\dfrac{e^{i\pi/4}}{\sqrt{2}}\!\begin{pmatrix}1&{+i}\\{+i}&1\end{pmatrix}$ &
$\begin{pmatrix}1&0&0&0\\0&0&0&{-1}\\0&0&1&0\\0&1&0&0\end{pmatrix}$ \\[14pt]

General retarder,\newline fast axis at $\chi$, retardation $\delta$ &
$\mathbf{R}_J(-\chi)\!\begin{pmatrix}1&0\\0&e^{i\delta}\end{pmatrix}\!\mathbf{R}_J(\chi)$,\newline see Eq.~(\ref{eq:Jret_general}) &
see Eq.~(\ref{eq:Mret_general}) \\[14pt]

\bottomrule
\end{tabular}
\caption{Jones and Mueller matrices of the general retarder and of the quarter-wave plate. The general retarder
entry uses the rotation sandwich
$\mathbf{R}_J(-\chi)\,\mathbf{J}^{(0)}\,\mathbf{R}_J(\chi)$
of Eq.~(\ref{eq:Jones_rotation}) (see Section~\ref{sec:rotation}). Jones
matrices are defined up to a global phase factor, which changes no
observable; the factor $e^{i\pi/4}$ of the QWP at $\pm45^\circ$ is kept so
that the entries equal Eq.~(\ref{eq:Jret_general}) exactly. The rotation matrices
$\mathbf{R}_J(\chi)$ and $\mathbf{R}_M(\chi)$ themselves are given in
Eqs.~(\ref{eq:Rchi}) and (\ref{eq:RM}).}
\label{tab:JM_elements}
\end{table}

\begin{table}[H]
\centering
\small
\renewcommand{\arraystretch}{1.8}
\begin{tabular}{p{3.5cm}p{4.5cm}p{5.0cm}}
\toprule
\textbf{Element} & \textbf{Jones matrix} $\mathbf{J}$ & \textbf{Mueller matrix} $\mathbf{M}$ \\
\midrule

HWP, fast axis at $\chi=0^\circ$ &
$\begin{pmatrix}1&0\\0&-1\end{pmatrix}$ &
$\begin{pmatrix}1&0&0&0\\0&1&0&0\\0&0&-1&0\\0&0&0&-1\end{pmatrix}$ \\[14pt]

HWP, fast axis at $\chi=22.5^\circ$\newline(H $\to$ $+45^\circ$) &
$\dfrac{1}{\sqrt{2}}\!\begin{pmatrix}1&1\\1&-1\end{pmatrix}$ &
$\begin{pmatrix}1&0&0&0\\0&0&1&0\\0&1&0&0\\0&0&0&-1\end{pmatrix}$ \\[14pt]

HWP, fast axis at $\chi=45^\circ$\newline(H $\leftrightarrow$ V) &
$\begin{pmatrix}0&1\\1&0\end{pmatrix}$ &
$\begin{pmatrix}1&0&0&0\\0&-1&0&0\\0&0&1&0\\0&0&0&-1\end{pmatrix}$ \\[14pt]

HWP, fast axis at $\chi$ &
$\begin{pmatrix}\cos 2\chi&\sin 2\chi\\\sin 2\chi&-\cos 2\chi\end{pmatrix}$ &
see Eq.~(\ref{eq:MHWP_general}) \\

\bottomrule
\end{tabular}
\caption{Jones and Mueller matrices of the half-wave plate ($\delta = \pi$,
Eqs.~\ref{eq:MHWP0}--\ref{eq:MHWP_general}). At $\chi$ the HWP turns linear
polarization at $\theta$ into linear polarization at $2\chi - \theta$, and
reverses the handedness of circular polarization.}
\label{tab:JM_HWP}
\end{table}

\section{Linear Polarizers}
\label{sec:general_polarizers}

A \textbf{linear polarizer} is the second basic building block of
polarization optics, alongside the retarder of
Section~\ref{sec:general_retarders}. Where a retarder is lossless and merely
dephases the two eigenpolarizations, an ideal polarizer transmits the field
component along its \textbf{transmission axis} unattenuated and completely
blocks the orthogonal component: it converts an arbitrary input into a
single, fixed polarization state, at the cost of discarding whatever
intensity does not project onto that axis. This section derives its aligned
and rotated Jones and Mueller matrices, following the same pattern used for
the retarder, for use in the polarimetry protocol of
Section~\ref{sec:experimental}. Figure~\ref{fig:polarizer} illustrates its
action.

\subsection{Jones and Mueller Matrix of an Aligned Linear Polarizer}
\label{sec:polarizer_aligned}

\begin{figure}[H]
\centering
\begin{tikzpicture}[>=stealth, scale=0.95]

\begin{scope}[shift={(1.5,0)}]
    \draw[->, gray] (-1.35,0) -- (1.35,0) node[right, font=\small] {$x$};
    \draw[->, gray] (0,-1.35) -- (0,1.4) node[above, font=\small] {$y$};
    \draw[<->, red, very thick] (-0.246,-0.918) -- (0.246,0.918);
    \draw[->] (0.5,0) arc (0:75:0.5);
    \node[font=\small] at (0.42,0.62) {$\alpha$};
    \node[font=\small, align=center] at (0,-2.1)
        {(a) in: linear\\at $\alpha$};
\end{scope}

\draw[->, very thick] (3.2,0) -- (4.3,0);

\begin{scope}[shift={(5.7,0)}]
    \draw[fill=blue!6, draw=black] (-0.72,-1.35) -- (0.72,-0.95)
        -- (0.72,1.4) -- (-0.72,1.0) -- cycle;
    \begin{scope}
        \clip (-0.72,-1.35) -- (0.72,-0.95) -- (0.72,1.4) -- (-0.72,1.0) -- cycle;
        \foreach \k in {-2.4,-2.0,...,2.4} {
            \draw[blue!35] (\k-1.5,-1.05) -- (\k+1.5,1.05);
        }
    \end{scope}
    \draw[<->, blue, very thick] (-0.62,-0.434) -- (0.62,0.434);
    \node[font=\small, align=center] at (0,-2.1)
        {(b) polarizer,\\axis at $\theta$};
\end{scope}

\draw[->, very thick] (7.9,0) -- (9.0,0);

\begin{scope}[shift={(10.4,0)}]
    \draw[->, gray] (-1.35,0) -- (1.35,0) node[right, font=\small] {$x$};
    \draw[->, gray] (0,-1.35) -- (0,1.4) node[above, font=\small] {$y$};
    \draw[blue!45, thick] (-1.15,-0.805) -- (1.15,0.805);
    \draw[<->, red, very thick] (-0.598,-0.419) -- (0.598,0.419);
    \draw[->] (0.5,0) arc (0:35:0.5);
    \node[font=\small] at (0.68,0.19) {$\theta$};
    \node[font=\small, align=center] at (0,-2.1)
        {(c) out: linear\\along the axis};
\end{scope}

\begin{scope}[shift={(5.9,-5.9)}]
    \draw[->, gray] (-2.1,0) -- (2.5,0) node[right, font=\small] {$x$};
    \draw[->, gray] (0,-1.5) -- (0,2.15) node[above, font=\small] {$y$};

    \draw[blue!70, thick] (-1.515,-1.062) -- (1.515,1.062);
    \node[blue!70, font=\small, anchor=west] at (1.50,1.20) {axis};

    \draw[->, red, very thick] (0,0) -- (0.440,1.642);
    \node[red, font=\small] at (0.26,1.90) {$\mathbf{E}$};

    \draw[gray, dashed] (0.440,1.642) -- (1.066,0.747);
    \draw[->, red!65!black, line width=2pt] (0,0) -- (1.066,0.747);
    \draw[gray] (0.960,0.672) -- (0.886,0.778) -- (0.992,0.853);
    \node[red!65!black, font=\small, anchor=west] at (1.16,0.40)
        {$|\mathbf{E}|\cos(\alpha\!-\!\theta)$};

    \draw[->] (0.50,0) arc (0:35:0.50);
    \node[font=\small] at (0.70,0.13) {$\theta$};
    \draw[->] (0.655,0.459) arc (35:75:0.80);
    \node[font=\small] at (0.70,0.74) {$\alpha\!-\!\theta$};

    \node[font=\small] at (0,-2.5)
        {(d) the transmitted field is the projection of $\mathbf{E}$ on the axis};
\end{scope}

\end{tikzpicture}
\caption{The ideal linear polarizer, drawn as in Figure~\ref{fig:retarder}.
Top row, the transformation: (a) an input linearly polarized at angle
$\alpha$; (b) the polarizer, whose transmission axis lies at $\theta$; (c) the
output, linearly polarized \emph{along the transmission axis} whatever the
input was, with its amplitude reduced to $|\mathbf{E}|\cos(\alpha-\theta)$.
Bottom, (d): the same statement as a construction --- the transmitted field is
the orthogonal projection of $\mathbf{E}$ onto the transmission axis, whence
the intensity $|\mathbf{E}|^2\cos^2(\alpha-\theta)$ of Malus' law. This is the
amplitude-level content of the projector $\mathbf{J}_{LP}$ of
Eqs.~(\ref{eq:JLP0}) and (\ref{eq:JLPtheta}). Because a projector is singular,
the discarded component is lost: unlike the retarder of
Figure~\ref{fig:retarder}, a polarizer is not a lossless element. The general
intensity transmitted by this element, for an arbitrary (and possibly
partially polarized) input, is Eq.~(\ref{eq:Itheta}).}
\label{fig:polarizer}
\end{figure}

With the transmission axis aligned along $\hat{x}$ ($\theta = 0$), the
aligned Jones matrix is the projector onto $\hat{x}$:
\begin{equation}
    \mathbf{J}_{LP}^{(0)} = \begin{pmatrix} 1 & 0 \\ 0 & 0 \end{pmatrix} .
    \label{eq:JLP0}
\end{equation}
Unlike the retarder of Eq.~(\ref{eq:Jret0}), which is unitary (with
$|\det| = 1$), hence lossless, $\mathbf{J}_{LP}^{(0)}$ is
\emph{idempotent}, $\bigl(\mathbf{J}_{LP}^{(0)}\bigr)^2 =
\mathbf{J}_{LP}^{(0)}$, and singular ($\det = 0$): it is a rank-1 projector,
not a unitary matrix, and is lossy for any input not already polarized
along its transmission axis: in general $S_0' \leq S_0$, with equality only
for a fully polarized input along that axis.

Applying the Pauli trace formula (Eq.~\ref{eq:PauliTrace}) to
$\mathbf{J}_{LP}^{(0)}$ gives the aligned Mueller matrix:
\begin{equation}
    \mathbf{M}_{LP}^{(0)} = \frac{1}{2}
    \begin{pmatrix}
        1 & 1 & 0 & 0 \\
        1 & 1 & 0 & 0 \\
        0 & 0 & 0 & 0 \\
        0 & 0 & 0 & 0
    \end{pmatrix}
    \label{eq:MLP0}
\end{equation}
which halves the intensity of unpolarized light ($S_1=S_2=S_3=0$) and passes
horizontally polarized light ($S_1=S_0$) unattenuated, as expected of an
ideal polarizer aligned with the input. Because the polarizer changes the
intensity, it is here that the Stokes parameters and the normalized
Poincar\'{e}--Stokes parameters of Eq.~(\ref{eq:Pdef}) must be kept apart.
For any input, $\mathbf{M}_{LP}^{(0)}\mathbf{S} = \tfrac{1}{2}S_0(1+P_1)\,
(1,1,0,0)^T$: the transmitted intensity, $S_0' = \tfrac{1}{2}S_0(1+P_1)$,
depends on the input polarization, while the output is always fully and
horizontally polarized (whenever $S_0' \neq 0$), $(P_1', P_2', P_3') = (1,0,0)$.

\subsection{Rotated Linear Polarizer at Angle $\theta$}
\label{sec:polarizer_rotated}

For a polarizer with transmission axis at angle $\theta$ to the horizontal
(measured in the same sense as $\chi$ in Section~\ref{sec:rotation}),
applying the rotation sandwich of Eq.~(\ref{eq:Jones_rotation}) to
$\mathbf{J}_{LP}^{(0)}$ gives:
\begin{equation}
    \mathbf{J}_{LP}(\theta) =
    \mathbf{R}_J(-\theta)\,\mathbf{J}_{LP}^{(0)}\,\mathbf{R}_J(\theta) =
    \begin{pmatrix}
        \cos^2\theta & \cos\theta\sin\theta \\
        \cos\theta\sin\theta & \sin^2\theta
    \end{pmatrix}
    \label{eq:JLPtheta}
\end{equation}
and, applying the Mueller rotation of Eq.~(\ref{eq:Mueller_rotation})
(equivalently, the Pauli trace formula) to $\mathbf{M}_{LP}^{(0)}$:
\begin{equation}
    \mathbf{M}_{LP}(\theta) = \frac{1}{2}
    \begin{pmatrix}
        1            &  \cos 2\theta   &  \sin 2\theta            & 0 \\
        \cos 2\theta &  \cos^2 2\theta & \sin 2\theta\cos 2\theta & 0 \\
        \sin 2\theta & \sin 2\theta\cos 2\theta & \sin^2 2\theta  & 0 \\
        0            &  0              &  0                       & 0
    \end{pmatrix}
    \label{eq:MLPtheta}
\end{equation}
One verifies immediately: $\theta = 0$ recovers Eq.~(\ref{eq:JLP0}) and
Eq.~(\ref{eq:MLP0}) \checkmark. These are the matrices used throughout
Section~\ref{sec:experimental} below.

\subsection{Real Polarizers: Leakage}
\label{sec:polarizer_leaky}

The polarizer of Eq.~(\ref{eq:JLP0}) is an idealized polarization filter: a
projector with eigenvalues 1 and 0. A real polarizer transmits a small
fraction of the orthogonal component as well. Its aligned Jones matrix is
\begin{equation}
    \mathbf{J}_{\varepsilon}^{(0)} =
    \begin{pmatrix} 1 & 0 \\ 0 & \varepsilon \end{pmatrix},
    \qquad |\varepsilon| \ll 1
    \label{eq:JLPleak}
\end{equation}
where the complex amplitude $\varepsilon$ is the leakage, and
$|\varepsilon|^2$ is the intensity leakage (the inverse of the extinction ratio as often defined; for a
crystal polarizer, $|\varepsilon|^2 = R_p/R_s$,
Section~\ref{sec:polarizer_brewster}, Eq.~\ref{eq:delta0_def}). Eq.~(\ref{eq:JLPleak}) is the crystal matrix of
Section~\ref{sec:crystal_plate} with $r_s = 1$ and $r_p = \varepsilon$, and
its Mueller matrix is therefore Eq.~(\ref{eq:M0_Pauli}) with $\bar{R} =
(1+|\varepsilon|^2)/2$, $\tilde{R} = (1-|\varepsilon|^2)/2$, $R_{sp} =
|\varepsilon|$ and $\delta = \arg\varepsilon$, i.e.\
$R_{sp}\cos\delta = \varepsilon'$ and $R_{sp}\sin\delta = \varepsilon''$,
with $\varepsilon = \varepsilon' + i\varepsilon''$.

\paragraph{Rotated leaky polarizer.} With the transmission axis at angle
$\theta$, the rotation sandwich of Eq.~(\ref{eq:Jones_rotation}) gives
the Jones matrix
\begin{equation}
    \mathbf{J}_{\varepsilon}(\theta) =
    \mathbf{R}_J(-\theta)\,\mathbf{J}_{\varepsilon}^{(0)}\,\mathbf{R}_J(\theta) =
    \begin{pmatrix}
        \cos^2\theta + \varepsilon\sin^2\theta & (1-\varepsilon)\cos\theta\sin\theta \\
        (1-\varepsilon)\cos\theta\sin\theta & \sin^2\theta + \varepsilon\cos^2\theta
    \end{pmatrix}
    \label{eq:JLPleak_theta}
\end{equation}
and Eq.~(\ref{eq:Mueller_final}) with $\chi = \theta$ and the values above
gives the Mueller matrix ($c_2 \equiv \cos 2\theta$, $s_2 \equiv \sin
2\theta$)
\begin{equation}
    \mathbf{M}_{\varepsilon}(\theta) = \frac{1}{2}
    \begin{pmatrix}
        1+|\varepsilon|^2 & (1-|\varepsilon|^2)\,c_2 & (1-|\varepsilon|^2)\,s_2 & 0 \\[4pt]
        (1-|\varepsilon|^2)\,c_2 & (1+|\varepsilon|^2)\,c_2^2 + 2\varepsilon'\,s_2^2
            & (1+|\varepsilon|^2-2\varepsilon')\,s_2c_2 & 2\varepsilon''\,s_2 \\[4pt]
        (1-|\varepsilon|^2)\,s_2 & (1+|\varepsilon|^2-2\varepsilon')\,s_2c_2
            & (1+|\varepsilon|^2)\,s_2^2 + 2\varepsilon'\,c_2^2 & -2\varepsilon''\,c_2 \\[4pt]
        0 & -2\varepsilon''\,s_2 & 2\varepsilon''\,c_2 & 2\varepsilon'
    \end{pmatrix}
    \label{eq:MLPleak_theta}
\end{equation}
Both reduce to Eqs.~(\ref{eq:JLPtheta})--(\ref{eq:MLPtheta}) for
$\varepsilon = 0$. Two consequences follow.

\paragraph{Repeated passes reduce the leakage.} $\mathbf{J}_{\varepsilon}^{(0)}$
is no longer idempotent: $N$ passes through the same polarizer give
$\bigl(\mathbf{J}_{\varepsilon}^{(0)}\bigr)^N = \mathrm{diag}(1, \varepsilon^N)$,
an intensity leakage $|\varepsilon|^{2N}$, whereas the ideal polarizer gains
nothing from a second pass. This is the principle of the multi-reflection
channel-cut polarizers of Section~\ref{sec:polarizer_brewster}, whose
leakage decreases roughly as $|\varepsilon|^{2N}$. It applies to the leakage
contained in each ray's own Jones matrix; the geometric leakage due to the
divergence of the beam (Eq.~\ref{eq:Plim}) is not reduced by additional
passes.

\paragraph{A leaky analyzer reduces the measured contrast.} Rotated to
$\theta$, the first row of its Mueller matrix (Eq.~\ref{eq:Mueller_final}
with the values above) gives the transmitted intensity
\begin{equation}
    I_{\varepsilon}(\theta) = \frac{1+|\varepsilon|^2}{2}\,S_0
    \left[1 + \frac{1-|\varepsilon|^2}{1+|\varepsilon|^2}
    \bigl(P_1\cos 2\theta + P_2\sin 2\theta\bigr)\right]
    \label{eq:Itheta_leak}
\end{equation}
which reduces to Eq.~(\ref{eq:Itheta}) of the next section for
$\varepsilon = 0$. A fit of measured intensities with the ideal formula
therefore returns $P_1$ and $P_2$ multiplied by $(1-|\varepsilon|^2)/
(1+|\varepsilon|^2)$, a relative correction of $-2|\varepsilon|^2$ to first order.

\section{Experimental Polarization Analysis}
\label{sec:experimental}

\subsection{Principle}

A linear polarizer placed in the beam transmits only the projection of the
electric field onto its transmission axis. By recording the transmitted intensity
$I(\theta)$ as a function of the polarizer rotation angle $\theta$, and by
optionally inserting a quarter-wave plate (QWP) analyzer before the polarizer,
all four Stokes parameters of an \emph{arbitrary} polarization state
$(S_0, S_1, S_2, S_3)$ can be determined.

\subsection{Transmitted Intensity Formula}

For a linear polarizer at angle $\theta$ to the horizontal, the Jones matrix
$\mathbf{J}_{LP}(\theta)$ and Mueller matrix $\mathbf{M}_{LP}(\theta)$ are
those of Eqs.~(\ref{eq:JLPtheta})--(\ref{eq:MLPtheta}) of
Section~\ref{sec:general_polarizers}. Applying these to an arbitrary incident Stokes vector
$\mathbf{S} = (S_0, S_1, S_2, S_3)^T$, the detected intensity is the first
component of $\mathbf{M}_{LP}(\theta)\,\mathbf{S}$:
\begin{equation}
    {
    I(\theta) = \frac{1}{2}\bigl(S_0 + S_1\cos 2\theta + S_2\sin 2\theta\bigr)
    }
    \label{eq:Itheta}
\end{equation}

This is a \textbf{sinusoidal function of $2\theta$} with mean $S_0/2$, cosine
amplitude $S_1/2$, and sine amplitude $S_2/2$. In terms of the
Poincar\'{e}--Stokes parameters of Eq.~(\ref{eq:Pdef}) it reads
\begin{equation}
    I(\theta) = \frac{S_0}{2}\bigl[1 + P_1\cos 2\theta + P_2\sin 2\theta\bigr]
    \label{eq:Itheta_P}
\end{equation}
which separates the intensity $S_0$ from the polarization. Writing
$P_1 = \mathrm{DOLP}\cos 2\theta_{\mathrm{in}}$ and $P_2 =
\mathrm{DOLP}\sin 2\theta_{\mathrm{in}}$, with $\mathrm{DOLP}$ of
Eq.~(\ref{eq:DOCP_def}) and $\theta_{\mathrm{in}}$ the angle of the linearly
polarized part of the beam to the horizontal ($\tan 2\theta_{\mathrm{in}} =
P_2/P_1$), it becomes
\begin{equation}
    I(\theta) = \frac{S_0}{2}\bigl[1 + \mathrm{DOLP}\,
    \cos 2(\theta - \theta_{\mathrm{in}})\bigr]
    \label{eq:Itheta_DOLP}
\end{equation}
the modulation depth of the scan is the degree of linear polarization, and
the analyzer angle of maximum transmission gives its direction,
$\theta_{\mathrm{in}}$. This is the form used at beamlines (e.g.\
Ref.~\cite{mazzoli2007}), with $S_0$ then the monitor-normalized intensity.
For a real analyzer with leakage,
$P_1$ and $P_2$ are reduced by the factor of Eq.~(\ref{eq:Itheta_leak}).

Note that $S_3$ does \emph{not} appear: a linear polarizer alone is
insensitive to circular polarization. In particular, it cannot distinguish
circularly polarized light from unpolarized light, or from any mixture of
the two with the same $S_0$: all give the same flat scan, $I(\theta) =
S_0/2$ (Fig.~\ref{fig:analyzer_scans}).

\subsection{Measurements at Key Angles: General Case}
\label{sec:key_angles}

Four orientations of the polarizer are sufficient to
determine $S_0$, $S_1$, and $S_2$ from Eq.~(\ref{eq:Itheta}).

\paragraph{$\theta = 0^\circ$ (Horizontal transmission axis)}
\begin{equation}
    I_H \equiv I(0^\circ) = \frac{1}{2}(S_0 + S_1)
    \label{eq:IH}
\end{equation}

\paragraph{$\theta = 90^\circ$ (Vertical transmission axis)}
\begin{equation}
    I_V \equiv I(90^\circ) = \frac{1}{2}(S_0 - S_1)
    \label{eq:IV}
\end{equation}

\paragraph{$\theta = 45^\circ$ ($+45^\circ$ transmission axis)}
\begin{equation}
    I_{45} \equiv I(45^\circ) = \frac{1}{2}(S_0 + S_2)
    \label{eq:I45}
\end{equation}

\paragraph{$\theta = 135^\circ$ ($-45^\circ$ transmission axis)}
\begin{equation}
    I_{135} \equiv I(135^\circ) = \frac{1}{2}(S_0 - S_2)
    \label{eq:I135}
\end{equation}

From these four measurements the first three Stokes parameters are recovered by
simple linear combinations:
\begin{align}
    {S_0 = I_H + I_V = I_{45} + I_{135}}
    \label{eq:S0rec}\\[6pt]
    {S_1 = I_H - I_V}
    \label{eq:S1rec}\\[6pt]
    {S_2 = I_{45} - I_{135}}
    \label{eq:S2rec}
\end{align}

Equations~(\ref{eq:S0rec})--(\ref{eq:S2rec}) are valid for \emph{any} incident
polarization state. They show that $S_1$ is the difference between the
horizontal and vertical transmissions, and $S_2$ the difference between the
two diagonal transmissions.

These four measurements fully define $S_0$, $S_1$ and $S_2$. Except for $S_0$, however, they provide no redundancy and no estimate of the measurement error. An alternative is therefore to measure $I(\theta)$ at a larger number of angles $\theta$, and to fit the results to Eq.~(\ref{eq:Itheta}) or (\ref{eq:Itheta_P}).

Note that the circular component, $S_3$, cannot be determined with a linear polarization analyzer. 
We can only set an upper limit for $\left|S_3\right|$ by noting that $\mathrm{DOP} \leq 1$ and therefore $S_3^2 \leq S_0^2-S_1^2-S_2^2$.

\subsection{Measuring $S_3$: Adding a QWP Analyzer}

Since a linear polarizer alone cannot detect $S_3$, a QWP must be inserted
upstream of the polarizer. Its role is to convert circular polarization
into linear polarization that the polarizer can then analyze.

\begin{center}
\begin{tikzpicture}[>=stealth, thick, node distance=4.2cm]
  \node (beam) [draw, rounded corners, fill=orange!15,
        minimum width=2.2cm, minimum height=0.8cm] {Beam ($S_0,S_1,S_2,S_3$)};
  \node (qwp)  [draw, rounded corners, fill=green!15,
        minimum width=2.2cm, minimum height=0.8cm, right of=beam]
        {QWP @ $\pm45^\circ$};
  \node (lp)   [draw, rounded corners, fill=blue!15,
        minimum width=2.2cm, minimum height=0.8cm, right of=qwp]
        {LP @ $\theta=0^\circ$};
  \node (det)  [draw, rounded corners, fill=gray!15,
        minimum width=2cm, minimum height=0.8cm, right of=lp] {Detector};
  \draw[->] (beam) -- (qwp);
  \draw[->] (qwp)  -- (lp);
  \draw[->] (lp)   -- (det);
\end{tikzpicture}
\end{center}

In the Mueller formalism the detected intensity is the first component of
the product of the matrices of the elements, taken in the order opposite to
that of the beam path:
\begin{equation}
    I_{\pm}(\theta) =
    \Bigl[\mathbf{M}_{LP}(\theta)\,\mathbf{M}_{\mathrm{QWP}}^{(\pm45^\circ)}\,
    \mathbf{S}\Bigr]_0
    = \frac{1}{2}\bigl(S_0 \pm S_3\cos 2\theta + S_2\sin 2\theta\bigr)
    \label{eq:Ipm_theta}
\end{equation}
with $\mathbf{M}_{LP}(\theta)$ of Eq.~(\ref{eq:MLPtheta}) and
$\mathbf{M}_{\mathrm{QWP}}^{(\pm45^\circ)}$ of Eqs.~(\ref{eq:MQWP45}) and
(\ref{eq:MQWPm45}). The QWP at $+45^\circ$ maps $(S_1, S_2, S_3)$ onto
$(S_3, S_2, -S_1)$, so that the analyzer now sees $S_3$ in the place of
$S_1$; at $-45^\circ$ it sees $-S_3$. For the polarizer at $\theta=0^\circ$,
explicitly,
\begin{equation}
    I_{\pm}(0) = \Biggl[\frac{1}{2}
    \begin{pmatrix} 1&1&0&0\\1&1&0&0\\0&0&0&0\\0&0&0&0 \end{pmatrix}
    \begin{pmatrix} 1&0&0&0\\0&0&0&\pm1\\0&0&1&0\\0&\mp1&0&0 \end{pmatrix}
    \begin{pmatrix} S_0\\S_1\\S_2\\S_3 \end{pmatrix}\Biggr]_0
    = \frac{1}{2}(S_0 \pm S_3)
    \label{eq:Ipm0_explicit}
\end{equation}
which gives the two measurements usually quoted:

\begin{align}
    I_{+} &\equiv I\!\left[\mathrm{QWP}(+45^\circ),\,\mathrm{LP}(0^\circ)\right]
            = \frac{1}{2}(S_0 + S_3) \label{eq:Iplus}\\[4pt]
    I_{-} &\equiv I\!\left[\mathrm{QWP}(-45^\circ),\,\mathrm{LP}(0^\circ)\right]
            = \frac{1}{2}(S_0 - S_3) \label{eq:Iminus}
\end{align}

The general relation for $S_3$ is:
\begin{equation}
    S_3 = I_{+} - I_{-}
    \label{eq:S3rec}
\end{equation}

or equivalently, normalized to the total intensity:
\begin{equation}
    \mathrm{DOCP} = \frac{S_3}{S_0} = \frac{I_+ - I_-}{I_+ + I_-}
    \label{eq:DOCP_exp}
\end{equation}

\paragraph{Effect of the analyzer leakage.} The formulas above assume an
ideal analyzer. With the leaky polarizer of Section~\ref{sec:polarizer_leaky},
only the first row of its Mueller matrix, Eq.~(\ref{eq:MLPleak_theta}),
enters the detected intensity, and it does not depend on the phase of
$\varepsilon$. Without QWP the scan is Eq.~(\ref{eq:Itheta_leak}); with the
QWP at $\pm45^\circ$, Eq.~(\ref{eq:Ipm_theta}) becomes
\begin{equation}
    I_{\pm,\varepsilon}(\theta) = \frac{1}{2}\Bigl[(1+|\varepsilon|^2)\,S_0
    + (1-|\varepsilon|^2)\bigl(\pm S_3\cos 2\theta + S_2\sin 2\theta\bigr)\Bigr]
    \label{eq:Ipm_theta_leak}
\end{equation}
The key-angle combinations therefore give $I_H + I_V = (1+|\varepsilon|^2)\,
S_0$ and $I_H - I_V = (1-|\varepsilon|^2)\,S_1$, and similarly for $S_2$
and, with the QWP, for $S_3$. The ratios used in practice,
$P_1 = (I_H - I_V)/(I_H + I_V)$, $P_2$, and $\mathrm{DOCP}$ of
Eq.~(\ref{eq:DOCP_exp}), are all multiplied by the same factor
$(1-|\varepsilon|^2)/(1+|\varepsilon|^2) \simeq 1 - 2|\varepsilon|^2$: the
ideal formulas underestimate every degree of polarization by the relative
amount $2|\varepsilon|^2$, and the direction $\theta_{\mathrm{in}}$ is not
affected. For a perfect-crystal analyzer at exactly $2\theta_B = 90^\circ$,
$|\varepsilon|^2$ is of order $10^{-6}$ or smaller
(Section~\ref{sec:polarizer_brewster}) and the correction is negligible.
Away from $90^\circ$ it grows quickly: the ratio of the integrated $\pi$ and
$\sigma$ reflectivities of the analyzer is $|\cos 2\theta_B|$ for a perfect
crystal and $\cos^2 2\theta_B$ for a mosaic one (effective, angle-averaged
values of $|\varepsilon|^2$, valid when the divergence and bandwidth of the
beam cover the $\sigma$ rocking curve, or when the analyzer is rocked and
the intensity integrated), i.e.\ $3.5\times10^{-2}$ and
$1.2\times10^{-3}$ at $2\theta_B = 88^\circ$, and the degrees of
polarization are then underestimated by about 7\% and 0.24\%, respectively.
In that case the scans should be fitted with
Eqs.~(\ref{eq:Itheta_leak}) and (\ref{eq:Ipm_theta_leak}), with
$|\varepsilon|^2$ calculated or measured with a beam of known polarization.

\begin{table}[H]
\centering
\small
\begin{tabular}{llll}
\toprule
\textbf{Configuration} & \textbf{Detected intensity} & \textbf{Stokes relation} & \textbf{RCP value} \\
\midrule
No optics                      & $I_0    = S_0$                   & $S_0 = I_0$                          & $I_0$   \\
LP at $\theta=0^\circ$         & $I_H    = \tfrac{1}{2}(S_0+S_1)$ & $S_1 = 2I_H - S_0$                   & $I_0/2$ \\
LP at $\theta=90^\circ$        & $I_V    = \tfrac{1}{2}(S_0-S_1)$ & $S_0 = I_H + I_V$                    & $I_0/2$ \\
LP at $\theta=45^\circ$        & $I_{45} = \tfrac{1}{2}(S_0+S_2)$ & $S_2 = 2I_{45} - S_0$                & $I_0/2$ \\
LP at $\theta=135^\circ$       & $I_{135}=\tfrac{1}{2}(S_0-S_2)$  & check: $S_0 = I_{45}+I_{135}$        & $I_0/2$ \\
QWP$(+45^\circ)$+LP$(0^\circ)$ & $I_+    = \tfrac{1}{2}(S_0+S_3)$ & $S_3 = I_+ - I_-$                    & $0$     \\
QWP$(-45^\circ)$+LP$(0^\circ)$ & $I_-    = \tfrac{1}{2}(S_0-S_3)$ & $\mathrm{DOCP} = (I_+-I_-)/(I_++I_-)$& $I_0$   \\
\bottomrule
\end{tabular}
\caption{General six-measurement (seven, counting the direct $S_0$
measurement) Stokes polarimetry protocol, valid for any incident
polarization state $(S_0,S_1,S_2,S_3)$; $\theta$ is the polarizer
transmission-axis angle measured from horizontal. The \textbf{RCP value}
column gives the numerical outcome of each measurement for the right
circularly polarized beam $\mathbf{S} = I_0(1,0,0,-1)^T$ produced by the
QWP at $\chi=+45^\circ$ (Section~\ref{sec:QWP45}), derived in
Section~\ref{sec:circular_case} below.}
\label{tab:general_protocol}
\end{table}

\subsection{Particular Case: Circular Polarization, and the Complete Protocol}
\label{sec:circular_case}

For the RCP beam produced by the QWP at $\chi=+45^\circ$
(Section~\ref{sec:QWP45}), the Stokes vector is
$\mathbf{S} = I_0(1,\,0,\,0,\,-1)^T$, i.e.\ $S_1 = S_2 = 0$ and $S_3 = -S_0 =
-I_0$. Substituting into the general formulas of Table~\ref{tab:general_protocol}:

\paragraph{Linear polarizer scan.}
\begin{equation}
    I(\theta) = \frac{1}{2}(I_0 + 0\cdot\cos 2\theta + 0\cdot\sin 2\theta)
    = \frac{I_0}{2}, \quad \forall\,\theta
    \label{eq:Icirc}
\end{equation}

\begin{figure}[H]
\centering
\includegraphics[width=\textwidth]{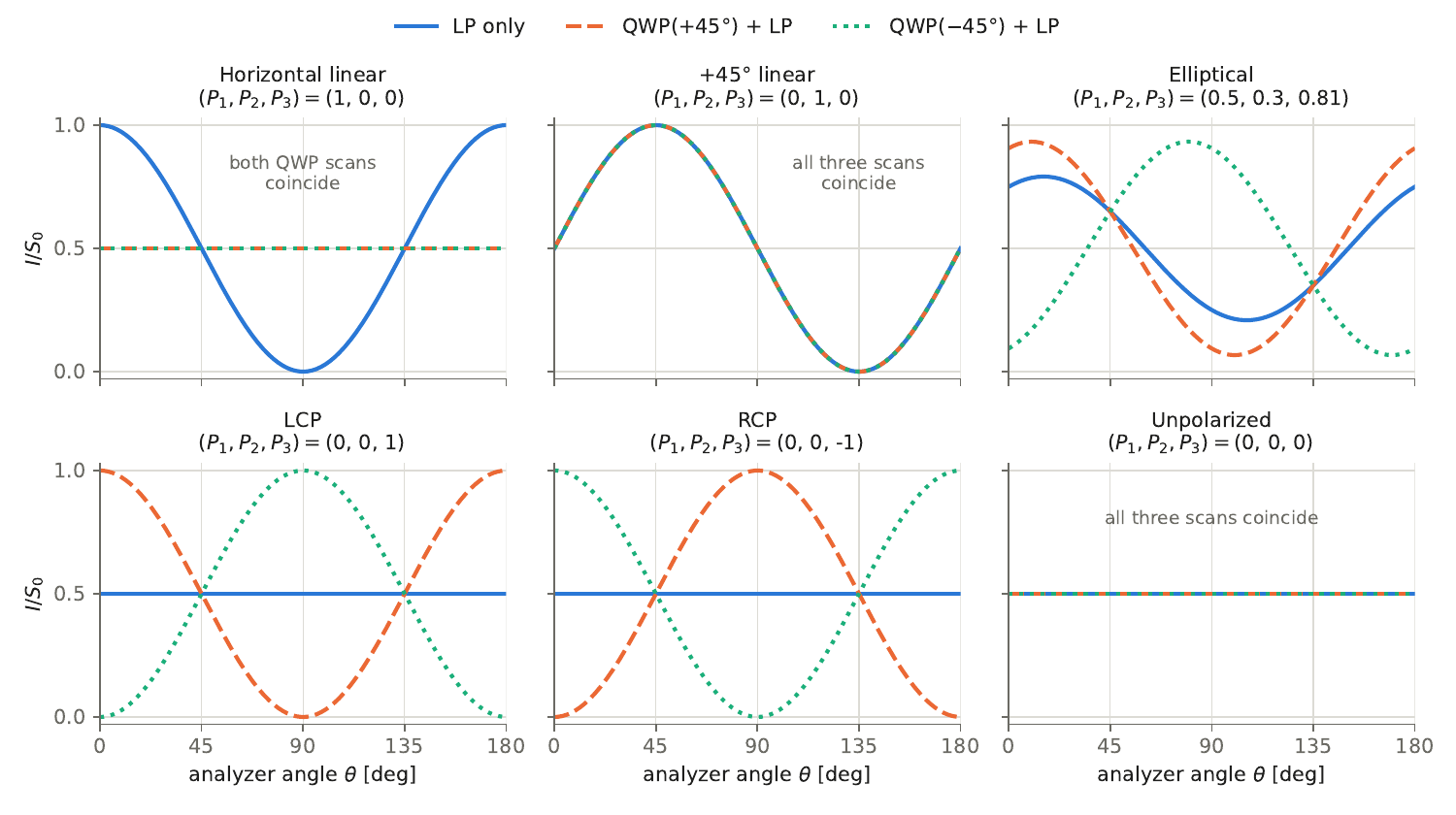}
\caption{Intensity behind a linear analyzer rotated by $\theta$, normalized
to $S_0$, for six polarization states (Poincar\'{e}--Stokes parameters in
each title), in the three scans of the protocol: linear analyzer only
(Eq.~\ref{eq:Itheta}), and with a QWP at $+45^\circ$ or $-45^\circ$ in
front of it (Eq.~\ref{eq:Ipm_theta}). The elliptical state has $(P_1, P_2)
= (0.5, 0.3)$ and $P_3 = \sqrt{1 - 0.5^2 - 0.3^2} = 0.81$. Without the QWP, LCP, RCP and unpolarized
light give the same flat scan. Computed with the Mueller matrices of
Sections~\ref{sec:general_retarders} and \ref{sec:general_polarizers}
(script \texttt{analyzer\_scans\_plot.py}).}
\label{fig:analyzer_scans}
\end{figure}

The intensity is \textbf{constant regardless of polarizer angle}: because
$S_1 = S_2 = 0$, the sinusoidal modulation of Eq.~(\ref{eq:Itheta}) vanishes
completely. A flat scan is necessary for circular polarization but not
sufficient: an unpolarized beam gives the same result
(Fig.~\ref{fig:analyzer_scans}), and only the QWP analyzer below tells the
two apart.

\paragraph{Key angle measurements.}
From Eqs.~(\ref{eq:IH})--(\ref{eq:I135}):
\begin{equation}
    I_H = I_V = I_{45} = I_{135} = \frac{I_0}{2}
\end{equation}
All four linear polarizer positions give the same reading, confirming $S_1=S_2=0$.

\paragraph{QWP analyzer for $S_3$.}
From Eqs.~(\ref{eq:Iplus})--(\ref{eq:Iminus}):
\begin{align}
    I_{+} &= \frac{1}{2}(I_0 - I_0) = 0   \quad \text{(minimum)} \\
    I_{-} &= \frac{1}{2}(I_0 + I_0) = I_0 \quad \text{(maximum)}
\end{align}

\begin{equation}
    {S_3 = I_+ - I_- = 0 - I_0 = -I_0, \qquad
    \mathrm{DOCP} = \frac{I_+ - I_-}{I_+ + I_-} = \frac{-I_0}{I_0} = -1} \quad\checkmark
\end{equation}

For the LCP beam produced at $\chi=-45^\circ$ all signs are reversed:
$I_+ = I_0$, $I_- = 0$, $\mathrm{DOCP} = +1$.

This six-measurement (or seven, counting $I_0$) protocol, summarized together
with its RCP-case values in Table~\ref{tab:general_protocol}, is the
idealized version of the \textbf{Stokes polarimetry} procedure used at X-ray
synchrotron beamlines, where the linear polarizer is a Bragg analyzer crystal
near $2\theta_B=90^\circ$ rotated around the beam, and $S_3$ is often not
measured directly but inferred, assuming full polarization, from
$|P_3| = \sqrt{1-P_1^2-P_2^2}$ (the bound of Section~\ref{sec:key_angles}).
In practice $S_0$ need not be measured separately since $S_0 = I_H + I_V =
I_{45} + I_{135} = I_+ + I_-$, providing two independent consistency checks
(three if $I_0$ is also measured).

\paragraph{A robust protocol: three analyzer scans.}
As noted above, the single-angle measurements provide little redundancy. A more robust
protocol records three full scans of the analyzer angle $\theta$: without
QWP, Eq.~(\ref{eq:Itheta}), and with the QWP at $+45^\circ$ and at
$-45^\circ$, Eq.~(\ref{eq:Ipm_theta}). Fitting each scan with a constant
and the two harmonics $\cos 2\theta$, $\sin 2\theta$ gives $S_0$ three times,
$S_2$ three times, $S_1$ once and $S_3$ twice (with opposite signs), so that
every Stokes parameter is obtained with an error estimate and with internal
consistency checks. Figure~\ref{fig:analyzer_scans} shows the three scans for
six polarization states. The linear scan alone is flat for LCP, RCP and
unpolarized light alike; the QWP scans separate them, and the two QWP scans
are interchanged between LCP and RCP. The $\sin 2\theta$ component, $S_2$, is the same in all three scans, because the QWP at
$\pm45^\circ$ leaves $S_2$ unchanged.

\section{Crystal Phase Plate in the X-ray regime}
\label{sec:crystal_plate}

\begin{figure}[H]
\centering
\begin{tikzpicture}[scale=1.6, >=stealth]
  \def\tb{28}
  \fill[gray!12] (-3.2,-1.1) rectangle (3.2,0);
  \draw[thick] (-3.2,0) -- (3.2,0);
  \foreach \y in {-0.25,-0.5,-0.75,-1.0} { \draw[gray!55] (-3.2,\y) -- (3.2,\y); }
  \node[gray!60!black, font=\small, anchor=west] at (-3.1,-0.88) {crystal (diffracting planes)};
  \draw[->, very thick] (180-\tb:3.1) -- (0,0);
  \node[above] at (180-\tb:3.1) {$\mathbf{k}_0$};
  \draw[->, very thick] (0,0) -- (\tb:3.1);
  \node[above] at (\tb:3.1) {$\mathbf{k}_h$};
  \draw[->, thick, dashed, gray] (0,0) -- (-\tb:1.6);
  \node[gray!60!black, font=\small, anchor=west] at (-\tb:1.6) {transmitted};
  \draw (1.0,0) arc (0:\tb:1.0);   \node at (\tb/2:1.25) {$\theta_B$};
  \draw (-1.0,0) arc (180:180-\tb:1.0); \node at (180-\tb/2:1.25) {$\theta_B$};
  \coordinate (A) at (180-\tb:2.0);
  \draw[fill=white] (A) circle (0.09); \fill (A) circle (0.03);
  \node[below left=1pt, blue] at (A) {$\boldsymbol{\epsilon}_\sigma$};
  \draw[->, red, thick] (A) -- ++(90-\tb:0.75) node[above right=-1pt, red] {$\boldsymbol{\epsilon}_\pi$};
  \coordinate (B) at (\tb:2.0);
  \draw[fill=white] (B) circle (0.09); \fill (B) circle (0.03);
  \node[below right=1pt, blue] at (B) {$\boldsymbol{\epsilon}'_\sigma$};
  \draw[->, red, thick] (B) -- ++(90+\tb:0.75) node[above left=-1pt, red] {$\boldsymbol{\epsilon}'_\pi$};
  \coordinate (C) at (-\tb:1.1);
  \draw[fill=white] (C) circle (0.09); \fill (C) circle (0.03);
  \node[below left=1pt, blue] at (C) {$\boldsymbol{\epsilon}_\sigma$};
  \draw[->, red, thick] (C) -- ++(90-\tb:0.5) node[right=1pt, red] {$\boldsymbol{\epsilon}_\pi$};
\end{tikzpicture}
\caption{Diffraction geometry. The page is the diffraction (scattering)
plane, containing the incident and diffracted wave vectors $\mathbf{k}_0$
and $\mathbf{k}_h$, at the Bragg angle $\theta_B$ to the diffracting planes.
The $\sigma$ polarization vectors ($\odot$, blue) are perpendicular to the
page and identical for the two beams, $\boldsymbol{\epsilon}'_\sigma =
\boldsymbol{\epsilon}_\sigma$; the $\pi$ vectors (red) lie in the page,
perpendicular to their wave vectors, and make the angle $2\theta_B$ with
each other, $\boldsymbol{\epsilon}'_\pi\cdot\boldsymbol{\epsilon}_\pi =
\cos 2\theta_B$. This gives the polarization factor of
Eq.~(\ref{eq:polfactor}). The forward-transmitted beam (dashed) keeps the
direction of $\mathbf{k}_0$, and with it the polarization vectors
$\boldsymbol{\epsilon}_\sigma$ and $\boldsymbol{\epsilon}_\pi$ of the
incident beam.}
\label{fig:sigma_pi}
\end{figure}
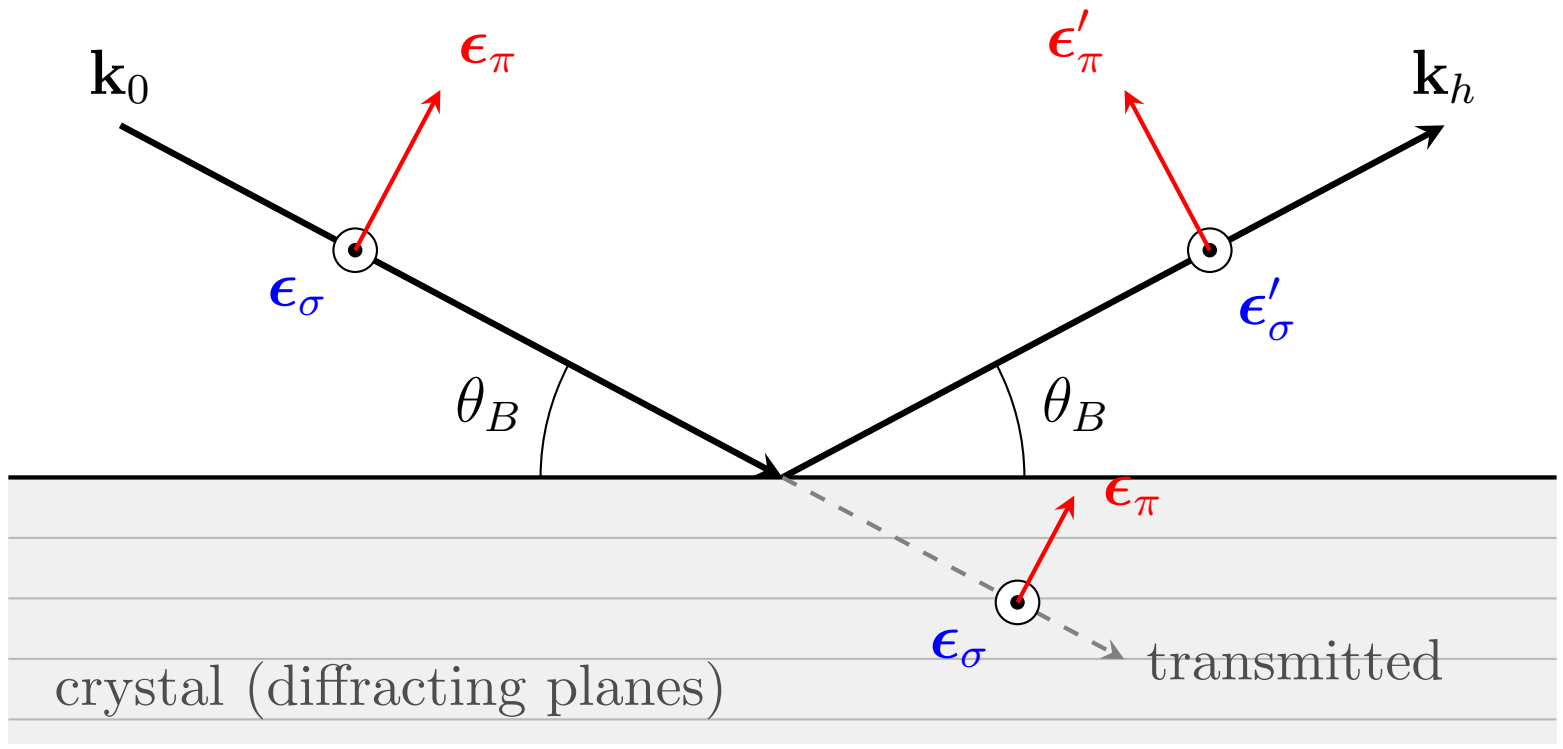

In the X-ray regime a birefringent crystal phase plate exploits the
difference in reflectivity, and in phase shift, between the two
eigenpolarizations of a perfect crystal near a Bragg
reflection~\cite{detlefs2012,belyakov1989}. These eigenpolarizations are
defined with respect to the diffraction (scattering) plane, the plane
containing the incident and diffracted wave vectors
(Fig.~\ref{fig:sigma_pi}): $\sigma$ (or $s$) polarization has the electric
field perpendicular to it, $\pi$ (or $p$) polarization has the field in it,
perpendicular to the wave vector. In the crystal frame used below,
$\sigma$ is along $\hat{x}$ and $\pi$ along $\hat{y}$. This section derives
the crystal's Jones and Mueller matrices in terms of the two complex
amplitudes $r_s$ and $r_p$, rotates the crystal to an arbitrary azimuth
$\chi$, and obtains the output polarization state, in particular the degree
of circular polarization, for a general incident beam.

\subsection{Jones and Mueller Matrix of the Crystal: $r_s$ and $r_p$ reflection or transmission amplitudes}

\paragraph{The general form and why it is diagonal.}
Within the dynamical theory of X-ray
diffraction~\cite{batterman1964,authier2001}, the most general linear
response of the crystal in the $(\sigma,\pi)$ eigenbasis is the $2\times2$
matrix of complex reflection amplitudes
\begin{equation}
    \mathbf{J}_e =
    \begin{pmatrix}
        r_{\sigma\sigma} & r_{\sigma\pi} \\
        r_{\pi\sigma}    & r_{\pi\pi}
    \end{pmatrix}
    \label{eq:Je_general}
\end{equation}
where $r_{ij}$ is the amplitude for an incident $j$-polarized
wave to be diffracted into the $i$ channel. It is a well-established
result~\cite{batterman1964,authier2001} that for an ordinary (non-magnetic)
crystal in the two-beam case the
\emph{cross terms vanish}, $r_{\sigma\pi} = r_{\pi\sigma} = 0$, so that
$\mathbf{J}_e$ is diagonal; the same holds for the transmission amplitudes
$t_{ij}$ of the forward-transmitted beam. The argument has two
steps.

\paragraph{Single scattering.} Charge (Thomson) scattering carries its
entire polarization dependence in the factor
$\boldsymbol{\epsilon}'^{\,*}\cdot\boldsymbol{\epsilon}$, where
$\boldsymbol{\epsilon}$ and $\boldsymbol{\epsilon}'$ are the incident and
scattered polarization unit vectors. By construction
$\boldsymbol{\epsilon}_\sigma = \boldsymbol{\epsilon}'_\sigma$ is
perpendicular to the scattering plane, while $\boldsymbol{\epsilon}_\pi$ and
$\boldsymbol{\epsilon}'_\pi$ both lie in it (Fig.~\ref{fig:sigma_pi}), so
\begin{equation}
    \begin{pmatrix}
        \boldsymbol{\epsilon}'_\sigma\cdot\boldsymbol{\epsilon}_\sigma &
        \boldsymbol{\epsilon}'_\sigma\cdot\boldsymbol{\epsilon}_\pi \\[2pt]
        \boldsymbol{\epsilon}'_\pi\cdot\boldsymbol{\epsilon}_\sigma &
        \boldsymbol{\epsilon}'_\pi\cdot\boldsymbol{\epsilon}_\pi
    \end{pmatrix}
    =
    \begin{pmatrix} 1 & 0 \\ 0 & \cos 2\theta_B \end{pmatrix}
    \label{eq:polfactor}
\end{equation}
The off-diagonal products vanish because a vector perpendicular to the
scattering plane is orthogonal to every vector lying in it: a single
scattering event never turns $\sigma$ into $\pi$ or vice versa.

\paragraph{Multiple scattering (dynamical theory).} In a perfect crystal the
field is not scattered once: the forward wave, with amplitude $D_0$ and wave
vector $\mathbf{K}_0$, and the diffracted wave, $D_h$ and $\mathbf{K}_h$, are
continuously scattered into each other, and the reflected and transmitted
amplitudes follow from the self-consistent solution of this coupling. With a
scalar susceptibility $\chi(\mathbf{r})$, of Fourier components $\chi_0$,
$\chi_h$, $\chi_{\bar{h}}$, Maxwell's equations in the two-beam case reduce
to the fundamental equations of the dynamical
theory~\cite{batterman1964,authier2001}
\begin{equation}
    \begin{aligned}
    \bigl[k^2(1+\chi_0) - \mathbf{K}_0\cdot\mathbf{K}_0\bigr]\,D_0
        + k^2\,C\,\chi_{\bar{h}}\,D_h &= 0 \\
    k^2\,C\,\chi_h\,D_0
        + \bigl[k^2(1+\chi_0) - \mathbf{K}_h\cdot\mathbf{K}_h\bigr]\,D_h &= 0
    \end{aligned}
    \label{eq:twobeam}
\end{equation}
where $k$ is the vacuum wave number and $C$ is the polarization factor of
Eq.~(\ref{eq:polfactor}): $C = 1$ for $\sigma$ and $C = \cos 2\theta_B$ for
$\pi$. Each coupling term is a single scattering event between the two
waves, with the geometry of Fig.~\ref{fig:sigma_pi}, and therefore carries
the factor $C$ of its own polarization state and no cross term. To all orders
of multiple scattering, Eq.~(\ref{eq:twobeam}) is two independent scalar
problems, one for $\sigma$ and one for $\pi$, which differ only through
$C$. Hence $r_{\sigma\pi} = r_{\pi\sigma} = 0$, for the diffracted and for
the forward-transmitted beam alike. The different value of $C$ is also the
origin of the different reflectivity and phase of the two eigenmodes that the
phase plate exploits (and, at $\theta_B = 45^\circ$, of the vanishing $\pi$
reflectivity used in polarization analyzers). In symmetry language, the
scattering plane is a mirror plane of the scalar problem, and $\sigma$ and
$\pi$ have opposite parity under that reflection, so no amplitude can connect
them.

Writing $r_s \equiv r_{\sigma\sigma}$ and $r_p \equiv r_{\pi\pi}$, and with
the crystal axes aligned along $\hat{x}$ ($\sigma$) and $\hat{y}$ ($\pi$),
Eq.~(\ref{eq:Je_general}) therefore reduces to the diagonal Jones matrix used
throughout this section:
\begin{equation}
    \mathbf{J}_0 = \begin{pmatrix} r_s & 0 \\ 0 & r_p \end{pmatrix}
    \label{eq:J0crystal}
\end{equation}
Note that $\mathbf{J}_0$ is diagonal only in the crystal's own
$(\sigma,\pi)$ frame: rotating the crystal by $\chi$ about the beam axis
generates off-diagonal elements in the laboratory frame
(Eq.~\ref{eq:Jchi_matrix} below), and it is exactly that
rotation-induced mixing --- not any intrinsic $\sigma\!-\!\pi$ coupling ---
which converts linear into circular polarization.

\paragraph{When the cross terms do \emph{not} vanish.}
Equation~(\ref{eq:polfactor}) relies on charge scattering from a scalar
susceptibility. Cross-polarized amplitudes do appear when that assumption is
broken, and each case is a well-studied effect in its own right:
\begin{itemize}
    \item \textbf{Magnetic scattering.} The magnetic (non-resonant and
          resonant) scattering amplitude contains terms built from
          $\boldsymbol{\epsilon}'^{\,*}\!\times\boldsymbol{\epsilon}$ and from the
          magnetization direction, which connect the two channels. The
          $\sigma\!\to\!\pi$ channel is in fact the standard experimental
          signature used to isolate magnetic from charge
          scattering~\cite{blume1988,blume1994,hill1996}.
    \item \textbf{Anisotropic (tensorial) susceptibility.} Near an
          absorption edge the atomic scattering factor becomes a tensor
          rather than a scalar (Templeton, or ATS,
          scattering)~\cite{templeton1980}, which produces cross-polarized
          amplitudes and gives rise to otherwise forbidden
          reflections~\cite{dmitrienko1983,belyakov1989} whenever the local
          tensor axes are not aligned with the scattering plane.
    \item \textbf{Multiple-beam ($n$-beam) diffraction.} When several
          reciprocal-lattice points are simultaneously excited, the
          scattering planes of the individual reflections differ and the
          $\sigma$ and $\pi$ channels mix~\cite{okitsu2006,stetsko2004}. The
          effect has been used to determine the complete polarization state
          of an X-ray beam, including its circular part~\cite{shen1992}, and to
          characterize the circular polarization delivered by a phase
          plate~\cite{lang1995}.
    \item \textbf{Optical activity} in chiral crystals, a much weaker effect
          at X-ray wavelengths~\cite{belyakov1989}.
\end{itemize}
For the phase plates treated here --- non-magnetic perfect crystals such as
diamond or silicon, operated in a two-beam setting and away from absorption
edges~\cite{hirano1993,giles1994} --- all of these contributions are
negligible, and the diagonal form of Eq.~(\ref{eq:J0crystal}) is an excellent
approximation.

\paragraph{Magnitude, phase, retardation and reflectivities.}
The two amplitudes are generally different in both magnitude and phase,
\begin{equation}
    r_s = |r_s|\,e^{i\phi_s}, \qquad r_p = |r_p|\,e^{i\phi_p}
    \label{eq:rsrp_polar}
\end{equation}
and the key quantity is their relative phase, the \textbf{retardation} of the
crystal,
\begin{equation}
    \delta = \phi_p - \phi_s = \arg(r_p) - \arg(r_s)
    \label{eq:Dphi_def}
\end{equation}
We use the same symbol as for the ideal retarder of
Section~\ref{sec:retarder} because it is the same quantity: up to the common
factor $r_s$, $\mathbf{J}_0 = r_s\,\mathrm{diag}\bigl(1,\,(|r_p|/|r_s|)\,
e^{i\delta}\bigr)$, which for $|r_s| = |r_p|$ is the retarder of
Eq.~(\ref{eq:Jret0}) with $\sigma$ ($\hat{x}$) as the fast axis when
$\delta > 0$. The crystal adds to the ideal retarder only a common
attenuation and a \emph{linear dichroism}, $|r_s| \neq |r_p|$ (we use the
term, which strictly refers to a difference in absorption, for any difference
between the $\sigma$ and $\pi$ reflectivities or transmittances; in
polarimetry it is called \emph{diattenuation}~\cite{chipman2018}). The Mueller
matrices and Stokes vectors below contain the amplitudes only through the
intensity reflectivities (or transmittances) and the cross product
$r_sr_p^*$. To keep the formulas readable we introduce
\begin{equation}
    \begin{gathered}
    R_s \equiv |r_s|^2, \qquad R_p \equiv |r_p|^2, \qquad
    \bar{R} \equiv \frac{R_s+R_p}{2}, \qquad
    \tilde{R} \equiv \frac{R_s-R_p}{2}, \\
    R_{sp} \equiv |r_s||r_p| = \sqrt{\bar{R}^2-\tilde{R}^2}
    \end{gathered}
    \label{eq:Rsp_def}
\end{equation}
($\bar{R}$, with a bar, is the mean reflectivity and $\tilde{R}$, with a
tilde, the dichroism; neither must be confused with the rotation
matrices $\mathbf{R}_J(\chi)$, $\mathbf{R}_M(\chi)$), and
the cross product carries the retardation:
\begin{equation}
    \mathrm{Re}(r_s r_p^*) = R_{sp}\cos\delta, \qquad
    \mathrm{Im}(r_s r_p^*) = R_{sp}\sin(\phi_s-\phi_p)
                           = -R_{sp}\sin\delta
    \label{eq:crossprod_phases}
\end{equation}
A perfect quarter-wave plate corresponds to $\tilde{R} = 0$ together with
$\delta = \pm\pi/2$. Finally, the whole formalism applies
verbatim to the transmitted (forward-diffracted) beam by replacing the
reflection amplitudes $r_s$, $r_p$ with the transmission amplitudes $t_s$,
$t_p$ (we keep the symbols $r_s$, $r_p$ for both).

Using the Pauli basis (Eq.~\ref{eq:Pauli_def}), the diagonal Jones matrix of
Eq.~(\ref{eq:J0crystal}) decomposes as:
\begin{equation}
    {\mathbf{J}_0 = \frac{r_s+r_p}{2}\,\boldsymbol{\sigma}_0
                        + \frac{r_s-r_p}{2}\,\boldsymbol{\sigma}_1}
    \label{eq:J0_Pauli}
\end{equation}

This is physically transparent: the crystal acts as an isotropic attenuator
(the $\boldsymbol{\sigma}_0$ term, proportional to the mean amplitude) plus
a linear H/V birefringence and dichroism (the
$\boldsymbol{\sigma}_1$ term, proportional to the complex amplitude
difference).

\subsubsection{Aligned Mueller Matrix ($\chi = 0$)}

Applying the Pauli trace formula (Eq.~\ref{eq:PauliTrace}) to $\mathbf{J}_0$
and using Eqs.~(\ref{eq:Rsp_def})--(\ref{eq:crossprod_phases}):
\begin{equation}
    {
    \mathbf{M}_0 =
    \begin{pmatrix}
        \bar{R}  &  \tilde{R}  &  0  &  0 \\
        \tilde{R}  &  \bar{R}  &  0  &  0 \\
        0  &  0  &  R_{sp}\cos\delta  &  -R_{sp}\sin\delta \\
        0  &  0  &  R_{sp}\sin\delta  &  R_{sp}\cos\delta
    \end{pmatrix}
    }
    \label{eq:M0_Pauli}
\end{equation}

The block structure is transparent: the upper-left block contains the mean
reflectivity $\bar{R}$ and the dichroism $\tilde{R}$, while the lower-right
$2\times2$ block is $R_{sp}$ times exactly the $(S_2,S_3)$ rotation by the
angle $\delta$ of the ideal retarder, Eq.~(\ref{eq:Mret0}).

\subsection{Crystal Oriented at Arbitrary Angle $\chi$}
\label{sec:crystal_chi}

\subsubsection{Jones Matrix}

Applying the rotation sandwich $\mathbf{J}^{(\chi)} =
\mathbf{R}_J(-\chi)\,\mathbf{J}_0\,\mathbf{R}_J(\chi)$ (Eq.~\ref{eq:Jones_rotation})
and using the rotation identity
$\mathbf{R}_J(-\chi)\,\boldsymbol{\sigma}_1\,\mathbf{R}_J(\chi)
= \cos 2\chi\,\boldsymbol{\sigma}_1 + \sin 2\chi\,\boldsymbol{\sigma}_2$,
the Pauli decomposition of the rotated Eq.~(\ref{eq:J0crystal}) is:

\begin{equation}
    {
    \mathbf{J}^{(\chi)} = \frac{r_s+r_p}{2}\,\boldsymbol{\sigma}_0
    + \frac{r_s-r_p}{2}\bigl(\cos 2\chi\,\boldsymbol{\sigma}_1
                             + \sin 2\chi\,\boldsymbol{\sigma}_2\bigr)
    }
    \label{eq:Jchi_Pauli}
\end{equation}

Rotation by $\chi$ simply rotates the birefringence axis from
$\boldsymbol{\sigma}_1$ (H/V) toward $\boldsymbol{\sigma}_2$ ($\pm45^\circ$)
by the angle $2\chi$ in Pauli/Stokes space. Expanded in matrix form:

\begin{equation}
    {
    \mathbf{J}^{(\chi)} =
    \frac{r_s+r_p}{2}\,\mathbf{I}
    + \frac{r_s-r_p}{2}
    \begin{pmatrix} \cos 2\chi & \sin 2\chi \\ \sin 2\chi & -\cos 2\chi \end{pmatrix}
    }
    \label{eq:Jchi_matrix}
\end{equation}

\paragraph{In terms of moduli and phases.}
Substituting the polar form of $r_s$ and $r_p$ (Eq.~\ref{eq:rsrp_polar})
directly into Eq.~(\ref{eq:Jchi_matrix}) gives the equivalent, entirely
explicit form:
\begin{equation}
    {
    \mathbf{J}^{(\chi)} =
    \begin{pmatrix}
        |r_s|\,e^{i\phi_s}\cos^2\!\chi + |r_p|\,e^{i\phi_p}\sin^2\!\chi
            & \bigl(|r_s|\,e^{i\phi_s} - |r_p|\,e^{i\phi_p}\bigr)\sin\chi\cos\chi \\[8pt]
        \bigl(|r_s|\,e^{i\phi_s} - |r_p|\,e^{i\phi_p}\bigr)\sin\chi\cos\chi
            & |r_s|\,e^{i\phi_s}\sin^2\!\chi + |r_p|\,e^{i\phi_p}\cos^2\!\chi
    \end{pmatrix}
    }
    \label{eq:Jchi_polar}
\end{equation}
each element mixing the two eigenamplitudes.

\subsubsection{Mueller Matrix}

Using the Mueller rotation matrix $\mathbf{R}_M(\chi)$ (Eq.~\ref{eq:RM}):
\begin{equation}
    \mathbf{M}^{(\chi)} = \mathbf{R}_M(-\chi)\cdot\mathbf{M}_0\cdot\mathbf{R}_M(\chi)
\end{equation}

The rotation fills the $S_1$ and $S_2$ rows and columns of the matrix, while
$M_{03}=M_{30}=0$ for all $\chi$: $S_0$ and $S_3$ are never coupled directly.
Carrying out the full matrix product with $c_2 \equiv \cos 2\chi$,
$s_2 \equiv \sin 2\chi$:

\begin{equation}
    {
    \mathbf{M}^{(\chi)} =
    \begin{pmatrix}
        \bar{R}
            & \tilde{R}\,c_2
            & \tilde{R}\,s_2
            & 0 \\[6pt]
        \tilde{R}\,c_2
            & \bar{R}\,c_2^2 + R_{sp}\cos\delta\,s_2^2
            & \bigl(\bar{R}-R_{sp}\cos\delta\bigr)s_2c_2
            & R_{sp}\sin\delta\,s_2 \\[6pt]
        \tilde{R}\,s_2
            & \bigl(\bar{R}-R_{sp}\cos\delta\bigr)s_2c_2
            & \bar{R}\,s_2^2 + R_{sp}\cos\delta\,c_2^2
            & -R_{sp}\sin\delta\,c_2 \\[6pt]
        0
            & -R_{sp}\sin\delta\,s_2
            & R_{sp}\sin\delta\,c_2
            & R_{sp}\cos\delta
    \end{pmatrix}
    }
    \label{eq:Mueller_final}
\end{equation}
In terms of the complex amplitudes, $R_{sp}\cos\delta = \mathrm{Re}(r_sr_p^*)$
and $R_{sp}\sin\delta = -\mathrm{Im}(r_sr_p^*)$
(Eq.~\ref{eq:crossprod_phases}).

$M_{00} = \bar{R}$ and $M_{33} = R_{sp}\cos\delta$ are independent of
$\chi$. The first row and column carry the dichroism $\tilde{R}$: for
equal reflectivities they reduce to $\bar{R}\,(1,0,0,0)$, and the
transmitted intensity no longer depends on the input polarization.

\subsection{Output Jones Vector for a General Incident Beam}

Let the incident beam be an arbitrary Jones vector:
\begin{equation}
    \mathbf{E}_{\mathrm{in}} = \begin{pmatrix} E_x^{(0)} \\ E_y^{(0)} \end{pmatrix}
    \label{eq:Ein_general}
\end{equation}

Applying the crystal Jones matrix (Eq.~\ref{eq:Jchi_matrix}):

\begin{equation}
    {
    \mathbf{E}_{\mathrm{out}} =
    \frac{r_s+r_p}{2}\begin{pmatrix}E_x^{(0)}\\E_y^{(0)}\end{pmatrix}
    + \frac{r_s-r_p}{2}
    \begin{pmatrix}
        \cos 2\chi\, E_x^{(0)} + \sin 2\chi\, E_y^{(0)} \\
        \sin 2\chi\, E_x^{(0)} - \cos 2\chi\, E_y^{(0)}
    \end{pmatrix}
    }
    \label{eq:Eout_general}
\end{equation}

The first term is an isotropic multiplication by the mean amplitude $(r_s+r_p)/2$;
the second introduces differential retardation and attenuation along the
rotated eigenmodes.
Equivalently, using Eq.~(\ref{eq:Jchi_Pauli}) the output can be written in the
Pauli basis as:
\begin{equation}
    \mathbf{E}_{\mathrm{out}} = \left[
    \frac{r_s+r_p}{2}\,\boldsymbol{\sigma}_0
    + \frac{r_s-r_p}{2}\bigl(\cos 2\chi\,\boldsymbol{\sigma}_1
                             + \sin 2\chi\,\boldsymbol{\sigma}_2\bigr)
    \right]\mathbf{E}_{\mathrm{in}}
    \label{eq:Eout_Pauli}
\end{equation}

\paragraph{Particular case: horizontal linear polarization.}
Setting $E_x^{(0)} = E_0$, $E_y^{(0)} = 0$ in Eq.~(\ref{eq:Eout_general}):
\begin{equation}
    \mathbf{E}_{\mathrm{out}}\big|_H = \frac{E_0}{2}
    \begin{pmatrix}
        (r_s+r_p) + (r_s-r_p)\cos 2\chi \\
        (r_s-r_p)\sin 2\chi
    \end{pmatrix}
    = E_0
    \begin{pmatrix}
        r_s\cos^2\!\chi + r_p\sin^2\!\chi \\
        (r_s-r_p)\sin\chi\cos\chi
    \end{pmatrix}
    \label{eq:Eout_chi}
\end{equation}

\paragraph{Particular case: vertical linear polarization.}
Setting $E_x^{(0)} = 0$, $E_y^{(0)} = E_0$:
\begin{equation}
    \mathbf{E}_{\mathrm{out}}\big|_V = \frac{E_0}{2}
    \begin{pmatrix}
        (r_s-r_p)\sin 2\chi \\
        (r_s+r_p) - (r_s-r_p)\cos 2\chi
    \end{pmatrix}
    = E_0
    \begin{pmatrix}
        (r_s-r_p)\sin\chi\cos\chi \\
        r_s\sin^2\!\chi + r_p\cos^2\!\chi
    \end{pmatrix}
\end{equation}

\subsection{Output Stokes Vector for a General Incident Beam}

Let the incident beam have an arbitrary Stokes vector
$\mathbf{S}_{\mathrm{in}} = (S_0^{\mathrm{in}},\,S_1^{\mathrm{in}},\,
S_2^{\mathrm{in}},\,S_3^{\mathrm{in}})^T$.
Applying the rotated Mueller matrix $\mathbf{M}^{(\chi)}$
(Eq.~\ref{eq:Mueller_final}):

\begin{equation}
    \mathbf{S}_{\mathrm{out}} = \mathbf{M}^{(\chi)}\,\mathbf{S}_{\mathrm{in}}
    \label{eq:Sout_general_op}
\end{equation}

Computing each component from Eq.~(\ref{eq:Mueller_final}),
with $c_2 = \cos 2\chi$, $s_2 = \sin 2\chi$:

\begin{align}
    {S_0^{\mathrm{out}}} &=
        \bar{R}\,S_0^{\mathrm{in}}
        + \tilde{R}\,\bigl(c_2\,S_1^{\mathrm{in}} + s_2\,S_2^{\mathrm{in}}\bigr)
    \label{eq:Sout0_gen}\\[6pt]
    {S_1^{\mathrm{out}}} &=
        \tilde{R}\,c_2\,S_0^{\mathrm{in}}
        + \bigl(\bar{R}\,c_2^2 + R_{sp}\cos\delta\,s_2^2\bigr)\,S_1^{\mathrm{in}}
        + \bigl(\bar{R} - R_{sp}\cos\delta\bigr)s_2c_2\,S_2^{\mathrm{in}}
        + R_{sp}\sin\delta\,s_2\,S_3^{\mathrm{in}}
    \label{eq:Sout1_gen}\\[6pt]
    {S_2^{\mathrm{out}}} &=
        \tilde{R}\,s_2\,S_0^{\mathrm{in}}
        + \bigl(\bar{R} - R_{sp}\cos\delta\bigr)s_2c_2\,S_1^{\mathrm{in}}
        + \bigl(\bar{R}\,s_2^2 + R_{sp}\cos\delta\,c_2^2\bigr)\,S_2^{\mathrm{in}}
        - R_{sp}\sin\delta\,c_2\,S_3^{\mathrm{in}}
    \label{eq:Sout2_gen}\\[6pt]
    {S_3^{\mathrm{out}}} &=
        -R_{sp}\sin\delta\,\bigl(s_2\,S_1^{\mathrm{in}} - c_2\,S_2^{\mathrm{in}}\bigr)
        + R_{sp}\cos\delta\,S_3^{\mathrm{in}}
    \label{eq:Sout3_gen}
\end{align}

\paragraph{Particular case: horizontal linear polarization input.}
$\mathbf{S}_{\mathrm{in}} = I_0(1,\,1,\,0,\,0)^T$.
Substituting into Eqs.~(\ref{eq:Sout0_gen})--(\ref{eq:Sout3_gen}):

\begin{align}
    {S_0^{\mathrm{out}}} &= I_0\bigl(\bar{R} + \tilde{R}\cos 2\chi\bigr)
                       = I_0\bigl(R_s\cos^2\!\chi + R_p\sin^2\!\chi\bigr)
    \label{eq:FinalS0H}\\[6pt]
    {S_1^{\mathrm{out}}} &= I_0\bigl(\tilde{R}\cos 2\chi
            + \bar{R}\cos^2\!2\chi + R_{sp}\cos\delta\,\sin^2\!2\chi\bigr)
    \label{eq:FinalS1H}\\[6pt]
    {S_2^{\mathrm{out}}} &= I_0\sin 2\chi\,\bigl[\tilde{R}
            + \bigl(\bar{R}-R_{sp}\cos\delta\bigr)\cos 2\chi\bigr]
    \label{eq:FinalS2H}\\[6pt]
    {S_3^{\mathrm{out}}} &= -I_0\,R_{sp}\sin\delta\,\sin 2\chi
    \label{eq:FinalS3H}
\end{align}

$S_3^{\mathrm{out}}$ (Eq.~\ref{eq:FinalS3H}) is non-zero and comes from
$M_{31} = -R_{sp}\sin\delta\sin 2\chi$ acting on $S_1^{\mathrm{in}} = I_0$.
For a QWP
($R_s = R_p = R_{sp} = r^2$, $\delta = \pm\pi/2$) at $\chi=45^\circ$,
Eqs.~(\ref{eq:FinalS0H})--(\ref{eq:FinalS3H}) give, for $\delta=+\pi/2$:
\begin{equation}
    \mathbf{S}_{\mathrm{out}} = I_0 r^2 \begin{pmatrix}1\\0\\0\\-1\end{pmatrix}
    \quad\text{(pure circular, $|P_C|=1$)}
\end{equation}
i.e.\ right circular polarization (RCP), in exact agreement with the ideal
QWP of Section~\ref{sec:QWP45}; $\delta = -\pi/2$ gives LCP.
For a HWP ($R_s = R_p = R_{sp} = r^2$, $\delta = \pm\pi$, $\cos\delta =
-1$), Eqs.~(\ref{eq:FinalS0H})--(\ref{eq:FinalS3H}) give instead
\begin{equation}
    \mathbf{S}_{\mathrm{out}} = I_0\,r^2
    \begin{pmatrix} 1 \\ \cos 4\chi \\ \sin 4\chi \\ 0 \end{pmatrix}
    \label{eq:Sout_H_HWP}
\end{equation}
i.e.\ linear polarization at the angle $2\chi$ to the horizontal (vertical
for $\chi = 45^\circ$), as for the ideal HWP of Section~\ref{sec:HWP}: a
crystal phase plate at the half-wave offset rotates the plane of linear
polarization.

\paragraph{Particular case: $+45^\circ$ linear polarization input.}
$\mathbf{S}_{\mathrm{in}} = I_0(1,\,0,\,1,\,0)^T$:

\begin{align}
    S_0^{\mathrm{out}} &= I_0\bigl(\bar{R} + \tilde{R}\sin 2\chi\bigr) \\[4pt]
    S_1^{\mathrm{out}} &= I_0\bigl[\tilde{R}\cos 2\chi
                         + \bigl(\bar{R}-R_{sp}\cos\delta\bigr)\sin 2\chi\cos 2\chi\bigr] \\[4pt]
    S_2^{\mathrm{out}} &= I_0\bigl[\tilde{R}\sin 2\chi
                         + \bar{R}\sin^2\!2\chi
                         + R_{sp}\cos\delta\cos^2\!2\chi\bigr] \\[4pt]
    S_3^{\mathrm{out}} &= I_0\,R_{sp}\sin\delta\cos 2\chi
\end{align}

For a QWP at $\chi=0^\circ$ ($\tilde{R} = 0$, $R_{sp} = r^2$, $\cos\delta = 0$):
$S_3^{\mathrm{out}} = I_0\,r^2\sin\delta = \pm I_0 r^2$. At $\chi=45^\circ$:
$S_3^{\mathrm{out}} = 0$.

\paragraph{Particular case: left circular polarization input.}
$\mathbf{S}_{\mathrm{in}} = I_0(1,\,0,\,0,\,1)^T$ (LCP):

\begin{align}
    S_0^{\mathrm{out}} &= I_0\,\bar{R} \\[4pt]
    S_1^{\mathrm{out}} &= I_0\bigl(\tilde{R}\cos 2\chi
                         + R_{sp}\sin\delta\sin 2\chi\bigr) \\[4pt]
    S_2^{\mathrm{out}} &= I_0\bigl(\tilde{R}\sin 2\chi
                         - R_{sp}\sin\delta\cos 2\chi\bigr) \\[4pt]
    S_3^{\mathrm{out}} &= I_0\,R_{sp}\cos\delta
\end{align}

For a QWP ($\cos\delta=0$): $S_3^{\mathrm{out}} = 0$, confirming
conversion to linear polarization for any $\chi$. The terms in
$\tilde{R}$ (dichroism) vanish for $|r_s|=|r_p|$; they come from
the coupling of $S_0^{\mathrm{in}}$ to $S_{1,2}^{\mathrm{out}}$, and are also
present for the $+45^\circ$ input above.

\subsection{Key Result: Degree of Circular Polarization for a General Input}

The \textbf{degree of circular polarization} of the output beam is
$P_C = S_3^{\mathrm{out}}/S_0^{\mathrm{out}}$ (Eq.~\ref{eq:DOCP_def}). Using
Eqs.~(\ref{eq:Sout3_gen}) and (\ref{eq:Sout0_gen}), and dividing numerator
and denominator by $S_0^{\mathrm{in}}$, the degree of circular polarization
for an arbitrary incident beam depends only on its Poincar\'{e}--Stokes
parameters $P_k^{\mathrm{in}}$ (Eq.~\ref{eq:Pdef}):

\begin{equation}
    {
    P_C\!\left(\chi,\mathbf{S}_{\mathrm{in}}\right)
    = \frac{-R_{sp}\sin\delta\,\bigl(s_2\,P_1^{\mathrm{in}}
                                       - c_2\,P_2^{\mathrm{in}}\bigr)
            + R_{sp}\cos\delta\,P_3^{\mathrm{in}}}
           {\bar{R}
            + \tilde{R}\,\bigl(c_2\,P_1^{\mathrm{in}}
                                                 + s_2\,P_2^{\mathrm{in}}\bigr)}
    }
    \label{eq:PC_general}
\end{equation}
with $c_2=\cos 2\chi$, $s_2=\sin 2\chi$. The numerator shows the three
roles of the crystal parameters: $\sin\delta$ converts the linear content of
the input, $P_1^{\mathrm{in}}$ and $P_2^{\mathrm{in}}$, into circular
polarization (for horizontal or vertical input only when $\chi \neq 0^\circ,
90^\circ$); $\cos\delta$ preserves an incident circular component
$P_3^{\mathrm{in}}$, and vanishes for an ideal QWP; and the unpolarized part
of the beam generates no circular polarization at all, since the matrix
element $M_{30}$ of Eq.~(\ref{eq:Mueller_final}) is zero. (All output
Stokes parameters are of course proportional to the input intensity; it is
the \emph{fraction} of circular polarization that depends only on
$P_k^{\mathrm{in}}$.)

\paragraph{Particular case: horizontal linear polarization input.}

For $\mathbf{S}_{\mathrm{in}} = I_0(1,1,0,0)^T$, using
Eqs.~(\ref{eq:FinalS0H}) and (\ref{eq:FinalS3H}):
\begin{equation}
    {
    P_C(\chi)\big|_H
    = \frac{-R_{sp}\sin\delta\,\sin 2\chi}
           {R_s\cos^2\!\chi + R_p\sin^2\!\chi}
    }
    \label{eq:PC_final}
\end{equation}

This agrees, up to an overall sign, with the equation in Giles' thesis~\cite{giles1995thesis},
written there as
\begin{equation}
    {
    P_C(\chi)\big|_H
    = - \frac{a \sin(\phi_s-\phi_p) \sin 2\chi}
           {\cos^2\!\chi + a^2\sin^2\!\chi}
    }
    \nonumber
\end{equation}
where $a=|r_p|/|r_s| = R_{sp}/R_s$. Since $\delta = \phi_p-\phi_s$,
$\sin(\phi_s-\phi_p) = -\sin\delta$; also $R_p/R_s = a^2$. Dividing
Eq.~(\ref{eq:PC_final}) through by $R_s$ therefore gives
\begin{equation}
    P_C(\chi)\big|_H = \frac{-a\,\sin\delta\,\sin 2\chi}{\cos^2\!\chi + a^2\sin^2\!\chi}
    \nonumber
\end{equation}
which is exactly $-1$ times Giles' expression: the overall sign difference
comes from a different sign convention (for the sign of the degree of circular
polarization, the time factor or the definition of the phase
difference), and has no physical consequence.

For a QWP ($R_s = R_p$, $\delta = +\pi/2$, i.e.\ $r_p=ir_s$)
at $\chi=45^\circ$: $P_C = -1$ (RCP), in agreement with Section~\ref{sec:QWP45}.

\paragraph{Particular case: $+45^\circ$ linear polarization input.}

For $\mathbf{S}_{\mathrm{in}} = I_0(1,0,1,0)^T$:
\begin{equation}
    P_C(\chi)\big|_{+45}
    = \frac{R_{sp}\sin\delta\,\cos 2\chi}
           {\bar{R} + \tilde{R}\sin 2\chi}
\end{equation}

For equal reflectivities ($\tilde{R} = 0$, $R_{sp} = \bar{R}$):
$P_C = \sin\delta\cos 2\chi$.

\paragraph{Particular case: left circular polarization input.}

For $\mathbf{S}_{\mathrm{in}} = I_0(1,0,0,1)^T$ (LCP):
\begin{equation}
    P_C(\chi)\big|_{\mathrm{LCP}}
    = \frac{R_{sp}\cos\delta}{\bar{R}}
\end{equation}

For a QWP ($\cos\delta=0$): $P_C = 0$ --- confirming conversion
to linear polarization for any $\chi$.

\begin{table}[H]
\centering
\renewcommand{\arraystretch}{1.8}
\small
\begin{tabular}{p{2.8cm}p{6.0cm}p{4.0cm}}
\toprule
\textbf{Input} $\mathbf{S}_{\mathrm{in}}$ &
\textbf{General $P_C$ (any $r_s$, $r_p$)} &
\textbf{Ideal QWP}\newline
($\tilde{R} = 0$, $\delta = \pm\pi/2$) \\
\midrule
H linear $(1,1,0,0)^T$ &
$\dfrac{-R_{sp}\sin\delta\,\sin 2\chi}{R_s\cos^2\!\chi+R_p\sin^2\!\chi}$ &
$|P_C|=1$ at $\chi=45^\circ$ \\[10pt]

$+45^\circ$ linear $(1,0,1,0)^T$ &
$\dfrac{R_{sp}\sin\delta\,\cos 2\chi}{\bar{R}+\tilde{R}\sin 2\chi}$ &
$P_C = \sin\delta\cos 2\chi$ \\[10pt]

LCP $(1,0,0,1)^T$ &
$\dfrac{R_{sp}\cos\delta}{\bar{R}}$ &
$P_C = 0$, $\forall\chi$ \\[10pt]

General $(S_0,S_1,S_2,S_3)^T$ &
\multicolumn{2}{l}{see Eq.~(\ref{eq:PC_general})} \\

\bottomrule
\end{tabular}
\caption{Degree of circular polarization $P_C = S_3^{\mathrm{out}}/S_0^{\mathrm{out}}$
in terms of the reflectivities and the retardation of the crystal
(Eq.~\ref{eq:Rsp_def}). $\sin\delta$ drives
conversion between linear and circular polarization, while $\cos\delta$
governs preservation of existing circular polarization.}
\label{tab:Pc_general}
\end{table}

\subsection{Conditions for Perfect Circular Polarization}

For a horizontal linear input (Eq.~\ref{eq:PC_final}), $|P_C|=1$ requires
$S_1^{\mathrm{out}} = S_2^{\mathrm{out}} = 0$ simultaneously
(Eqs.~\ref{eq:FinalS1H}--\ref{eq:FinalS2H}), which leads to
two independent conditions:
\begin{enumerate}
    \item \textbf{Phase condition:} $\cos\delta = 0$, i.e.\ $\delta = \pm\pi/2$:
          the crystal acts as a quarter-wave retarder.
    \item \textbf{Amplitude-angle condition:}
          $|r_s|\cos\chi = |r_p|\sin\chi$, i.e.\
          $\tan\chi = |r_s|/|r_p|$.
\end{enumerate}

For equal reflectivities $|r_s| = |r_p|$, condition~2 reduces to $\chi =
45^\circ$. For unequal reflectivities, the crystal must be rotated in
azimuth to compensate; $\chi=-\arctan(|r_s|/|r_p|)$ also gives $|P_C|=1$,
with the opposite handedness.

For a $+45^\circ$ linear input, condition~1 ($\delta=\pm\pi/2$) together
with equal reflectivities $|r_s|=|r_p|$ gives $|P_C|=1$ at $\chi=0^\circ$ (or
$90^\circ$, with the opposite handedness). For a general input
$(S_0,S_1,S_2,S_3)^T$, perfect circular polarization is achieved whenever the
numerator of Eq.~(\ref{eq:PC_general}) equals $\pm$ its denominator.

\begin{table}[H]
\centering
\small
\renewcommand{\arraystretch}{1.3}
\begin{tabular}{llllll}
\toprule
\textbf{Input} & $\chi$ & $\delta$ & $|r_s|$ vs $|r_p|$ & $P_C$ & \textbf{Output} \\
\midrule
\multirow{5}{*}{H linear}
& $0^\circ$ or $90^\circ$ & any          & any              & $0$       & Linear (no conversion) \\
& $45^\circ$              & $\pm\pi/2$   & $|r_s|=|r_p|$    & $\mp1$    & Pure RCP/LCP \\
& $45^\circ$              & $\pm\pi/2$   & $|r_s|\neq|r_p|$ & $|P_C|<1$ & Elliptical \\
& $45^\circ$              & $0$ or $\pi$ & any              & $0$       & Linear \\
& $\arctan(|r_s|/|r_p|)$ & $\pm\pi/2$   & any              & $\mp1$    & Pure RCP/LCP \\
\midrule
$+45^\circ$ linear & $0^\circ$  & $\pm\pi/2$ & $|r_s|=|r_p|$ & $\pm1$ & Pure LCP/RCP \\
$+45^\circ$ linear & $90^\circ$ & $\pm\pi/2$ & $|r_s|=|r_p|$ & $\mp1$ & Pure RCP/LCP \\
$+45^\circ$ linear & any & $0$        & any & $0$    & Linear$^\dagger$ \\
\midrule
LCP & any & $0$      & $|r_s|=|r_p|$ & $+1$ & LCP (unchanged) \\
LCP & any & $\pm\pi/2$  & $|r_s|=|r_p|$ & $0$ & Linear (orientation set by $\chi$) \\
\bottomrule
\end{tabular}
\caption{Output polarization state for selected input polarizations, crystal
orientations and parameters. The H linear results use
Eq.~(\ref{eq:PC_final}); others use the general Eq.~(\ref{eq:PC_general}).
$^\dagger$Unchanged for $|r_s|=|r_p|$; otherwise the dichroism rotates the
plane of polarization.}
\label{tab:Pc_cases}
\end{table}

\newpage
\subsection{Summary}

\begin{table}[H]
\renewcommand{\arraystretch}{2.2}
\setlength{\tabcolsep}{5pt}
\small
\begin{tabular}{p{2.3cm}p{4.4cm}p{3.4cm}p{4.5cm}}
\toprule
\textbf{Stage / case}
  & \textbf{Jones vector} $\mathbf{E}$
  & \textbf{Pauli decomp.\ of} $\mathbf{J}$
  & \textbf{Stokes vector} / \textbf{Mueller matrix} \\
\midrule

Input\newline(horiz.\ linear) &
$E_0\begin{pmatrix}1\\0\end{pmatrix}$ &
--- &
$\mathbf{S}_{\mathrm{in}} = I_0\begin{pmatrix}1\\1\\0\\0\end{pmatrix}$ \\[8pt]

\hline

Ideal QWP\newline($\chi=45^\circ$,\newline$r_s=r$, $r_p=ir$,\newline$\delta=\pi/2$) &
$\dfrac{E_0 r\,e^{i\pi/4}}{\sqrt{2}}\begin{pmatrix}1\\-i\end{pmatrix}$ &
$r\,\mathbf{J}_{\mathrm{QWP}}^{(45^\circ)} = \dfrac{r(1+i)}{2}\boldsymbol{\sigma}_0$\newline
$+\,\dfrac{r(1-i)}{2}\boldsymbol{\sigma}_2$ &
$\mathbf{S}_{\mathrm{out}} = I_0 r^2\begin{pmatrix}1\\0\\0\\-1\end{pmatrix}$~(RCP)\newline\newline
$r^2\mathbf{M}_{\mathrm{QWP}}^{(45^\circ)} = r^2\begin{pmatrix}1&0&0&0\\0&0&0&1\\0&0&1&0\\0&{-1}&0&0\end{pmatrix}$ \\[10pt]

\bottomrule
\end{tabular}
\caption{Unified summary. $\delta=\phi_p-\phi_s$ is the
retardation between the $\pi$ and $\sigma$ eigenmodes (Eq.~\ref{eq:Dphi_def}).}
\label{tab:summary}
\end{table}

\begin{table}[H]
\renewcommand{\arraystretch}{2.2}
\setlength{\tabcolsep}{5pt}
\small
\begin{tabular}{p{2.3cm}p{4.4cm}p{3.4cm}p{4.5cm}}
\toprule
\textbf{Stage / case}
  & \textbf{Jones vector} $\mathbf{E}$
  & \textbf{Pauli decomp.\ of} $\mathbf{J}$
  & \textbf{Stokes vector} / \textbf{Mueller matrix} \\
\midrule

Crystal @ $\chi=0$\newline($r_s,r_p$ general) &
$E_0\begin{pmatrix}r_s\\0\end{pmatrix}$ &
$\mathbf{J}_0 = \dfrac{r_s+r_p}{2}\boldsymbol{\sigma}_0$\newline
$+\,\dfrac{r_s-r_p}{2}\boldsymbol{\sigma}_1$ &
$\mathbf{M}_0$: see Eq.~(\ref{eq:M0_Pauli})\newline\newline
$P_C = 0$ (no circular polarization) \\[10pt]

\hline

Crystal @\newline arb.\ $\chi$,\newline$r_s,r_p$ general &
$E_0\begin{pmatrix}r_s\cos^2\!\chi+r_p\sin^2\!\chi\\(r_s-r_p)\sin\chi\cos\chi\end{pmatrix}$ &
$\mathbf{J}^{(\chi)} = \dfrac{r_s+r_p}{2}\boldsymbol{\sigma}_0$\newline
$+\,\dfrac{r_s-r_p}{2}\bigl(\cos 2\chi\,\boldsymbol{\sigma}_1$\newline
$+\sin 2\chi\,\boldsymbol{\sigma}_2\bigr)$ &
$\mathbf{M}^{(\chi)}$: see Eq.~(\ref{eq:Mueller_final})\newline\newline
$S_0 = I_0(R_s\cos^2\!\chi+R_p\sin^2\!\chi)$\newline
$S_3 = -I_0\,R_{sp}\sin\delta\,\sin 2\chi$\newline
$P_C$: see Eq.~(\ref{eq:PC_final}) \\

\bottomrule
\end{tabular}
\caption{Continuation of Table~\ref{tab:summary}, with $R_s$, $R_p$, $R_{sp}$, $\bar{R}$, $\tilde{R}$ of Eq.~(\ref{eq:Rsp_def}).}
\label{tab:summary2}
\end{table}

\section{Practical Implementation of Polarizers and Phase Shifters in the
X-ray Regime Using Perfect Crystal Diffraction}
\label{sec:practical}

Sections~\ref{sec:retarder}--\ref{sec:crystal_plate} treated ideal retarders
and the crystal phase plate, the latter as an element characterized by two complex amplitudes
$r_s$, $r_p$. This final section connects that formalism to how polarizers
and phase shifters are actually realized with perfect-crystal optics at
synchrotron and X-ray free-electron-laser beamlines, and to the tools used
to evaluate their real, imperfect performance. The first part, below, treats
\textbf{polarizers} operated at the X-ray Brewster angle; a second part,
treating \textbf{phase shifters} (crystal quarter-wave plates), follows.

\subsection{X-ray Polarizers at the Brewster Angle}
\label{sec:polarizer_brewster}

Section~\ref{sec:crystal_plate} showed that the two eigenpolarizations couple
to the crystal with different strength through the polarization factor of
Eq.~(\ref{eq:polfactor}), $C_\sigma = 1$ and $C_\pi = \cos 2\theta_B$. At the
particular Bragg angle
\begin{equation}
    \theta_B = 45^\circ \quad\Longrightarrow\quad C_\pi = \cos 90^\circ = 0,
\end{equation}
the $\pi$-channel coupling to the crystal potential vanishes identically: in
the idealized two-beam dynamical theory the $\pi$-reflectivity curve
collapses to zero width and zero peak height, so $r_p \to 0$ while $r_s$
retains its full (near-unity, for a strong reflection) value. A crystal set
to this condition therefore acts as a near-perfect \textbf{linear polarizer
(or analyzer)}, reflecting essentially only $\sigma$-polarized light ---
the X-ray analogue of the optical Brewster angle, and the mechanism
exploited for high-precision X-ray polarimetry, e.g., in \cite{marx2013}. In the
notation of Eq.~(\ref{eq:Rsp_def}), the polarizer's residual leakage is the
ratio of the $\pi$ and $\sigma$ intensities after the polarizer
\begin{equation}
    |\varepsilon|^2 = \frac{R_p}{R_s} \;
    \longrightarrow\; 0 \qquad (\theta_B \to 45^\circ),
    \label{eq:delta0_def}
\end{equation}
the intensity leakage of Eq.~(\ref{eq:JLPleak}) (called $\delta_0$ in
Ref.~\cite{marxglowna2026}; we keep $|\varepsilon|^2$ to avoid confusion
with the retardation $\delta$).
Modelling such a device --- its reflectivity curve, and the transfer matrix
of a stack of crystal layers or of the multiple internal reflections inside
a monolithic channel-cut polarizer --- is exactly the calculation
implemented, from the Takagi--Taupin equations, in the open-source
\texttt{crystalpy} library~\cite{Guigay2024}.

\paragraph{Why the leakage is never exactly zero.} Equation~(\ref{eq:delta0_def})
assumes an infinitely narrow, perfectly collimated, monochromatic incident
beam probing the exact Bragg condition. A real beam has finite angular
divergence and finite energy bandwidth: rays of slightly different energy
or direction are reflected at Bragg angles slightly different from
$45^\circ$, where $C_\pi = \cos 2\theta_B \neq 0$ and the $\pi$-reflectivity
curve, though narrow, is not exactly zero. The
polarization purity actually achievable is then set not by
Eq.~(\ref{eq:delta0_def}) alone. Three mechanisms contribute. The first is
dynamical: the overlap of the crystal's narrow but nonzero $\pi$ rocking curve
with the beam's divergence and bandwidth. The second is geometric: a ray
whose direction differs from the central one has its own scattering plane, so its
$\sigma$ and $\pi$ directions are rotated with respect to those of the
polarizer by an angle of the order of its divergence, and part of the
$\sigma$-polarized field is projected onto the $\pi$ direction (an intensity
of the order of the square of that angle). For a well-collimated beam the
geometric term dominates~\cite{schulze2018,yu2025} and gives, to leading
order, the fundamental-limitation formula of \cite{yu2025} (after
\cite{schulze2018}), in which the divergences in and out of the diffraction
plane enter with the same weight,
\begin{equation}
    P_{\lim} = \sigma_h^2 + \sigma_v^2,
    \label{eq:Plim}
\end{equation}
with $\sigma_h$, $\sigma_v$ the rms horizontal and vertical divergence on the
polarizer. It contains no crystal parameter: the residual impurity grows
quadratically with beam divergence, and vanishes only for a perfectly
collimated beam. The third is \emph{multiple-beam} ($n$-beam) diffraction
(Section~\ref{sec:crystal_plate}): when another reciprocal-lattice point
is excited simultaneously, its own scattering plane mixes the $\sigma$ and
$\pi$ channels. It depends on the azimuthal orientation of the crystal and
of the beam, and it can limit the purity even when the azimuth is carefully
chosen to avoid strong multiple-beam reflections; this was apparently the
case in Refs.~\cite{marx2013,marxglowna2026}. The two-beam dynamical
theory of \texttt{crystalpy}, and therefore the ray tracing described
below, does not include this contribution.

\paragraph{Multiple reflections.} A single Bragg reflection near
$\theta_B=45^\circ$ is typically limited to a polarization impurity (residual
$\pi/\sigma$ intensity ratio) of $10^{-6}$--$10^{-8}$. Since successive reflections inside a monolithic
channel-cut crystal keep $\sigma$ on its Bragg peak while $\pi$ is
re-diffracted from its own, much weaker, off-peak tail each time, the
dynamical leakage is reduced multiplicatively at every bounce, roughly as
$|\varepsilon|^{2N}$ for $N$ reflections (the leaky polarizer of
Section~\ref{sec:polarizer_leaky}). The geometric term of
Eq.~(\ref{eq:Plim}) is not reduced by additional reflections, and it sets
the limit of multi-bounce polarizers. Silicon channel-cut polarizers with six
$90^\circ$ reflections have reached purities of $2.4\times10^{-10}$ and
$5.7\times10^{-10}$ at $6.457$ and $12.914$\,keV~\cite{marx2013},
and $(1.2\pm0.5)\times10^{-8}$ at $22.494$\,keV with the Si\,(5\,7\,11)
reflection, whose Bragg angle at this energy is $45.1^\circ$, in an
asymmetrically cut channel-cut polarizer~\cite{marxglowna2026}. This residual polarization impurity is critical
whenever the signal of interest is extremely faint compared to the residual
$\pi$-leakage: nuclear resonant (M\"ossbauer) scattering, where a
high-purity polarizer combined with a crossed analyzer rejects the
non-resonant prompt background~\cite{marxglowna2026}, and the detection of
vacuum birefringence at X-ray free-electron lasers, which requires linear
polarization purity below the expected signal, of order
$10^{-10}$--$10^{-13}$~\cite{yu2025}.

\paragraph{The role of ray tracing.} Because the achievable purity depends on
the full phase space of the source and beamline optics upstream of the
polarizer --- not just on the crystal's intrinsic rocking curve --- it must
generally be evaluated numerically rather than from Eq.~(\ref{eq:Plim})
alone. A ray-tracing simulation represents the beam as a multitude of
individual rays, each carrying its own energy and direction (divergence);
every ray is diffracted through each crystal reflection using the Jones
(Eq.~\ref{eq:Jchi_matrix}) or Mueller (Eq.~\ref{eq:Mueller_final}) matrix
appropriate to its own angle of incidence, with the amplitudes $r_s$, $r_p$
calculated by \texttt{crystalpy}~\cite{Guigay2024}; the implementation in
\texttt{SHADOW4} is described in \cite{SanchezdelRio2026efields}. Reference~\cite{yu2025} used this
approach, with an upgraded \texttt{SHADOW3}/\texttt{SHADOW4} implementation,
to simulate a complete vacuum-birefringence beamline and verified
Eq.~(\ref{eq:Plim}) down to simulated purities of $10^{-14}$ for realistic
(Gaussian) source divergences and $10^{-30}$--$10^{-32}$ for near-zero
divergence point-source limits, confirming that the fundamental limitation is
set by beam divergence rather than by any approximation in the crystal
reflectivity calculation itself.

\subsection{X-ray Phase Shifters Using the Transmitted Beam}
\label{sec:phase_shifter_geometry}

A crystal phase shifter needs both usable intensity and a large,
well-controlled phase retardation $\delta = \arg(r_p) - \arg(r_s)$
(Eq.~\ref{eq:Dphi_def}) between the two eigenpolarizations. A perfect
crystal offers four combinations of geometry and beam: the \textbf{Bragg
case} or the \textbf{Laue case}, each usable either in the
\textbf{diffracted} beam or in the \textbf{forward-transmitted} beam. The
two cases are distinguished by the surface through which the diffracted
beam leaves the crystal. With $\alpha$ the angle between the diffracting
planes and the crystal surface (the asymmetry angle, $\alpha = 0$ for a
symmetric Bragg and $\alpha = 90^\circ$ for a symmetric Laue cut, as in
\texttt{crystalpy}; its sign distinguishes the two orientations of an
asymmetric plate, the incidence angle on the surface being
$\theta_B + \alpha$), the diffracted beam leaves through the entrance surface
when $\theta_B > |\alpha|$ (Bragg case) and through the back surface when
$\theta_B < |\alpha|$ (Laue case). The same plate and reflection can
therefore work in the Bragg case at low energy and in the Laue case at high
energy: for example, the (111) planes of a diamond plate with a (110)
surface are at $\alpha = 35.3^\circ$, and the 111 reflection is in the Bragg
case below 5.2\,keV. Hirano et al.~\cite{hirano1993nim} worked out the
reflectivity (or transmittance) and the phase difference for exactly these
four cases as a function of the angular deviation from the Bragg angle, in
units of the Darwin half-width. We plot them versus the angular deviation
$\Delta\theta = \theta - \theta_B$ in microradians; the Darwin half-width
($\sigma$ polarization) is $10.0\,\mu$rad for Si 220 at 10\,keV, so that
``$n$ Darwin half-widths'' below means $n \times 10\,\mu$rad.

We reproduce these four cases with \texttt{crystalpy}~\cite{Guigay2024},
using the same dynamical-theory transfer matrices as in
Section~\ref{sec:polarizer_brewster}, for the Si 220 reflection at
10\,keV in all four cases (Hirano et al.\ used Ge 220 for the Bragg
diffraction case). The qualitative
behavior, in each case, is shown in Figs.~\ref{fig:hirano_intensity} and
\ref{fig:hirano_phase}:

\begin{figure}[H]
\centering
\includegraphics[width=\textwidth]{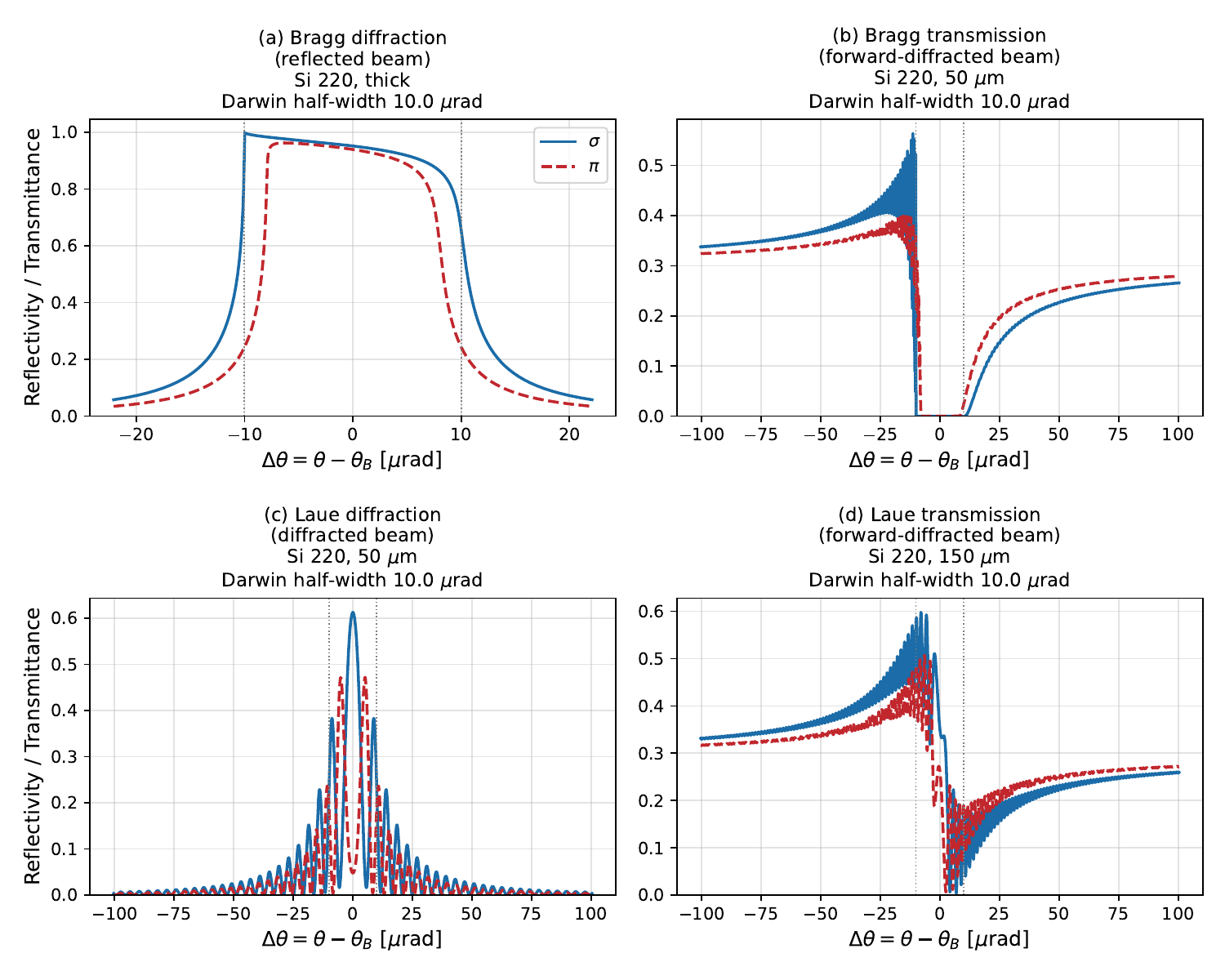}
\caption{Reflectivity and transmittance
($\sigma$, solid; $\pi$, dashed) versus angular deviation from the Bragg
angle, $\Delta\theta$, for the four geometry/beam combinations of
\cite{hirano1993nim}: (a) Bragg diffraction, Si 220, thick crystal;
(b) Bragg transmission, Si 220, 50\,$\mu$m; (c) Laue diffraction,
Si 220, 50\,$\mu$m; (d) Laue transmission, Si 220,
150\,$\mu$m. Panel (a) uses Si 220, instead of the Ge 220 of case (a) of
Ref.~\cite{hirano1993nim}, so that the four panels are for the same
reflection. Dotted vertical lines mark $\pm1$ Darwin half-width ($\sigma$
polarization), $10.0\,\mu$rad for Si 220, also given in each panel. Computed with \texttt{crystalpy}~\cite{Guigay2024}
at 10\,keV (script \texttt{hirano\_phase\_shifter\_plots.py}).}
\label{fig:hirano_intensity}
\end{figure}

\begin{figure}[H]
\centering
\includegraphics[width=\textwidth]{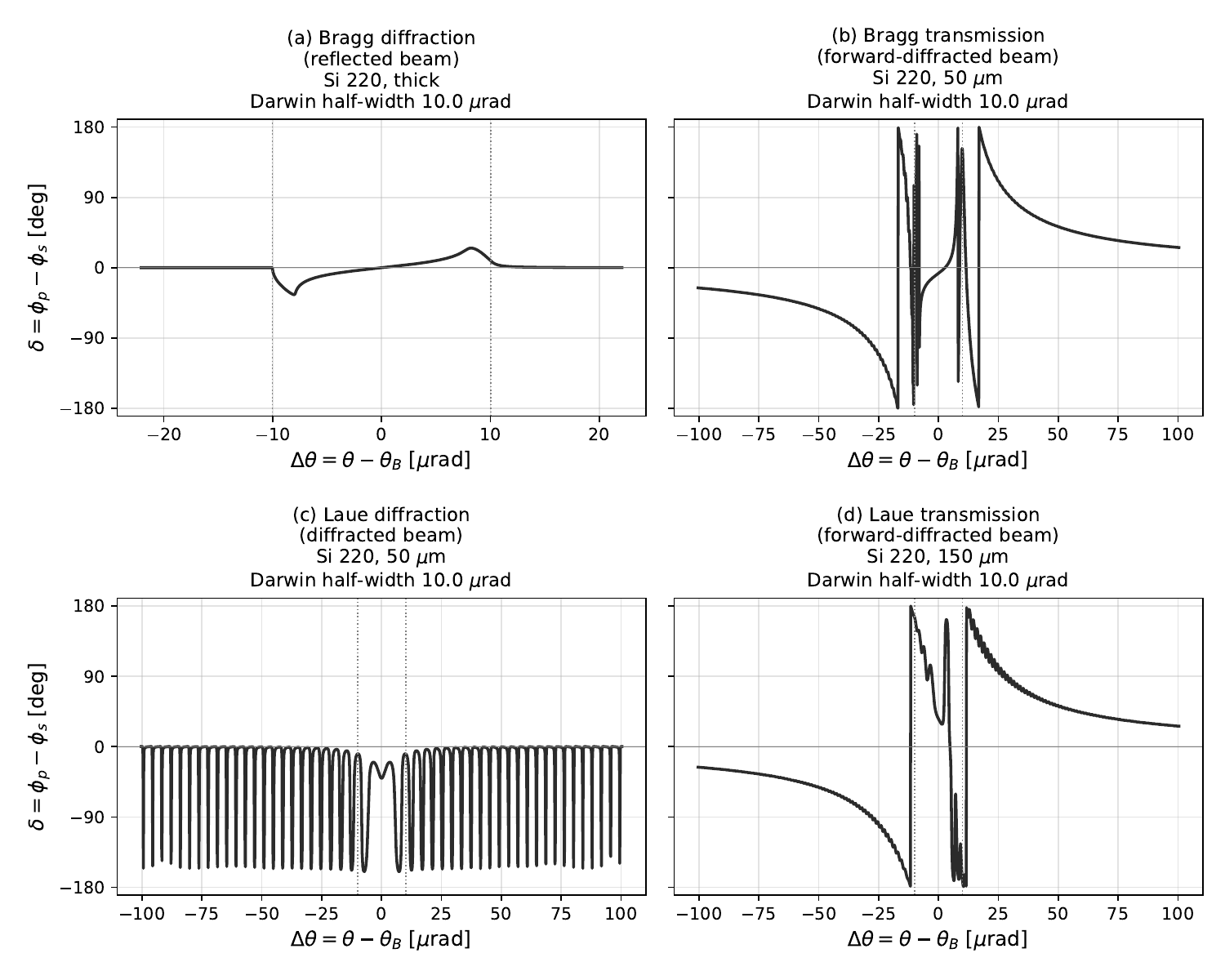}
\caption{Retardation $\delta = \phi_p - \phi_s$
(Eq.~\ref{eq:Dphi_def}, time factor $e^{-i\omega t}$) corresponding to
Fig.~\ref{fig:hirano_intensity}, for the same four cases. Only the two
transmission cases, (b) and (d), combine a large, smoothly varying
$\delta$ with substantial intensity (Fig.~\ref{fig:hirano_intensity}) a
few Darwin half-widths off Bragg. Note that Hirano et al.~\cite{hirano1993nim} plot the
phase retardations $\delta_h=\mathrm{Arg}(R_{\sigma\sigma}/R_{\pi\pi})$ and
$\delta_0=\mathrm{Arg}(T_{\sigma\sigma}/T_{\pi\pi})$, both equal to
$\phi_s-\phi_p=-\delta$ (their Eqs.~7 and 8), so that their curves have the opposite
sign; their case (a) is for Ge 220 and is the phase accumulated after four
successive reflections, while a single Si 220 reflection is shown here. (Script
\texttt{hirano\_phase\_shifter\_plots.py}.)}
\label{fig:hirano_phase}
\end{figure}

\paragraph{Bragg diffraction (reflected beam).} The reflectivity is close to
unity over the $\sim\!1$-Darwin-width-wide plateau (narrower for $\pi$, by
the polarization factor $|\cos2\theta_B|$ of Eq.~\ref{eq:polfactor}), but
$\delta$ is small over that plateau: it changes sign across it, reaching
only $-35^\circ$ and $+25^\circ$ for a single Si 220 reflection close to its low- and
high-angle edges, precisely where the reflectivity is falling steeply, and
vanishing altogether outside it (several successive reflections are needed to
accumulate a quarter-wave retardation~\cite{hirano1993nim}). The largest phase shifts are therefore available only at the very
edges of an extremely narrow angular range --- of the order of ten
microradians for a hard X-ray reflection --- demanding a narrow, carefully aligned reflection
and a beam of correspondingly high angular collimation.

\paragraph{Bragg transmission (forward-diffracted beam of a thin Bragg-cut
crystal).} Close to the Bragg position the transmitted intensity is
strongly suppressed, as most flux is diverted into the diffracted beam, but
it recovers to $35\%$--$43\%$ at about 2--5 Darwin half-widths off Bragg on
the low-angle side (where the transmission is enhanced; on the high-angle side
it is lower, $10\%$--$25\%$). At those detunings $|\delta|$ is still
large, from about $140^\circ$ at 2 half-widths to $50^\circ$ at 5 half-widths,
with opposite signs on the two sides of the reflection ($\delta<0$ on the
low-angle side). Away from the reflection $|\delta|$ decreases (as the
inverse of the deviation). On the low-angle side the transmission also
decreases slowly, from its enhanced value toward the normal absorption of the
plate (30\% far from the reflection); it increases only on the high-angle
side. The operating point is therefore a compromise between the size of the
phase difference and its sensitivity to the angle: close to the reflection
$\delta$ is large but changes rapidly (from $-143^\circ$ at 2 to
$-91^\circ$ at 3 half-widths), so that the divergence and bandwidth of the
beam spread it, while farther away it is smaller but flatter. As both the
phase and the transmission vary smoothly, a wide range of practically usable
operating points is available, requiring only modest, not extreme, angular
precision.

\paragraph{Laue diffraction (diffracted beam of a Laue-cut crystal,
$\alpha\approx90^\circ$).} The diffracted intensity oscillates
(Pendell\"osung fringes) over many Darwin widths. It is large only in the
central peak (up to $\approx60\%$ for $\sigma$ within about $\pm0.3$
Darwin half-width, where $\pi$ is nearly extinguished), where $\delta$ is
only $-20^\circ$ to $-40^\circ$. Beyond $\pm2$ half-widths the fringes stay
below $\approx11\%$, and $\delta$ oscillates rapidly from
fringe to fringe (down to $\approx-160^\circ$ near the minima of the
intensity): a large phase shift and significant intensity never coexist.

\paragraph{Laue transmission (forward-diffracted beam of the same Laue-cut
crystal).} Quantitatively similar to Bragg transmission, because the
150\,$\mu$m Laue plate has a path $t/\cos\theta_B = 159\,\mu$m close to the
155\,$\mu$m of the 50\,$\mu$m Bragg plate --- transmission
$35\%$--$43\%$ on the low-angle side with $|\delta| \sim 50^\circ$--$120^\circ$ at
2--5 Darwin half-widths off Bragg. It uses diffracting planes at a large
angle to the surface (perpendicular in the symmetric Laue case, but not
necessarily: e.g., the (111) planes of a (100)-cut diamond plate are at $54.7^\circ$
from the surface, so that the 111 reflection of such a plate is in the Laue
case above 3.7\,keV).

\paragraph{Angular tolerance.} In all four geometries the large
retardations ($|\delta| \gtrsim 90^\circ$) occur within about $30\,\mu$rad
(three Darwin half-widths) of the Bragg angle
(Figs.~\ref{fig:hirano_intensity} and \ref{fig:hirano_phase}). A phase
plate therefore requires a perfect, or very nearly perfect, crystal and a
beam with tightly controlled divergence and bandwidth. These conditions are
much relaxed for the polarization analyzers of
Section~\ref{sec:polarizer_brewster}: their suppression of $\pi$ comes from
the polarization factor $\cos 2\theta_B$, which varies only slowly around
$2\theta_B = 90^\circ$, and not from the fine structure of the rocking
curve.

\begin{figure}[H]
\centering
\includegraphics[width=\textwidth]{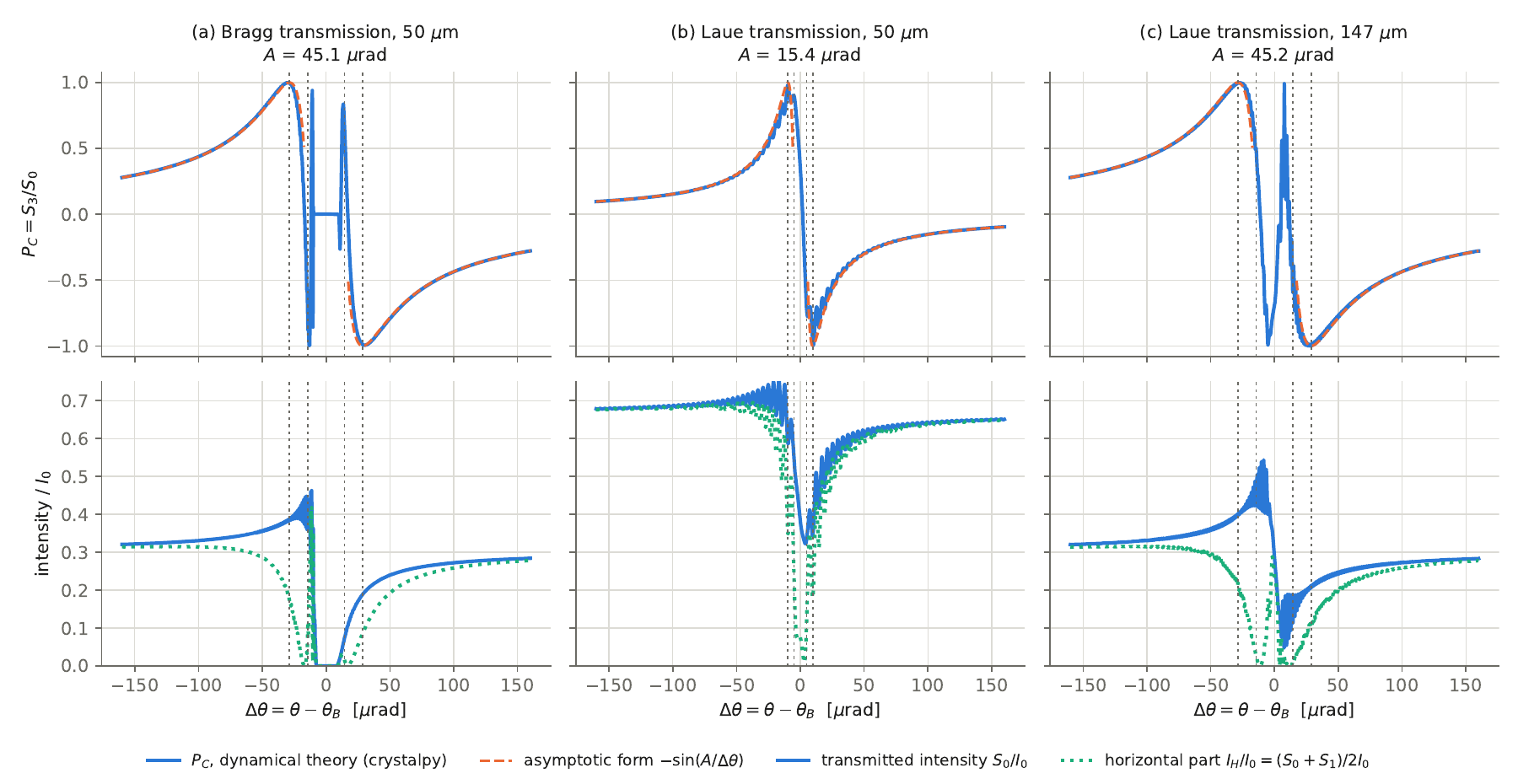}
\caption{Si 220 phase plate at 10\,keV, rotated to $\chi = 45^\circ$,
horizontally polarized input, versus the angular offset $\Delta\theta$ from
the Bragg angle: (a) symmetric Bragg transmission, 50\,$\mu$m; (b)
symmetric Laue transmission, 50\,$\mu$m; (c) symmetric Laue transmission,
147\,$\mu$m, the same path length $t/|\gamma_0|$ as in (a). Top: degree of
circular polarization of the transmitted beam, from the \texttt{crystalpy}
amplitudes inserted in Eq.~(\ref{eq:Mueller_final}) (solid), and the
asymptotic form $-\sin(A/\Delta\theta)$ of
Eq.~(\ref{eq:Dphi_asymptotic}) (dashed), which fails only within about two
to three Darwin half-widths of the reflection. Bottom: transmitted intensity $S_0$
and its horizontally polarized part $I_H = (S_0+S_1)/2$, the quantity
recorded by a polarization monitor (Section~\ref{sec:practical_other}). The
dotted lines mark the quarter- and half-wave offsets $\pm 2A/\pi$ and $\pm
A/\pi$ of Eq.~(\ref{eq:Dphi_asymptotic}). (Script
\texttt{phase\_plate\_scan\_plot.py}.)}
\label{fig:phase_plate_scan}
\end{figure}

\paragraph{Conclusion.} Two of the four combinations are impractical: Bragg
diffraction confines a large $\delta$ to the very edge of an extremely
narrow angular acceptance, and Laue diffraction gives either a small
$\delta$ (in the central peak) or a small and rapidly oscillating
intensity and phase (outside it). The two transmission
configurations, by contrast, give a $\delta$ that varies smoothly with
the angle, with a transmission of the order of the normal absorption of the
plate, so a large phase shift and workable transmission are obtained
together over a wide angular range. Both transmission geometries are used in
practice, in the Bragg
case~\cite{giles1994,giles1995rsi,lang1995,bouchenoire2012,francoual2013} and in
the Laue
case~\cite{giles1995rsi,bouchenoire2003,mazzoli2007,scagnoli2009,boada2010,suzuki2013,francoual2013},
often with the same plates at different energies or
reflections~\cite{francoual2013}. They have
the practical advantage of not deflecting the beam: the phase plate can be
moved in and out of the beam without realigning the rest of the
experiment, which is convenient, for instance, for the measurements with and
without a QWP of Section~\ref{sec:circular_case}. In both, the crystal is a
thin plate detuned by a few Darwin half-widths from the exact Bragg
condition, and it is described, as in
Sections~\ref{sec:retarder}--\ref{sec:crystal_plate} of the present paper,
purely by its two forward (transmission) complex amplitudes $r_s$, $r_p$
near (but not exactly at) the Bragg peak. We use Bragg transmission as the
main example and show the Laue case alongside it
(Fig.~\ref{fig:phase_plate_scan}).

\paragraph{Retardation far from the reflection.} A few Darwin
half-widths away from the reflection the dynamical-theory retardation takes
a simple asymptotic form~\cite{hirano1993nim,giles1994}. Far from the
Bragg condition each eigenmode of the forward-transmitted wave acquires a
small correction to its refractive index, proportional to $C^2
\chi_h\chi_{\bar{h}}$ (with $C$ of Eq.~\ref{eq:twobeam}) and inversely
proportional to the angular deviation $\Delta\theta = \theta - \theta_B$.
The difference between the two modes is proportional to $1 - \cos^2 2\theta_B
= \sin^2 2\theta_B$; since the deviation parameter is $\Delta\theta\sin
2\theta_B$, one power of $\sin 2\theta_B$ cancels, and over the path
$t/|\gamma_0|$ through a plate of
thickness $t$ it gives
\begin{equation}
    \delta \simeq \frac{A}{\Delta\theta}, \qquad
    A = \frac{r_e^2\,\lambda^3\,\mathrm{Re}(F_hF_{\bar h})\,\sin 2\theta_B}
             {2\pi V^2}\,\frac{t}{|\gamma_0|}
    \label{eq:Dphi_asymptotic}
\end{equation}
with $r_e$ the classical electron radius, $\lambda$ the wavelength, $F_h$,
$F_{\bar h}$ the structure factors, $V$ the unit-cell volume and $\gamma_0$
the direction cosine of the incident beam with the surface normal,
$|\gamma_0| = |\sin(\theta_B + \alpha)|$ with the asymmetry angle $\alpha$
defined above ($\sin\theta_B$ for symmetric Bragg and $\cos\theta_B$ for
symmetric Laue geometry). The geometry enters only through the path length
$L = t/|\gamma_0|$: $A/L$ is a property of the material and the reflection,
the same in the Bragg and in the Laue case (we verified this with
\texttt{crystalpy} to about $0.3$--$0.4\%$ for Si 220 at 10\,keV, for symmetric and
asymmetric cuts in both cases). The sign agrees with the curves of
Fig.~\ref{fig:hirano_phase}: $\delta < 0$ on the low-angle side. For the Si
220, 50\,$\mu$m plate of Figs.~\ref{fig:hirano_intensity}(b) and
\ref{fig:hirano_phase}(b), $A = 45.1\,\mu$rad, and the full dynamical
calculation gives the same value to within $0.3\%$ far from the reflection.
Three practical rules follow.
\begin{itemize}
    \item \textbf{Quarter-wave condition.} $|\delta| = \pi/2$ at
          $|\Delta\theta_{\mathrm{QWP}}| = 2A/\pi$, i.e.\ $28.7\,\mu$rad (2.9
          Darwin half-widths) for the plate above. Since $A \propto t$, a
          thicker plate works farther from the reflection.
    \item \textbf{Degree of circular polarization.} For a horizontal input
          and $\chi = 45^\circ$, Eq.~(\ref{eq:PC_final}) with $|r_s| \simeq
          |r_p|$ gives $P_C \simeq -\sin(A/\Delta\theta)$, i.e.\ the
          retardation of Giles et al.~\cite{giles1994} inserted in
          Eq.~(\ref{eq:PC_final}) (Fig.~\ref{fig:phase_plate_scan}).
    \item \textbf{Sensitivity.} $d\delta/d\Delta\theta = -\delta/\Delta\theta$:
          an angular spread $\sigma_\theta$ of the rays (beam divergence, and
          the angular equivalent $\tan\theta_B\,\Delta E/E$ of the
          bandwidth) spreads the retardation by $|\delta|\,\sigma_\theta/
          |\Delta\theta|$. At the quarter-wave point, however, $P_C =
          -\sin\delta$ is stationary, so the loss of circular polarization is
          only of second order, $1 - |P_C| \simeq \frac{1}{2}(\frac{\pi}{2}
          \frac{\sigma_\theta}{\Delta\theta_{\mathrm{QWP}}})^2$.
\end{itemize}

\paragraph{Operating point.} One ingredient is still missing: the
orientation $\chi$. A synchrotron beam is, to good approximation, linearly
polarized in the horizontal plane, along $\hat{x}$ (Section~\ref{sec:introduction}), and converting
linear into circular polarization requires exciting both crystal
eigenmodes equally, which --- as established in
Section~\ref{sec:crystal_chi} --- needs the birefringence axis rotated to
$\chi=45^\circ$ with respect to the incident polarization (or $\tan\chi =
|r_s|/|r_p|$ when the dichroism is not negligible). The practical
recipe for an X-ray phase shifter therefore combines all of the above: a
\textbf{thin crystal, used in transmission (Bragg or Laue case), set at the quarter-wave
offset $\Delta\theta_{\mathrm{QWP}} = \pm 2A/\pi$ of
Eq.~(\ref{eq:Dphi_asymptotic}) (a few Darwin half-widths off the exact Bragg
angle), and rotated to $\chi=45^\circ$}. Under these conditions the crystal
is fully described by the general (arbitrary-$\chi$) Jones and Mueller
matrices of Eqs.~(\ref{eq:Jchi_matrix}), (\ref{eq:Jchi_polar}) and
(\ref{eq:Mueller_final}) of Section~\ref{sec:crystal_chi}, evaluated at
$\chi=45^\circ$ with the dynamical $r_s$, $r_p$ at that offset: close to, but
not exactly at, the ideal QWP of Section~\ref{sec:QWP45}
($|r_s| = |r_p|$, $\delta = \pm\pi/2$), and close enough to deliver a large,
tunable degree of circular polarization with workable transmitted flux.
Fig.~\ref{fig:phase_plate_scan} shows the result for Si 220 at 10\,keV in
three configurations.
(a) In symmetric Bragg transmission, 50\,$\mu$m: at $\Delta\theta = \mp
28.7\,\mu$rad, $P_C = +0.995$ (transmission 38\%) and $-0.989$ (19\%); the
low-angle side gives the larger transmission.
(b) In symmetric Laue transmission, with the same thickness: the path
length $t/\cos\theta_B$ is shorter by the factor $\tan\theta_B = 0.34$, and
so are $A = 15.4\,\mu$rad and the quarter-wave offset, $9.8\,\mu$rad, i.e.\
only about one Darwin half-width. The transmission is higher (59\% and 35\%
at the two quarter-wave offsets), and the degree of circular polarization
there is still $+0.980$ and $-0.977$, but the operating point is close to the
reflection, where the retardation is most sensitive to the divergence and
bandwidth of the beam.
(c) In symmetric Laue transmission with the same path length as in (a),
$t = 147\,\mu$m, the curves are very close to those of (a): $A = 45.2\,\mu$rad
and, at $\Delta\theta = \mp 28.8\,\mu$rad, $P_C = +0.995$ (40\%) and
$-0.985$ (21\%). Far from the reflection the two geometries are equivalent
at equal path length, as stated by Eq.~(\ref{eq:Dphi_asymptotic}); they
differ only near the reflection, where the Laue case shows Pendell\"osung
oscillations instead of the total-reflection gap of the Bragg case.

\paragraph{Role of ray tracing.} As for the polarizer of
Section~\ref{sec:polarizer_brewster}, the numbers above hold ray by ray:
rays of different direction and energy see different $r_s$, $r_p$ and hence
a different retardation $\delta$. The polarization actually delivered
follows from ray tracing, applying to each ray's Jones (Eq.~\ref{eq:Ein_general})
or Stokes vector the matrix of Eq.~(\ref{eq:Jchi_matrix}) or
(\ref{eq:Mueller_final}) at $\chi=45^\circ$ with that ray's own amplitudes
from \texttt{crystalpy}~\cite{Guigay2024}, and averaging the output Stokes
vectors over all rays (the implementation is described in
Ref.~\cite{SanchezdelRio2026efields}).

\subsection{Other practical aspects}
\label{sec:practical_other}

Equation~(\ref{eq:Dphi_asymptotic}) also organizes the remaining choices
made when designing a phase plate.

\paragraph{Thickness: polarization versus flux.} Since $A \propto t$, a
thicker plate reaches the quarter-wave condition farther from the
reflection, where the relative spread $\sigma_\theta/\Delta\theta$ caused by
the divergence and bandwidth of the beam, and by any residual mosaicity of
the crystal, is smaller. A thinner plate absorbs less, with transmission
$\exp(-\mu t/|\gamma_0|)$, and gives better counting statistics in the
experiment. The thickness is chosen to balance the two.

\begin{figure}[htbp]
\centering
\includegraphics[width=0.7\textwidth]{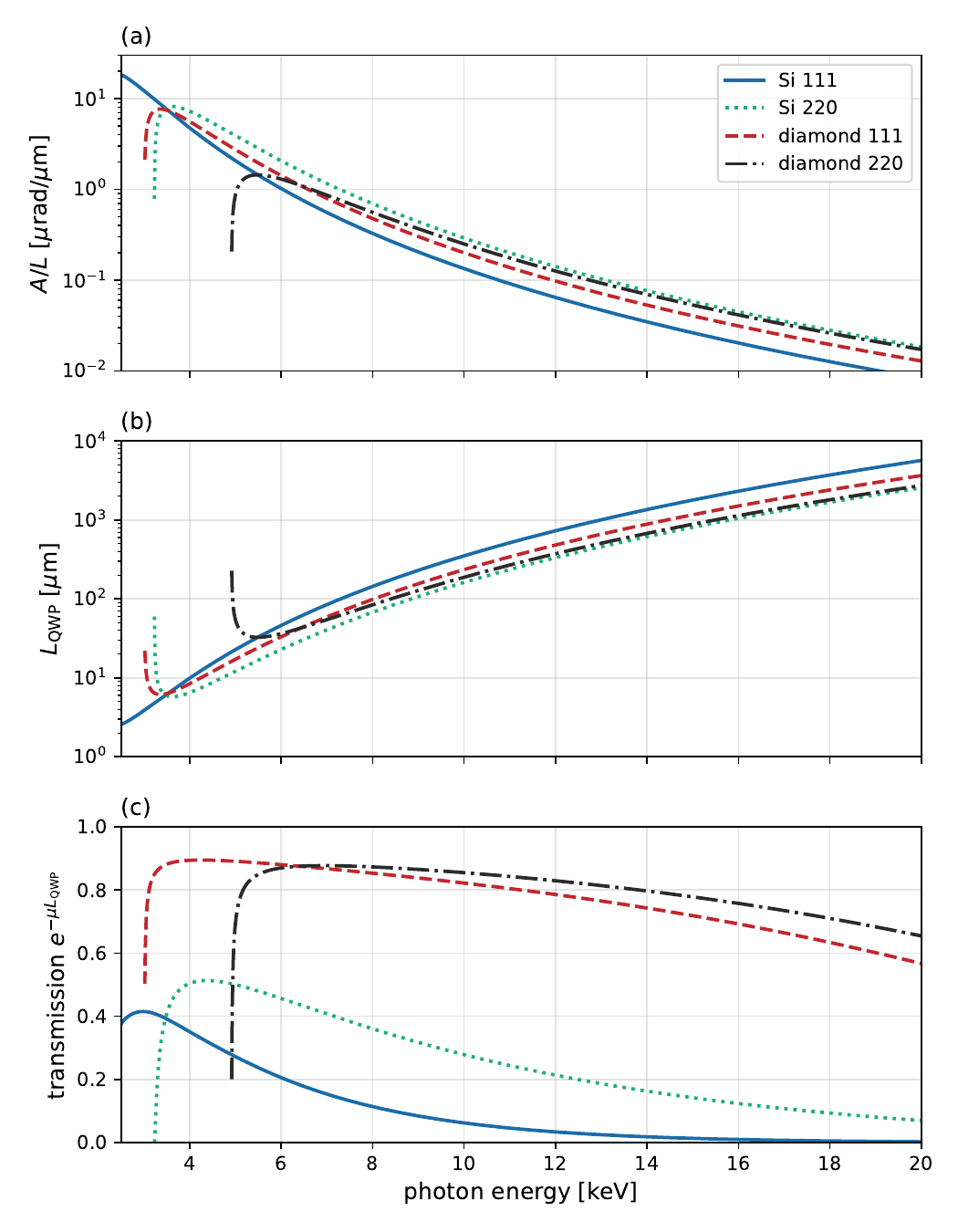}
\caption{Material and reflection of a transmission phase plate, from
Eq.~(\ref{eq:Dphi_asymptotic}), as a function of the photon energy: (a) $A/L$, with $L =
t/|\gamma_0|$ the path length through the plate; (b) the path length for a
quarter-wave plate at the offset $|\Delta\theta| = 30\,\mu$rad, $A =
(\pi/2)\times 30\,\mu$rad; (c) the transmission of that plate. The cutoff
energies $hc/2d$, where $\sin 2\theta_B \to 0$, are 1.98 (below the plotted
range), 3.23, 3.01 and 4.92\,keV for Si 111, Si 220, diamond 111 and diamond
220. Below about 5\,keV the offset of $30\,\mu$rad is within two Darwin
half-widths of the reflections, where Eq.~(\ref{eq:Dphi_asymptotic}) is not
accurate, and panels (b) and (c) are only indicative there. (Script
\texttt{phase\_plate\_materials\_plot.py}.)}
\label{fig:materials}
\end{figure}

\paragraph{Material and reflection.} The plate must be a near-perfect
crystal of low absorption: diamond and silicon are the usual choices.
Diamond (111) can only be used above $hc/2d_{111} = 3.01$\,keV; at lower
energies silicon (Si 111 down to $1.98$\,keV) is used. The reflection is
chosen with a figure of merit that combines the structure factor and the
Bragg angle, the quantity $\mathrm{Re}(F_hF_{\bar h})\sin 2\theta_B/V^2$ of
Eq.~(\ref{eq:Dphi_asymptotic}) per unit absorption.
Figure~\ref{fig:materials} compares Si 111, Si 220, diamond 111 and
diamond 220.
Both $A$ and the absorption are proportional to the path length $L =
t/|\gamma_0|$, so the comparison per unit path length does not depend on
the geometry: (a) $A/L$, (b) the path length $L_{\mathrm{QWP}}$ needed for
a quarter-wave plate at $|\Delta\theta| = 30\,\mu$rad, and (c) its
transmission. Diamond transmits far more than silicon at the same
retardation (at 10\,keV, 82\% and 86\% for diamond 111 and 220, against 6\%
and 28\% for Si 111 and 220). For silicon, 220 is better than 111 above
3.5\,keV.
Diamond 111 is the better reflection at low energy, but diamond 220 takes
over above 6.4\,keV, because the gain in $\sin 2\theta_B$ makes up for the
smaller structure factor (at 10\,keV, $|F_h|^2\sin 2\theta_B$ is about 171
for (111) and 214 for (220); see also the similar comparison of
Ref.~\cite{scagnoli2009}): at 10\,keV, a quarter-wave plate needs a path
of 235\,$\mu$m in diamond 111 and 188\,$\mu$m in diamond 220.

\paragraph{Double phase plates.} Splitting a given retardation between two
plates in series does not by itself reduce the total thickness: two plates
each giving half the retardation at the same offset are each half as
thick, with the same total absorption. Their advantage is that, set so that their dependences on
angle and energy partly cancel, they compensate much of the spread of the
retardation caused by the divergence and bandwidth of the beam (and keep
the same transmitted intensity for both helicities~\cite{francoual2013}).
The higher polarization purity obtained in this way allows the compromise
between purity and transmission to be moved toward thinner plates, which
matters most at low energy (around 4\,keV), where absorption is the
limit~\cite{scagnoli2009}.

\paragraph{Geometry.} Besides the symmetric Laue and Bragg cuts, an
asymmetric (miscut) crystal changes $|\gamma_0| = |\sin(\theta_B+\alpha)|$
and hence the path length $t/|\gamma_0|$ entering
Eq.~(\ref{eq:Dphi_asymptotic}), which gives an additional handle on $A$.
Since $A$ and the absorption depend only on the path length, the choice
between Bragg and Laue transmission, and of the cut, is mostly a practical
one: the available plate thickness (at low energy, where $\theta_B$ is
large and a very short path is needed, the symmetric Laue path
$t/\cos\theta_B$ would require an extremely thin plate, while the Bragg
path $t/\sin\theta_B$ is close to $t$), the energy range to be covered by
one plate, and the behavior close to the reflection
(Fig.~\ref{fig:phase_plate_scan}).

\paragraph{Polarization monitor and alignment.} A simple polarization
monitor is a thin polymer film in the beam, with one or two detectors
recording the Thomson-scattered intensity at $90^\circ$. A detector in the
vertical plane sees only the horizontally polarized part of the beam,
$I_H = (S_0+S_1)/2$, and one in the horizontal plane only the vertical part,
$I_V = (S_0-S_1)/2$, so that together they measure $P_1 = (I_H-I_V)/(I_H+I_V)$.
For a phase plate at $\chi = 45^\circ$ with $|r_s| \simeq |r_p|$,
Eq.~(\ref{eq:FinalS1H}) gives $P_1 \simeq \cos\delta(\Delta\theta)$. A scan of
the plate angle through the reflection therefore produces the characteristic
curve of Fig.~\ref{fig:phase_plate_scan} (bottom, sometimes called the
``Batman'' curve for its shape): $I_H$ equals the transmitted intensity far
from the reflection, where $\delta \to 0$, and drops toward zero near the
half-wave offsets. The scan locates the Bragg angle, and the quarter-wave
setting is where $P_1 = 0$ ($I_H = I_V$); a fit of $\cos(A/\Delta\theta)$ to
the wings calibrates $A$. Close to the reflection, within about two to three Darwin
half-widths, the asymptotic form is not accurate, so the calibration should
use the wings.

\section*{Appendix~A: Numerical and Symbolic Verification}
\addcontentsline{toc}{section}{Appendix A: Numerical and Symbolic Verification}
\label{sec:verification}

The formulas of this paper are verified in two independent ways: by a
Python script (SymPy and NumPy), described in Appendix~A.1, and by a
\emph{Mathematica} notebook, described in Appendix~A.2.

\subsection*{A.1\quad Python verification script}
\addcontentsline{toc}{subsection}{A.1\quad Python verification script}
\label{sec:verification_python}

Because the results collected here depend on a chain of sign and handedness
conventions (Section~\ref{sec:sign_convention}), and because a single
transcription slip in a $4\times4$ Mueller matrix is both easy to make and hard
to see, the matrix identities and closed-form results of this paper ---
every labelled identity of Table~\ref{tab:verification} (definitions are not
checked) --- are
checked by a companion script, \texttt{check\_paper\_v12.py}, distributed with
the source of this paper.\footnote{\url{https://github.com/srio/paper-polarisation}}

The verification uses two complementary methods. \emph{Symbolically}, with
SymPy, the identities are established exactly for arbitrary symbolic $\chi$,
$\delta$, $r_s$ and $r_p$ --- for instance the Pauli decompositions
Eqs.~(\ref{eq:J0_Pauli}) and (\ref{eq:Jchi_Pauli}), the rotation sandwich
Eq.~(\ref{eq:Jones_rotation}), and the Pauli form of the rotated retarder
given after Eq.~(\ref{eq:Jret_general}). \emph{Numerically}, each closed-form expression printed
in the text is compared against the matrix rebuilt from first principles ---
the Pauli trace formula Eq.~(\ref{eq:PauliTrace}) applied to an explicitly
constructed Jones matrix --- over several hundred pseudo-random parameter
sets per identity. The two routes through the formalism are evaluated
independently and compared: an output Stokes vector obtained from the Jones
vector via Eqs.~(\ref{eq:S0def})--(\ref{eq:S3def}) must agree with the one
obtained by applying the Mueller matrix to the input Stokes vector, so an
error in either route would show up as a disagreement.

Table~\ref{tab:verification} lists what is covered.

\begin{table}[H]
\centering
\renewcommand{\arraystretch}{1.3}
\begin{tabular}{@{}>{\raggedright\arraybackslash}p{4.1cm}
                  >{\raggedright\arraybackslash}p{10.5cm}@{}}
\toprule
\textbf{Group} & \textbf{Equations verified} \\
\midrule
Formalism and conventions (Section~\ref{sec:sign_convention}) &
Eqs.~(\ref{eq:physical_wave}), (\ref{eq:real_components}),
(\ref{eq:S0def})--(\ref{eq:S3def}), (\ref{eq:Pdef}),
(\ref{eq:S0def2})--(\ref{eq:S3def2}), (\ref{eq:Pauli_def}),
(\ref{eq:Sk_trace}), (\ref{eq:rho_from_Stokes}), (\ref{eq:PauliTrace}) \\
Rotations (Section~\ref{sec:rotation}) &
Eqs.~(\ref{eq:Rchi})--(\ref{eq:Mueller_rotation}), including
$\mathbf{R}_M(\chi)$ as the Mueller matrix of $\mathbf{R}_J(\chi)$, the
period $180^\circ$, the multiplicativity of the Pauli trace, the opposite
ordering of the sandwich and the polarizer at $\pm45^\circ$ \\
Retarders (Section~\ref{sec:general_retarders}) &
Eqs.~(\ref{eq:Jret0})--(\ref{eq:MQWP45}), (\ref{eq:MQWPm45}), (\ref{eq:Sin}),
(\ref{eq:Eout_RCP}), (\ref{eq:rho_RCP}), (\ref{eq:rho_LCP}),
(\ref{eq:MHWP0})--(\ref{eq:MHWP_general}), the QWP on linear and circular
input, two QWPs in series, and Tables~\ref{tab:JM_elements} and
\ref{tab:JM_HWP} \\
Linear polarizers (Section~\ref{sec:general_polarizers}) &
Eqs.~(\ref{eq:JLP0})--(\ref{eq:Itheta_leak}), including the projector
properties $\mathbf{J}_{LP}^2 = \mathbf{J}_{LP}$, $\det\mathbf{J}_{LP}=0$,
Malus' law as drawn in Figure~\ref{fig:polarizer}, and the leaky polarizer \\
Polarimetry (Section~\ref{sec:experimental}) &
Eqs.~(\ref{eq:Itheta})--(\ref{eq:DOCP_exp}), (\ref{eq:Icirc}), the
RCP column of Table~\ref{tab:general_protocol}, and the three-scan protocol
of Figure~\ref{fig:analyzer_scans} \\
Crystal phase plate (Section~\ref{sec:crystal_plate}) &
Eqs.~(\ref{eq:Je_general})--(\ref{eq:Jchi_matrix}), (\ref{eq:M0_Pauli}),
(\ref{eq:Jchi_polar}), (\ref{eq:Mueller_final}),
(\ref{eq:Eout_general})--(\ref{eq:Eout_chi}),
(\ref{eq:Sout0_gen})--(\ref{eq:FinalS3H}), the output Stokes vectors for
$+45^\circ$ and LCP inputs, (\ref{eq:PC_general}), (\ref{eq:PC_final}),
Tables~\ref{tab:Pc_general} and \ref{tab:Pc_cases}, and the geometry of
Figure~\ref{fig:sigma_pi} with the coupling of Eq.~(\ref{eq:twobeam}) \\
Phase plates in practice (Section~\ref{sec:practical}) &
The consequences of Eq.~(\ref{eq:Dphi_asymptotic}): quarter-wave offset,
$P_C \simeq -\sin(A/\Delta\theta)$, second-order loss at the quarter-wave
point, and the polarization monitor; the value of $A$ is checked against
the full dynamical calculation (\texttt{crystalpy}) by the script that
produces Figure~\ref{fig:phase_plate_scan} \\
\bottomrule
\end{tabular}
\caption{Groups of results verified by the companion script. Within each
group every labelled identity is checked, symbolically where the identity
holds for arbitrary parameters and numerically over randomized parameter sets
otherwise.}
\label{tab:verification}
\end{table}

Three checks deserve separate mention. First, the convention chain of
Section~\ref{sec:sign_convention} is verified \emph{as a chain}: the time
factor of Eq.~(\ref{eq:physical_wave}), the Stokes definitions
Eqs.~(\ref{eq:S0def})--(\ref{eq:S3def}), the Pauli assignment
Eq.~(\ref{eq:Pauli_def}) and the anchor state
$(1,i)^T/\sqrt{2} \mapsto S_3 = +1$ are imposed together, and the script
confirms that the handedness of every state quoted in the text follows from
them. Second, the decoupling of the $\sigma$ and $\pi$ channels
(Section~\ref{sec:crystal_plate}) is checked by building the polarization
factor of Eq.~(\ref{eq:polfactor}) from explicit scattering geometry, which
returns $\mathrm{diag}(1,\cos 2\theta_B)$ with strictly vanishing cross terms.
Third, the equivalence claimed in Section~\ref{sec:crystal_plate} between the
crystal and an ideal retarder is verified directly: for $|r_s| = |r_p|$ the
lower-right block of Eq.~(\ref{eq:M0_Pauli}) coincides with the $(S_2,S_3)$
rotation of Eq.~(\ref{eq:Mret0}) with the same retardation $\delta$.

Finally, the script is kept in step with the text without manual
bookkeeping. Its checks are keyed to the \LaTeX{} \texttt{\symbol{92}label} names rather than to equation numbers, and the current numbers are
resolved at run time from \texttt{main12.aux}. Renaming or deleting a labelled
equation therefore produces an explicit report of the missing label rather
than a silently stale cross-reference, and the equation numbers quoted in this
appendix, being \texttt{\symbol{92}ref}s, follow the document
automatically.

\subsection*{A.2\quad Mathematica notebook}
\addcontentsline{toc}{subsection}{A.2\quad Mathematica notebook}
\label{sec:verification_mathematica}

The notebook \texttt{HalfWavePlate3.nb}, distributed with the source of this
paper, re-derives the formalism in Jones, density-matrix and Mueller
notation with \emph{Mathematica}, independently of the Python script of
Appendix~A.1: the expressions are written separately and simplified by a
different computer-algebra system. It follows the order of the paper. It
starts from the definitions (Stokes parameters from the Jones vector,
Eqs.~\ref{eq:S0def}--\ref{eq:S3def}; the density matrix and the Pauli
basis; the Pauli trace formula, Eq.~\ref{eq:PauliTrace}) and goes through
the rotations of Section~\ref{sec:rotation}, the retarders, quarter- and
half-wave plates of Section~\ref{sec:general_retarders}, the ideal and leaky
polarizers of Section~\ref{sec:general_polarizers}, the polarimetry of
Section~\ref{sec:experimental}, the crystal of
Section~\ref{sec:crystal_plate} and the practical consequences of
Eq.~(\ref{eq:Dphi_asymptotic}).

Each printed equation is compared with its derivation by a single function,
\texttt{check}, which records whether the difference between the derived
and the printed expression vanishes identically. It first tries an exact
simplification and falls back on a numerical test only if that is
inconclusive. Complex amplitudes are written as modulus times phase
factor, so that all symbols can be treated as real. The checks are labelled
by equation number and \LaTeX{} label, and a summary table at the end of the
notebook lists them all; in the present version all 147 pass.

The notebook is generated from a plain-text source
(\texttt{HalfWavePlate3.src}) by a script (\texttt{build\_HalfWavePlate3.wl})
that evaluates every cell, so it can be rebuilt, and re-checked, without
opening the \emph{Mathematica} front end; a PDF export is provided for
reading without \emph{Mathematica}.

\section*{Data availability}
\addcontentsline{toc}{section}{Data availability}

The supplementary material of this paper is available on Zenodo,
\url{https://doi.org/10.5281/zenodo.24466587}. It contains the \LaTeX{}
source of the paper with its figures, the Python scripts that produce the
figures and the numbers of Sections~\ref{sec:experimental} and
\ref{sec:practical}, the Python verification script of Appendix~A.1, and
the \emph{Mathematica} notebook of Appendix~A.2 with its source and a PDF
export.

\bibliographystyle{unsrtnat}
\bibliography{xray_polarization}

\begin{thebibliography}{41}
\providecommand{\natexlab}[1]{#1}
\providecommand{\url}[1]{\texttt{#1}}
\expandafter\ifx\csname urlstyle\endcsname\relax
  \providecommand{\doi}[1]{doi: #1}\else
  \providecommand{\doi}{doi: \begingroup \urlstyle{rm}\Url}\fi

\bibitem[Blume and Gibbs(1988)]{blume1988}
M.~Blume and D.~Gibbs.
\newblock Polarization dependence of magnetic {X}-ray scattering.
\newblock \emph{Physical Review B}, 37\penalty0 (4):\penalty0 1779--1789, 1988.
\newblock \doi{10.1103/PhysRevB.37.1779}.

\bibitem[Blume(1985)]{blume1994}
M.~Blume.
\newblock Magnetic scattering of {X}-rays.
\newblock \emph{Journal of Applied Physics}, 57\penalty0 (8):\penalty0
  3615--3618, 1985.
\newblock \doi{10.1063/1.335023}.

\bibitem[Detlefs et~al.(2012)Detlefs, {Sanchez del Rio}, and
  Mazzoli]{detlefs2012}
C.~Detlefs, M.~{Sanchez del Rio}, and C.~Mazzoli.
\newblock X-ray polarization: General formalism and polarization analysis.
\newblock \emph{The European Physical Journal Special Topics}, 208\penalty0
  (1):\penalty0 359--371, 2012.
\newblock \doi{10.1140/epjst/e2012-01630-3}.

\bibitem[Hirano et~al.(1991)Hirano, Izumi, Ishikawa, Annaka, and
  Kikuta]{hirano1993}
K.~Hirano, K.~Izumi, T.~Ishikawa, S.~Annaka, and S.~Kikuta.
\newblock An {X}-ray phase plate using {Bragg}-case diffraction.
\newblock \emph{Japanese Journal of Applied Physics}, 30\penalty0
  (3B):\penalty0 L407--L410, 1991.
\newblock \doi{10.1143/JJAP.30.L407}.

\bibitem[Giles et~al.(1994)Giles, Malgrange, Goulon, de~Bergevin, Vettier,
  Dartyge, Fontaine, Giorgetti, and Pizzini]{giles1994}
C.~Giles, C.~Malgrange, J.~Goulon, F.~de~Bergevin, C.~Vettier, E.~Dartyge,
  A.~Fontaine, C.~Giorgetti, and S.~Pizzini.
\newblock Energy-dispersive phase plate for magnetic circular dichroism
  experiments in the {X}-ray range.
\newblock \emph{Journal of Applied Crystallography}, 27\penalty0 (3):\penalty0
  232--240, 1994.
\newblock \doi{10.1107/S0021889893007976}.

\bibitem[Bouchenoire et~al.(2003)Bouchenoire, Brown, Thompson, Duffy, Taylor,
  and Cooper]{bouchenoire2003}
L.~Bouchenoire, S.~D. Brown, P.~Thompson, J.~A. Duffy, J.~W. Taylor, and M.~J.
  Cooper.
\newblock Performance of phase plates on the {XM}a{S} beamline at the {ESRF}.
\newblock \emph{Journal of Synchrotron Radiation}, 10\penalty0 (2):\penalty0
  172--176, 2003.
\newblock \doi{10.1107/S0909049502018654}.

\bibitem[Lang and Srajer(1995)]{lang1995}
J.~C. Lang and G.~Srajer.
\newblock Bragg transmission phase plates for the production of circularly
  polarized {X} rays.
\newblock \emph{Review of Scientific Instruments}, 66\penalty0 (2):\penalty0
  1540--1542, 1995.
\newblock \doi{10.1063/1.1145902}.

\bibitem[Suzuki et~al.(2013)Suzuki, Kawamura, Mizumaki, Terada, Uruga,
  Fujiwara, Yamazaki, Yumoto, Koyama, Senba, Takeuchi, Ohashi, Nariyama,
  Takeshita, Kimura, Matsushita, Furukawa, Ohata, Kondo, Ariake, Richter, Fons,
  Sekizawa, Ishiguro, Tada, Goto, Yamamoto, Takata, and Ishikawa]{suzuki2013}
M.~Suzuki, N.~Kawamura, M.~Mizumaki, Y.~Terada, T.~Uruga, A.~Fujiwara,
  H.~Yamazaki, H.~Yumoto, T.~Koyama, Y.~Senba, T.~Takeuchi, H.~Ohashi,
  N.~Nariyama, K.~Takeshita, H.~Kimura, T.~Matsushita, Y.~Furukawa, T.~Ohata,
  Y.~Kondo, J.~Ariake, J.~Richter, P.~Fons, O.~Sekizawa, N.~Ishiguro, M.~Tada,
  S.~Goto, M.~Yamamoto, M.~Takata, and T.~Ishikawa.
\newblock A hard {X}-ray nanospectroscopy station at {SPring-8 BL39XU}.
\newblock \emph{Journal of Physics: Conference Series}, 430:\penalty0 012017,
  2013.
\newblock \doi{10.1088/1742-6596/430/1/012017}.

\bibitem[Apseros et~al.(2024)Apseros, Scagnoli, Holler, Guizar-Sicairos, Gao,
  Appel, Heyderman, Donnelly, and Ihli]{apseros2024}
A.~Apseros, V.~Scagnoli, M.~Holler, M.~Guizar-Sicairos, Z.~Gao, C.~Appel, L.~J.
  Heyderman, C.~Donnelly, and J.~Ihli.
\newblock {X}-ray linear dichroic tomography of crystallographic and
  topological defects.
\newblock \emph{Nature}, 636:\penalty0 354--360, 2024.
\newblock \doi{10.1038/s41586-024-08233-y}.

\bibitem[{Diamond Light Source}()]{diamondI16}
{Diamond Light Source}.
\newblock {I16 Materials and Magnetism beamline: beamline features}.
\newblock
  \url{https://www.diamond.ac.uk/Instruments/Magnetic-Materials/I16/Beamline-Guide/Beamline-Features.html}.
\newblock Accessed 23 November 2026.

\bibitem[Francoual et~al.(2013)Francoual, Strempfer, Reuther, Shukla, and
  Skaugen]{francoual2013}
S.~Francoual, J.~Strempfer, D.~Reuther, D.~K. Shukla, and A.~Skaugen.
\newblock Double phase-retarder set-up at beamline {P}09 at {PETRA} {III}.
\newblock \emph{Journal of Physics: Conference Series}, 425\penalty0
  (13):\penalty0 132010, 2013.
\newblock \doi{10.1088/1742-6596/425/13/132010}.

\bibitem[Jones(1941)]{jones1941}
R.~C. Jones.
\newblock A new calculus for the treatment of optical systems. {I}.
  {Description} and discussion of the calculus.
\newblock \emph{Journal of the Optical Society of America}, 31\penalty0
  (7):\penalty0 488--493, 1941.
\newblock \doi{10.1364/JOSA.31.000488}.

\bibitem[Born and Wolf(2002)]{born2002}
M.~Born and E.~Wolf.
\newblock \emph{Principles of {Optics}}.
\newblock Cambridge University Press, 7th (expanded) edition, 2002.
\newblock ISBN 978-0-521-64222-4.

\bibitem[Collett(1993)]{collett1993}
E.~Collett.
\newblock \emph{Polarized Light: Fundamentals and Applications}.
\newblock Marcel Dekker, New York, 1993.

\bibitem[Mazzoli et~al.(2007)Mazzoli, Wilkins, {Di Matteo}, Detlefs, Detlefs,
  Scagnoli, Paolasini, and Ghigna]{mazzoli2007}
C.~Mazzoli, S.~B. Wilkins, S.~{Di Matteo}, B.~Detlefs, C.~Detlefs, V.~Scagnoli,
  L.~Paolasini, and P.~Ghigna.
\newblock Disentangling multipole resonances through a full {X}-ray
  polarization analysis.
\newblock \emph{Physical Review B}, 76\penalty0 (19):\penalty0 195118, 2007.
\newblock \doi{10.1103/PhysRevB.76.195118}.

\bibitem[Jackson(1999)]{jackson1999}
J.~D. Jackson.
\newblock \emph{Classical Electrodynamics}.
\newblock Wiley, New York, 3rd edition, 1999.
\newblock ISBN 978-0-471-30932-1.
\newblock See Sec.~7.2 for the helicity vs.\ optics handedness naming.

\bibitem[McMaster(1961)]{mcmaster1961}
W.~H. McMaster.
\newblock Matrix representation of polarization.
\newblock \emph{Reviews of Modern Physics}, 33\penalty0 (1):\penalty0 8--28,
  1961.
\newblock \doi{10.1103/RevModPhys.33.8}.

\bibitem[Hamaker and Bregman(1996)]{hamaker1996}
J.~P. Hamaker and J.~D. Bregman.
\newblock Understanding radio polarimetry. {III}. {Interpreting} the {IAU/IEEE}
  definitions of the {Stokes} parameters.
\newblock \emph{Astronomy and Astrophysics Supplement Series}, 117:\penalty0
  161--165, 1996.
\newblock \doi{10.1051/aas:1996146}.

\bibitem[Mueller(1948)]{mueller1948}
H.~Mueller.
\newblock The foundation of optics.
\newblock \emph{Journal of the Optical Society of America}, 38:\penalty0 661,
  1948.
\newblock Abstract only; full lecture presented at the Spring Meeting of the
  Optical Society of America, 1943.

\bibitem[Chipman et~al.(2018)Chipman, Lam, and Young]{chipman2018}
R.~A. Chipman, W.-S.~T. Lam, and G.~Young.
\newblock \emph{Polarized Light and Optical Systems}.
\newblock CRC Press, Boca Raton, 2018.
\newblock \doi{10.1201/9781351129121}.

\bibitem[Belyakov and Dmitrienko(1989)]{belyakov1989}
V.~A. Belyakov and V.~E. Dmitrienko.
\newblock Polarisation phenomena in {X}-ray optics.
\newblock \emph{Soviet Physics Uspekhi}, 32\penalty0 (8):\penalty0 697--719,
  1989.
\newblock \doi{10.1070/PU1989v032n08ABEH002748}.

\bibitem[Batterman and Cole(1964)]{batterman1964}
B.~W. Batterman and H.~Cole.
\newblock Dynamical diffraction of {X} rays by perfect crystals.
\newblock \emph{Reviews of Modern Physics}, 36\penalty0 (3):\penalty0 681--717,
  1964.
\newblock \doi{10.1103/RevModPhys.36.681}.

\bibitem[Authier(2001)]{authier2001}
A.~Authier.
\newblock \emph{Dynamical Theory of {X}-ray Diffraction}.
\newblock Number~11 in IUCr Monographs on Crystallography. Oxford University
  Press, Oxford, 2001.
\newblock ISBN 0-19-855960-7.
\newblock Revised edition 2004.

\bibitem[Hill and McMorrow(1996)]{hill1996}
J.~P. Hill and D.~F. McMorrow.
\newblock {X}-ray resonant exchange scattering: Polarization dependence and
  correlation functions.
\newblock \emph{Acta Crystallographica Section A}, 52\penalty0 (2):\penalty0
  236--244, 1996.
\newblock \doi{10.1107/S0108767395012670}.

\bibitem[Templeton and Templeton(1980)]{templeton1980}
D.~H. Templeton and L.~K. Templeton.
\newblock Polarized {X}-ray absorption and double refraction in vanadyl
  bisacetylacetonate.
\newblock \emph{Acta Crystallographica Section A}, 36\penalty0 (2):\penalty0
  237--241, 1980.
\newblock \doi{10.1107/S0567739480000472}.

\bibitem[Dmitrienko(1983)]{dmitrienko1983}
V.~E. Dmitrienko.
\newblock Forbidden reflections due to anisotropic {X}-ray susceptibility of
  crystals.
\newblock \emph{Acta Crystallographica Section A}, 39\penalty0 (1):\penalty0
  29--35, 1983.
\newblock \doi{10.1107/S0108767383000057}.

\bibitem[Okitsu et~al.(2006)Okitsu, Yoda, Imai, Ueji, Urano, and
  Zhang]{okitsu2006}
K.~Okitsu, Y.~Yoda, Y.~Imai, Y.~Ueji, Y.~Urano, and X.~Zhang.
\newblock Polarization-dependent six-beam {X}-ray pinhole topographs.
\newblock \emph{Acta Crystallographica Section A}, 62\penalty0 (4):\penalty0
  237--247, 2006.
\newblock \doi{10.1107/S0108767306015443}.

\bibitem[Stetsko et~al.(2004)Stetsko, Lee, Tang, and Chang]{stetsko2004}
Y.~P. Stetsko, Y.-R. Lee, M.-T. Tang, and S.-L. Chang.
\newblock Phase-dependent polarization aspects of three-wave {X}-ray
  diffraction: an iterative {Born} approximation.
\newblock \emph{Acta Crystallographica Section A}, 60\penalty0 (1):\penalty0
  64--74, 2004.
\newblock \doi{10.1107/S0108767303025017}.

\bibitem[Shen and Finkelstein(1992)]{shen1992}
Q.~Shen and K.~D. Finkelstein.
\newblock Complete determination of {X}-ray polarization using multiple-beam
  {Bragg} diffraction.
\newblock \emph{Physical Review B}, 45\penalty0 (9):\penalty0 5075--5078, 1992.
\newblock \doi{10.1103/PhysRevB.45.5075}.

\bibitem[Giles(1995)]{giles1995thesis}
C.~Giles.
\newblock \emph{{\'E}tude de lames quart d'onde pour les rayons {X} et
  application {\`a} un montage dispersif en {\'e}nergie}.
\newblock PhD thesis, Universit{\'e} Paris 7, 1995.
\newblock \url{https://cloud.esrf.fr/s/WpeQwESMBodiKsW}.

\bibitem[Marx et~al.(2013)Marx, Schulze, Uschmann, K{\"a}mpfer, L{\"o}tzsch,
  Wehrhan, Wagner, Detlefs, Roth, H{\"a}rtwig, F{\"o}rster, St{\"o}hlker, and
  Paulus]{marx2013}
B.~Marx, K.~S. Schulze, I.~Uschmann, T.~K{\"a}mpfer, R.~L{\"o}tzsch,
  O.~Wehrhan, W.~Wagner, C.~Detlefs, T.~Roth, J.~H{\"a}rtwig, E.~F{\"o}rster,
  T.~St{\"o}hlker, and G.~G. Paulus.
\newblock High-precision {X}-ray polarimetry.
\newblock \emph{Physical Review Letters}, 110:\penalty0 254801, 2013.
\newblock \doi{10.1103/PhysRevLett.110.254801}.

\bibitem[Marx-Glowna et~al.(2026)Marx-Glowna, Uschmann, L{\"o}tzsch, Hippler,
  G{\"o}rs, Wunderlich, Sadashivaiah, Tolkiehn, Paulus, and
  R{\"o}hlsberger]{marxglowna2026}
B.~Marx-Glowna, I.~Uschmann, R.~L{\"o}tzsch, W.~Hippler, J.~G{\"o}rs,
  T.~Wunderlich, S.~Sadashivaiah, M.~Tolkiehn, G.~G. Paulus, and
  R.~R{\"o}hlsberger.
\newblock High-purity {X}-ray polarimetry at high energies for {M{\"o}ssbauer}
  spectroscopy.
\newblock \emph{Journal of Synchrotron Radiation}, 33\penalty0 (6):\penalty0
  1751--1758, 2026.
\newblock \doi{10.1107/S1600577526008702}.

\bibitem[Guigay and {Sanchez del Rio}(2024)]{Guigay2024}
J.-P. Guigay and M.~{Sanchez del Rio}.
\newblock Formulation of perfect-crystal diffraction from {Takagi--Taupin}
  equations: numerical implementation in the {crystalpy} library.
\newblock \emph{Journal of Synchrotron Radiation}, 31\penalty0 (6):\penalty0
  1469--1480, 2024.
\newblock \doi{10.1107/S160057752400924X}.

\bibitem[Schulze(2018)]{schulze2018}
K.~S. Schulze.
\newblock Fundamental limitations of the polarization purity of {X} rays.
\newblock \emph{APL Photonics}, 3\penalty0 (12):\penalty0 126106, 2018.
\newblock \doi{10.1063/1.5061807}.

\bibitem[Yu et~al.(2025)Yu, Zhang, Ji, Chi, Wee, Breese, and {Sanchez del
  Rio}]{yu2025}
X.~Yu, L.~Zhang, L.~Ji, X.~Chi, A.~T.~S. Wee, M.~B.~H. Breese, and M.~{Sanchez
  del Rio}.
\newblock High accuracy simulations of extreme polarization purity of an
  {X}-ray beam for {B}rewster crystal diffraction.
\newblock \emph{Journal of Physics: Conference Series}, 3010:\penalty0 012068,
  2025.
\newblock \doi{10.1088/1742-6596/3010/1/012068}.

\bibitem[{Sanchez del Rio} et~al.(2026){Sanchez del Rio}, Detlefs, Fiole,
  Francoual, Rebuffi, Reyes-Herrera, and Yu]{SanchezdelRio2026efields}
M.~{Sanchez del Rio}, C.~Detlefs, D.~Fiole, S.~Francoual, L.~Rebuffi,
  J.~Reyes-Herrera, and X.~Yu.
\newblock Ray tracing of polarized light in {SHADOW4}: application to
  polarizers and phase shifters for {X}-rays and visible light.
\newblock In preparation, 2026.

\bibitem[Hirano et~al.(1993)Hirano, Ishikawa, and Kikuta]{hirano1993nim}
K.~Hirano, T.~Ishikawa, and S.~Kikuta.
\newblock Perfect crystal {X}-ray phase retarders.
\newblock \emph{Nuclear Instruments and Methods in Physics Research Section A},
  336:\penalty0 343--353, 1993.
\newblock \doi{10.1016/0168-9002(93)91119-8}.

\bibitem[Giles et~al.(1995)Giles, Vettier, de~Bergevin, Malgrange, Gr{\"u}bel,
  and Grossi]{giles1995rsi}
C.~Giles, C.~Vettier, F.~de~Bergevin, C.~Malgrange, G.~Gr{\"u}bel, and
  F.~Grossi.
\newblock X-ray polarimetry with phase plates.
\newblock \emph{Review of Scientific Instruments}, 66\penalty0 (2):\penalty0
  1518--1521, 1995.
\newblock \doi{10.1063/1.1145896}.

\bibitem[Bouchenoire et~al.(2012)Bouchenoire, Morris, and
  Hase]{bouchenoire2012}
L.~Bouchenoire, R.~J.~H. Morris, and T.~P.~A. Hase.
\newblock A silicon $\langle111\rangle$ phase retarder for producing circularly
  polarized {X}-rays in the 2.1--3~ke{V} energy range.
\newblock \emph{Applied Physics Letters}, 101:\penalty0 064107, 2012.
\newblock \doi{10.1063/1.4740080}.

\bibitem[Scagnoli et~al.(2009)Scagnoli, Mazzoli, Detlefs, Bernard, Fondacaro,
  Paolasini, Fabrizi, and de~Bergevin]{scagnoli2009}
V.~Scagnoli, C.~Mazzoli, C.~Detlefs, P.~Bernard, A.~Fondacaro, L.~Paolasini,
  F.~Fabrizi, and F.~de~Bergevin.
\newblock Linear polarization scans for resonant {X}-ray diffraction with a
  double-phase-plate configuration.
\newblock \emph{Journal of Synchrotron Radiation}, 16\penalty0 (6):\penalty0
  778--787, 2009.
\newblock \doi{10.1107/S0909049509035006}.

\bibitem[Boada et~al.(2010)Boada, Laguna-Marco, Gallastegui, Castro, and
  Chaboy]{boada2010}
R.~Boada, M.~A. Laguna-Marco, J.~A. Gallastegui, G.~R. Castro, and J.~Chaboy.
\newblock X-ray magnetic circular dichroism measurements using an {X}-ray phase
  retarder on the {BM25 A-SpLine} beamline at the {ESRF}.
\newblock \emph{Journal of Synchrotron Radiation}, 17\penalty0 (3):\penalty0
  308--313, 2010.
\newblock \doi{10.1107/S0909049510005881}.

\end{thebibliography}

\end{document}